\documentclass[preprint,amsmath,amssymb,aps,superscriptaddress]{revtex4-1}
\usepackage{graphicx}
\usepackage{dcolumn}
\usepackage{bm}
\usepackage{caption}
\usepackage{subcaption}
\usepackage{multirow}
\usepackage{amsmath}

\begin{document}

\preprint{APS/123-QED}

\title{Comparison of the LBE and DUGKS  methods for DNS of decaying homogeneous isotropic turbulence}
\author{Peng Wang}
\affiliation{State Key Laboratory of Coal Combustion, Huazhong University of Science and Technology, Wuhan 430074, China}
\author{Lian-Ping Wang}
\affiliation{State Key Laboratory of Coal Combustion, Huazhong University of Science and Technology, Wuhan 430074, China}
\affiliation{Department of Mechanical Engineering, University of Delaware, Newark, DE 19716, USA}
\author{Zhaoli Guo}%
\email{zlguo@hust.edu.cn}
\affiliation{State Key Laboratory of Coal Combustion, Huazhong University of Science and Technology, Wuhan 430074, China}
\date{\today}
\begin{abstract}
The main objective of this work is to perform a detailed comparison of the lattice Boltzmann equation (LBE) and the recently developed discrete unified gas-kinetic scheme (DUGKS) methods for direct numerical simulation (DNS) of the decaying homogeneous isotropic turbulence (DHIT) in a periodic box. The flow fields and key statistical quantities computed by both methods are compared with those from pseudo-spectral (PS) method. The results show that the LBE and DUGKS have almost the same accuracy when the flow field is well-resolved, and that the LBE is less dissipative and is slightly more efficient than the DUGKS, but the latter has a superior numerical stability, particularly for high Reynolds number flows. Therefore, the DUGKS method can be viewed as a viable tool for DNS of turbulent flows. It should be emphasized that the main advantage of the DUGKS when compared with the LBE method is its feasibility in adopting non-uniform meshes, which is critical for wall-bounded turbulent flows. The present work provides a basis for further applications of DUGKS in studying the physics of the turbulent flows.
\begin{description}
\item[PACS numbers]
47.11.St, 47.45.-n, 47.61.-k
\end{description}
\end{abstract}

\maketitle


\section{Introduction}

In the study of turbulent flows, the ultimate objective is to obtain accurate coarse-grained quantitative theories or models. However, experience over more than a century has shown it to be notoriously difficult \cite{pope2000turbulent}. Fortunately, the ever-increasing power of computers makes it possible to calculate relevant properties of turbulent flows by direct numerical simulation (DNS). Significant insight into turbulence physics has been gained from the DNS of some idealized flows that cannot be easily obtained in the laboratory \cite{le1997direct,moin1998direct,moser1999direct}. The conventional DNS is based on the Navier-Stokes equations (NSEs), which are a set of second-order nonlinear partial-differential equations (PDE). However it is usually involute and computationally expensive to deal with the nonlinear and non-local convection term and pressure-gradient term in the NSEs \cite{pope2000turbulent}. Therefore, it is desirable to find an alternative numerical method for DNS which not only can accurately capture all the scales of turbulence, but is simpler and more efficient.

Recently, Boltzmann equation based kinetic schemes have received particular attentions as alternative solvers to the NSEs due to some distinctive features. Different from the NSEs, the Boltzmann equation is a first-order linear PDE, and the nonlinearity locally resides in its collision term; both make such schemes to be easily realized and parallelized to have a high computational efficiency. It has been argued that the kinetic equation with local nonlinearity is more feasible to handle the discontinuities or unresolved flow regions \cite{van2001computational}.
Furthermore, the Boltzmann equation provides a theoretical foundation for the hydrodynamic description from the underlying microscopic physics, and describes the phenomenon of fluid flows in the statistical mechanics framework. This physical mechanism is inherently consistent with the physical process of the turbulent flows which are characterized by its statistical behavior \cite{xu2009lattice}. Therefore, the kinetic schemes based on the Boltzmann equation have a great potential for DNS of turbulent flows  \cite{chen2003extended}.

In recent years,  some kinetic schemes have been utilized to simulate turbulent flows, such as the lattice Boltzmann equation (LBE) methods \cite{chen1992recovery,martinez1994comparison,hou1996lattice,eggels1996direct,amati1997massively,d2002multiple,yu2005dns,yu2005lattice,yu2006turbulent,chikatamarla2010lattice,peng2010comparison,gao2013lattice,bosch2015entropic,wang2015lattice} and the gas kinetic schemes \cite{liao2008comparative,liao2009agas,liao2009bgas,kumar2013weno,righi2014gas}. Particularly, the LBE methods have been successfully applied to complex and multiscale flows due to its simplicity in formulation and versatility \cite{chen1998lattice,succi2001lattice,aidun2010lattice,guoadvances}. The potential of the LBE methods for DNS of the turbulent flows has been demonstrated shortly after its emergence by comparing with pseudo-spectral (PS) simulations of the decaying homogeneous isotropic turbulence (DHIT) \cite{chen1992recovery,martinez1994comparison} and turbulence shear flows \cite{hou1996lattice,eggels1996direct}. An appealing feature of the LBE methods in turbulence simulations, as a scheme of second-order spatial accuracy, is that it has very low numerical dissipation compared to the second-order conventional Computational Fluids Dynamics (CFD) methods \cite{luo2010lattice}. It has been demonstrated that the larger numerical dissipation in second-order accurate conventional CFD translates into the greater resolution requirements \cite{moin1998direct}.

Recently, starting from the Boltzmann equation, a discrete unified gas-kinetic scheme (DUGKS) has been proposed for flows in all Knudsen regimes \cite{guo2013discrete,guo2015discrete}. Although sharing a common kinetic origin, there are some distinctive differences between DUGKS and LBE methods. In the standard LBE, the phase space and time step are coupled due to the particle motion from one node to another one within a time step \cite{guoadvances}, but the DUGKS has no such a restriction and the time step is independently determined by Courant-Friedrichs-Lewy (CFL) condition \cite{guo2013discrete}. In addition, the streaming process in LBE makes it difficult to be extended to non-uniform mesh, while the DUGKS can use arbitrary meshes \cite{zhu2015discrete}. Although some efforts have been made to release the close coupling between the mesh and discrete velocities \cite{cao1997physical,guo2003explicit,peng1999finite,rossi2005unstructured,ubertini2005recent,ubertini2008generalised,lee2001characteristic}, the decoupling also destroys the nice features of the standard LBE. For example, many of the existing finite volume (FV) LBE methods suffer from large numerical dissipation and poor numerical stability \cite{ubertini2005recent,ubertini2008generalised}. More importantly, there are modeling difference in LBE and DUGKS in the treatment of particle evolution. In the LBE, the particle streaming and collision processes are splitted. But, these two processes are fully coupled in DUGKS. It has been demonstrated that such a strategy ensures a low numerical dissipation feature \cite{ohwada2002construction,chen2015comparative}. These dynamic differences between the LBE and DUGKS methods determine the quality of solution in flow simulations. A comparative study of the LBE and DUGKS methods for laminar flows in the nearly incompressible limit has been made recently \cite{wang2015comparative}, which demonstrates that the DUGKS has the same accuracy as the LBE, but exhibits a superior numerical stability.  The superiority of the DUGKS compared to the LBE methods for laminar flows motivates us to make a  further comparative study of DUGKS and LBE methods for turbulent flows.

Our long term goal concentrates on providing some insights into the physics of complex turbulent flows by using DUGKS as a DNS tool. At a first step, the validation of the DUGKS for simulating simple turbulent flows must be undertaken. The DHIT is one of such basic flows in turbulence study, and also a canonical case to validate a numerical scheme for DNS of turbulent flows. The objective of this work is to make a detailed comparison of the LBE and DUGKS methods by simulating the DHIT in a periodic box.
To date, the pseudo-spectral (PS) method is well-established as the most accurate numerical tool for DNS of the DHIT. Therefore the DUGKS numerical results will be validated against those from the pseudo-spectral (PS) method. In addition, we use the LBE with the multiple relaxation time (MRT-LBE) collision model in this work due to its superiority to the single relaxation collision model \cite{wang2015comparative}.  The comparative study covers the following aspects of the simulated flows: $(\romannumeral 1)$ the instantaneous velocity and vorticity fields; $(\romannumeral 2)$ the evolutions of kinetic energy and dissipation rate; $(\romannumeral 3)$ the energy and the dissipation rate spectra; $(\romannumeral 4)$ the evolutions of the Kolmogorov length scale and the Taylor microscale length; and $(\romannumeral 5)$  the evolutions of the averaged velocity-derivative skewness and flatness.

The remainder of this paper is organized as follows: in Sec.~\ref{numerical_method}, we provide a brief introduction of the DUGKS and MRT-LBE methods; Sec.~\ref{DHIT} introduces the DHIT, and the quantities to be computed; Sec.~\ref{numerical} presents the numerical results followed by a summary of conclusions.

\section{Numerical Methods}
\label{numerical_method}
In this section, the essentials of  DUGKS and MRT-LBE will be introduced briefly first.
A more detailed description can be found in the references \cite{d2002multiple,guo2013discrete}.

\subsection{The DUGKS method}
The DUGKS is based on the BGK collision model \cite{bhatnagar1954model}, which begins with the model Boltzmann equation,
\begin{equation}\label{BGK}
 \frac{\partial f}{\partial t}+{\bm \xi}\cdot\nabla_x f=\Omega\equiv\frac{f^{eq}-f}{\tau},
\end{equation}
where $f=f(\bm{x},\bm{\xi},t)$ is the particle distribution function with particle velocity $\bm{\xi}$ at position $\bm{x}$ and time $t$,
and $f^{eq}$ is the Maxiwellian equilibrium distribution function,
\begin{equation}\label{MAX}
f^{eq}=\frac{\rho}{{(2\pi R T)}^{D/2}}\exp\left(-\frac{({\bm \xi}-{\bm u})^{2}}{2RT}\right),
\end{equation}
where $R$ is the gas constant, $D$ is the spatial dimension, $\rho$ is the density, $\bm{u}$ is the fluid velocity, and $T$ is the temperature.
 It should be noted that the dimensions of $f$ and $f^{eq}$ are both $kg/\left[m^D\cdot (m/s)^D\right]$.
For incompressible flow (i.e., when the Mach number $\text{Ma}$ is small), the Maxwellian distribution can be approximated by its Taylor expansion around zero particle velocity. As a result, the expanded equilibrium distribution function becomes
 \begin{equation}\label{equilibrium}
{f^{eq}}=\frac{\rho}{{(2\pi RT)}^{D/2}}\exp\left(-\frac{{\mid{\bm \xi}\mid}^{2}}{2RT}\right)\left[1+\frac{\bm{\xi}\cdot\bm{u}}{RT}+ \frac{(\bm{\xi} \cdot\bm{u})^2}{2(RT)^2}-\frac{\mid\bm{u}\mid^2}{2RT} \right].
\end{equation}
In order to obtain the correct NSEs in the limit of low Mach number, the discrete velocity set should be chosen so that the following quadratures of the expanded
equilibrium distribution function hold exactly
\begin{equation}\label{EDF}
\int \bm{\xi}^k f^{eq}d\bm{\xi}=\sum_i \omega_i \bm{\xi}_i^k f^{eq}(\bm{\xi}_i), \quad 0\leq k \leq3
\end{equation}
where $\omega_i$ and $\bm{\xi}_i$ are the weights and points of the numerical quadrature rule. Based on the formulation of Eq.~\eqref{equilibrium}, it is natural to choose a Guassian quadrature with $\omega_i=W_i(2 \pi RT)^{D/2}\text{exp}\left( \frac{\mid {\xi}_i \mid^2}{2RT}\right)$, in which $W_i$ is the weight
coefficient corresponding to the particle velocity $\bm{\xi}_i$.

In the present study, we use the nineteen velocities in three dimensions, i.e., the D3Q19 model, for both the DUGKS and LBE, where
\begin{equation}
\bm{\xi}_i=
\begin{cases}
         (0,0)                                            & i=0\\
\left(\pm 1, 0, 0\right)c, \left(0, \pm 1, 0\right)c, \left(0, 0, \pm 1\right)c            & i=1-6,\\
\left(\pm 1, \pm 1, 0\right)c, \left(\pm 1, 0,\pm 1\right)c, \left(0,\pm 1, \pm 1 \right)c & i=7-18,
\end{cases}
\end{equation}
where $c=\sqrt{3RT} $, and the corresponding weight coefficients are $W_0=1/3$, $W_{1,...,6}=1/18$ and $W_{7,...,18}=1/36$.

Once the quadrature rule is chosen, we can define a discrete distribution function, $f_i(\bm{x},t)=\omega_i f(\bm{x},\bm{\xi}_i,t)$, which satisfies the following equation
\begin{equation}\label{BGK_discrete}
 \frac{\partial f_i}{\partial t}+{\bm \xi_i}\cdot\nabla_x f_i=\Omega_i\equiv\frac{f_i^{eq}-f_i}{\tau}.
\end{equation}
where $f^{eq}_i=\omega_i f^{eq}(\bm{\xi}_i)$ is the discrete expanded equilibrium distribution function that can be written as
\begin{equation}\label{equilibrium_dis}
f^{eq}_i=W_i\left[\delta\rho+\rho_0\left(\frac{\bm{\xi}_i\cdot\bm{u}}{RT}+ \frac{(\bm{\xi}_i \cdot\bm{u})^2}{2(RT)^2}-\frac{\mid\bm{u}\mid^2}{2RT}\right) \right],
\end{equation}
 where the density has been expressed as $\rho=\delta\rho+\rho_0$, in which $\delta\rho$ is the density fluctuation, $\rho_0$ is the constant mean density of the fluid which is usually set to be 1. It should be emphasized that with the discrete velocity set, the dimensions of $f_i$ and $f^{eq}_i$ are both $kg/m^D$.  Then, the fluid density and velocity can be obtained from the discrete distribution function,
\begin{equation}\label{eq:macro}
  \rho=\rho_0+\delta\rho,\quad \delta\rho = \sum_i f_i, \quad \rho_0\bm u = \sum_i\bm \xi_i f_i
\end{equation}

The DUGKS is a finite-volume scheme in which the computational domain is divided into a set of control volumes.
Then integrating Eq.~\eqref{BGK_discrete} over a control volume $V_j$ centered at $\bm x_j$ from $t_{n}$ to $t_{n+1}$ (the time step $\Delta t=t_{n+1}-t_n$ is assumed to be a constant in the present work),
and using the midpoint rule for the integration of the flux term at the cell boundary and trapezoidal rule for the collision term inside each cell \cite{guo2013discrete}, we can get the evolution equation of DUGKS
\begin{equation}\label{update}
\tilde{f}^{n+1}_{i,j}=\tilde{f}^{+,n}_{i,j}-\frac{\Delta t}{|V_j|}F^{n+1/2}_i,
\end{equation}
where
\begin{equation}\label{microflux}
 F^{n+1/2}_i=\int_{\partial V_j}\left(\bm{\xi}_i \cdot \bm n\right)f_i\left(\bm x,t_{n+1/2}\right)d{\bm S} ,
\end{equation}
is the flux across the cell interface, and
\begin{equation}\label{aux}
\tilde{f}_i=f_i-\frac{\Delta t}{2}\Omega_i,\hspace{2mm} \tilde{f}^+_i=f_i+\frac{\Delta t}{2}\Omega_i.
\end{equation}
Based on the compatibility condition and the relationship between $f_i$ and $\tilde{f}_i$, the density $\rho$ and velocity $\bm u$ can be computed by
\begin{equation}
  \rho=\rho_0+\delta\rho,\quad \delta\rho = \sum_i \tilde{f}_i, \quad \rho_0\bm u = \sum_i\bm \xi_i \tilde{f}_i
  \label{eq:macro}.
\end{equation}

The key ingredient in updating $\tilde{f}_i$ is to evaluate the interface flux $F^{n+1/2}_i$,
which is solely determined by the distribution function $f_i(\bm x,t_{n+1/2})$ there.
In DUGKS, after integrating Eq.~\eqref{BGK_discrete} along a particle path within a half time step $(h=\Delta t/2)$,
the evaluation of the distribution function $f_i(\bm x,t_{n+1/2})$ at the cell interface can be traced back to the interior of neighboring cells,
 \begin{equation}\label{BGK_face3}
{\bar{f}_i}({\bm x}_b,t_n+h)={\bar{f}_i}^+\left({\bm x}_b,t_n\right)-h{\bm \xi}_i\cdot{\bm \sigma}_b,
\end{equation}
 where
 \begin{equation}\label{aux2}
 \bar{f_i}=f_i-\frac{h}{2}\Omega_i, \hspace{2mm}{\bar f_i}^+=f_i+\frac{h}{2}\Omega_i,
\end{equation}
$\bar{f_i}^+({\bm x}_b,t_n)$ and the gradient ${{\bm \sigma}}_b=\nabla \bar{f_i}^+({\bm x}_b,t_n) $ can be approximated by linear interpolation. For example, in the one dimensional case, the reconstructions become
\begin{equation}\label{inter1}
\bar{f_i}^+({x}_{j+1/2},t_n)={\bar{f_i}}^+({ x}_{j},t_n)+{\sigma}_{j+1/2}(x_{j+1/2}-x_j),
\end{equation}
where
\begin{equation}\label{inter}
{\sigma}_{j+1/2}=\frac{\bar{f_i}^+({ x}_{j+1},t_n)-\bar{f_i}^+({ x}_{j},t_n)}{x_{j+1}-x_{j}}.
\end{equation}

Note that the particle collision effect from $t_n$ to $t_{n+1}$ is included in the above reconstruction of the interface distribution function.
This is the key for the success of the DUGKS. Owing to the coupled treatment of the particle collision and transport
process in the reconstruction of the distribution function at cell interfaces,
DUGKS is a self-adaptive scheme for different flow regimes. It has been
shown in Ref.~\cite{guo2013discrete} that the reconstructed distribution function reduces to
the Chapman-Enskog one approximation at the Navier-Stokes level in the continuum limit, and
to the free-transport approximation in the free-molecular limit.

Based on the compatibility condition and the relationship between $f_i$ and $\bar{f_i}$,
the density $\rho$ and velocity $\bm u$ at the cell interface can be obtained,
  \begin{equation}
  \rho=\rho_0+\delta\rho,\quad \delta\rho = \sum_i \bar{f}_i, \quad \rho_0\bm u = \sum_i\bm \xi_i \bar{f}_i
  \label{eq:macro}
\end{equation}
from which the equilibrium distribution function $f^{eq}_i\left({\bm x}_b,t^n+h\right)$ at the cell interface can be obtained.
Therefore, based on Eq.~\eqref{aux2} and the obtained equilibrium state,  the real distribution function at the cell interface can be determined from $\bar{f_i}$ as,
\begin{equation}\label{original}
f_i(\bm{x}_b, t_n+h)=\frac{2\tau}{2\tau+h}\bar{f}_i\left(\bm{x}_b,t_n+h\right)+\frac{h}{2\tau+h}f^{eq}_i\left(\bm{x}_b,t_n+h\right),
\end{equation}
from which the interface flux term can be evaluated.

In computation, we only need to follow the evolution of $\tilde{f_i}$ in Eq.~\eqref{update}. The required variables for its evolution are determined by \cite{guo2013discrete}
\begin{equation}\label{relationv1}
\bar{f_i}^+=\frac{2\tau-h}{2\tau+\Delta t}\tilde{f_i}+\frac{3h}{2\tau+\Delta t}f^{eq}_i,
\end{equation}
\begin{equation}\label{relationv3}
\tilde{f_i}^+=\frac{4}{3}\bar{f_i}^+-\frac{1}{3}\tilde{f_i}.
\end{equation}

\subsection{The MRT-LBE method}

  In this work, we use the LBE with multiple-relaxation time  collision model (MRT-LBE) and the D3Q19 discrete velocity sets. The evolution equation of the MRT-LBE is
\begin{equation}
  {\bf f}({\bm x} + {\bf \xi}_i\Delta t, t_n+\Delta t)={\bf f}(\bm{x}, t_n)-{\bf M}^{-1}{\mathbf S}\left[{\bf m}({\bm x},t)-{\bf m}^{eq}({\bm x},t)\right],
  \label{eq:lbgk}
\end{equation}
 where ${\bf M}$ is an orthogonal transformation matrix converting the distribution function ${\bf f}$ from discrete velocity space to the moment space ${\bf m}$, in which the collision relaxation is performed.

The basic idea of MRT-LBE is that the streaming sub-step is handled in the microscopic lattice-velocity space but
the collision sub-step is performed in the moment space. The transformation between the microscopic velocity space and the moment space is carried out by matrix operations as  ${\bf m} = {\bf M} \cdot {\bf f},  \  {\bf f} = {\bf M^{-1}} \cdot {\bf m}$. The diagonal relaxation matrix ${\bf S}$ specifies the relaxation rates for the
non-conserved moments.

The macroscopic hydrodynamic variables, including the density $\rho$ and momentum, are obtained from the moments of the mesoscopic distribution function ${\bf f}$.
In the nearly incompressible formulation~\cite{he1997lattice}
\begin{equation}\label{drho}
\rho=\rho_0+\delta\rho,  \ \ \rho_0 = 1; \hspace{5mm}\delta\rho=\sum\limits_{i}{f}_i,\hspace{5mm} \rho_0 \bm u=(j_x,j_y,j_z)^T=\sum\limits_{i}\bm{\xi}_i{f}_i.
\end{equation}

For the D3Q19 velocity model, the corresponding 19 orthogonal moments
\begin{equation*}
{\bf m} = \left( \delta{\rho}, e, \varepsilon, j_x, q_x, j_y, q_y, j_z, q_z, 3p_{xx}, 3\pi_{xx}, p_{ww},
\pi_{ww}, p_{xy}, p_{yz}, p_{xz}, m_x, m_y, m_z \right)^T
\end{equation*}
are defined through the element of the transformation matrix (each subscript runs from 0 to 18) as
\begin{eqnarray*}
&& M_{0,\alpha}=||{\bf \xi}_{\alpha}||^0, \  \  M_{1,\alpha}=19 ||{\bf \xi}_{\alpha}||^2 - 30 , \   \
M_{2,\alpha}= \left(  21 ||{\bf \xi}_{\alpha}||^4 - 53||{\bf \xi}_{\alpha}||^2  +  24 \right)/2   \nonumber \\
&&    M_{3,\alpha}=\xi_{\alpha x}, \  \  M_{5,\alpha}=\xi_{\alpha y}, \   \
M_{7,\alpha}=\xi_{\alpha z},   \nonumber \\
&& M_{4,\alpha}= \left( 5 ||{\bf \xi}_{\alpha}||^2 - 9 \right) \xi_{\alpha x}, \  \
M_{6,\alpha}= \left( 5 ||{\bf \xi}_{\alpha}||^2 - 9 \right) \xi_{\alpha y}, \  \
M_{8,\alpha}= \left( 5 ||{\bf \xi}_{\alpha}||^2 - 9 \right) \xi_{\alpha z},  \\
&&  M_{9,\alpha}=3 \xi^2_{\alpha x} -  ||{\bf \xi}_{\alpha}||^2, \  \
M_{11,\alpha}= \xi^2_{\alpha y} -  \xi^2_{\alpha z},  \\
&&  M_{13,\alpha}= \xi_{\alpha x}\xi_{\alpha y}, \  \
M_{14,\alpha}= \xi_{\alpha y}\xi_{\alpha z}, \  \
M_{15,\alpha}= \xi_{\alpha x}\xi_{\alpha z},  \\
&&  M_{10,\alpha}= \left( 3 ||{\bf \xi}_{\alpha}||^2 - 5  \right)  \left(   3 \xi^2_{\alpha x} - ||{\bf \xi}_{\alpha}||^2 \right)  , \  \
M_{12,\alpha}= \left( 3 ||{\bf \xi}_{\alpha}||^2 - 5  \right)  \left(    \xi^2_{\alpha y} - \xi^2_{\alpha z}  \right) ,  \\
&&  M_{16,\alpha}=\left(    \xi^2_{\alpha y} - \xi^2_{\alpha z}  \right) \xi_{\alpha x}  , \  \
M_{17,\alpha}= \left(    \xi^2_{\alpha z} - \xi^2_{\alpha x}  \right) \xi_{\alpha y}  , \  \
M_{18,\alpha}= \left(    \xi^2_{\alpha x} - \xi^2_{\alpha y}  \right) \xi_{\alpha z}    .
\end{eqnarray*}
The equilibrium moments are defined as
\begin{eqnarray*}
&&  \widetilde{\rho}^{(eq)} =  \widetilde{\rho} =  \delta \rho, \  \
e^{(eq)} = -11 \delta \rho + {19\over \rho_0} \left( j_x^2 +j_y^2  + j_z^2 \right), \  \
 \varepsilon^{(eq)} =\omega_{\varepsilon} \delta \rho + { \omega_{\varepsilon j } \over \rho_0} \left( j_x^2 +j_y^2  + j_z^2 \right),    \nonumber \\
 &&  j_x^{(eq)} = j_x=  \rho_0u_x, \  \
j_y^{(eq)} = j_y=  \rho_0u_y, \  \
j_z^{(eq)} = j_z=  \rho_0u_z,    \nonumber \\
&&  q_x^{(eq)} = -{2\over 3} j_x, \ \
q_y^{(eq)} = -{2\over 3} j_y, \  \
q_z^{(eq)} = -{2\over 3} j_z,  \\
&&  p_{xx}^{(eq)} = {1\over 3\rho_0} \left[2j_x^2 - \left( j_y^2 + j_z^2 \right) \right] , \ \
p_{ww}^{(eq)} = {1\over \rho_0} \left[j_y^2 - j_z^2 \right] ,  \\
&&  p_{xy}^{(eq)} = {1\over \rho_0} j_x j_y , \ \
p_{yz}^{(eq)} = {1\over \rho_0} j_y j_z , \ \
p_{xz}^{(eq)} = {1\over \rho_0} j_x j_z ,   \\
&&  \pi_{xx}^{(eq)} = \omega_{xx}p_{xx}^{(eq)}  , \ \
\pi_{ww}^{(eq)} =  \omega_{xx}p_{ww}^{(eq)} ,  \\
&& m_x^{(eq)} = m_y^{(eq)} = m_z^{(eq)} = 0,
\end{eqnarray*}
with the following relaxation parameters
\begin{equation*}
{\bf S} = {\rm diag} \left( 0, s_1, s_2, 0, s_4, 0, s_4, 0, s_4, s_9, s_{10}, s_9, s_{10}, s_{13}, s_{13},
s_{13}, s_{16}, s_{16},s_{16} \right).
\end{equation*}
The kinematic viscosity $\nu$ and bulk viscosity $\zeta$ are related to the relaxation rates $s_{9}$ and $s_1$, respectively, where
\begin{equation}
\nu=\frac{1}{3}\left(\frac{1}{s_{9}}-\frac{1}{2}\right)c\Delta x,
\end{equation}
\begin{equation}
\zeta=\frac{5-9c_s^2}{9}\left(\frac{1}{s_{1}}-\frac{1}{2}\right)c\Delta x,
\end{equation}
where $c_s^2=RT$ is the speed of sound.

It is noted that some of the relaxation parameters do not affect the simulated flow, but may affect the numerical stability of the code. Specifically, $s_1$ determines the bulk viscosity which could absorb
low-amplitude acoustic oscillations.

\section{Decaying Homogeneous Isotropic Turbulence}
\label{DHIT}
The DHIT in a three-dimensional box with periodic boundary conditions in all three directions is a standard test case to validate numerical scheme for DNS. At the initial time, a random flow field is introduced with the kinetic energy contained only in the large eddies ( i.e., at low wave numbers). This initial flow is unstable and large eddies will break up, transferring their energy successively to  smaller and smaller eddies with high wave numbers until the eddy scale is sufficiently small, in which the eddy motions are stable and the viscosity is effective in dissipating the kinetic energy. After some time, a realistic DHIT will develop with some larger eddies supply kinetic energy for smaller eddies and the viscous action controls the size of the small eddies.

In the present work, the incompressible initial velocity field  $\bm{u}_0$ $(\nabla \cdot\bm{u}_0=0 )$ is specified by a Gaussian field with a prescribed kinetic energy spectrum \cite{peng2010comparison}:
\begin{equation}\label{Eq:spectrum_inital}
E_{0}(k):=E(k,t=0)=Ak^4e^{-0.14k^2}, \hspace{15mm} k\in[k_{min},k_{max}],
\end{equation}
where $k$ is the wavenumber, the magnitude $A$ and the range of the initial energy spectrum $[k_{min},k_{max}]$ determines the total initial kinetic energy $K_0$ in the simulation. The kinetic energy $K$ and dissipation rate $\epsilon$ are given by

\begin{equation}\label{Eq:ed}
K(t)=\int E(\bm{k},t)d\bm{k}, \hspace{15mm} \epsilon(t)=2\nu\int k^2E(\bm{k},t)d\bm{k},
\end{equation}
where $\nu$ is the kinematic viscosity, and
\begin{equation}\label{Eq:spectrum}
E(\bm{k},t)=\frac{1}{2}\hat{\bm{u}}(\bm{k},t)\hat{\bm{u}}^*(\bm{k},t),
\end{equation}
where  $\hat{\bm{u}}$ and $\hat{\bm{u}}^*$ are velocity and its complex conjugate in the spectral space. The DHIT is typically characterized by the Taylor microscale Reynolds number
\begin{equation}\label{Eq:Re}
Re_{\lambda}=\frac{u'\lambda}{\nu}
\end{equation}
where $u'$ is the root mean squared (rms) value  of the turbulent fluctuating velocity $\bm{u}$ in a given spatial direction and is defined by
\begin{equation}
u'=\frac{1}{\sqrt{3}}\sqrt{\langle\bm{u}\cdot\bm{u}\rangle},
\end{equation}
here $\langle\cdot\rangle$ designates the volume average;  $\lambda$ is the transverse Taylor microscale length
\begin{equation}
\lambda=\sqrt{\frac{15\nu}{\epsilon}}u'.
\end{equation}
The other statistical quantities of interest are as follows:
 \begin{subequations}\label{static}
 \begin{equation}
\eta=\sqrt[4]{{\nu}^3/\epsilon}
 \end{equation}
  \begin{equation}
D(k,t)=2\nu k^2 E(k,t),
\end{equation}
\begin{equation}
S(t)=\frac{\langle(\partial_x u)^3\rangle+\langle(\partial_y v)^3\rangle+\langle(\partial_z w)^3\rangle}{3\left[\langle(\partial_xu)^2\rangle^{3/2}+\langle(\partial_y v)^2\rangle^{3/2}+\langle(\partial_z w)^2\rangle^{3/2}\right]}
\end{equation}
\begin{equation}
F(t)=\frac{\langle(\partial_x u)^4\rangle+\langle(\partial_y v)^4\rangle+\langle(\partial_z w)^4\rangle}{3\left[\langle(\partial_xu)^2\rangle^{2}+\langle(\partial_y v)^2\rangle^{2}+\langle(\partial_z w)^2\rangle^{2}\right]}
\end{equation}
\end{subequations}
 where $\eta$ is the Kolmogorov length and $D(k,t)$ is the energy dissipation rate spectrum; $S(t)$ and $F(t)$ are the velocity-derivative skewness and flatness averaged over three directions, respectively.

\section{Numerical Results}
\label{numerical}
\subsection{Initial conditions}
We perform the simulations of DHIT in a periodic box with the domain size $L^3$ using the LBE, DUGKS and PS methods. The focus is on the comparison of LBE and DUGKS results with those from the PS method which is used as a benchmark due to its superior spatial accuracy. The PS method is same as in Ref.~\cite{peng2010comparison}. The units of LBE and DUGKS are converted back to the spectral units to allow for a direct comparison. The conversion requires a velocity scale $V_s$ which is the ratio of the fluid velocity magnitude in LBE or DUGKS units to the velocity magnitude in spectral units.

In the PS simulation, the domain size is set to be $L^3=(2\pi)^3$; for the initial energy spectrum $E_0(k)$ given by Eq.~\eqref{Eq:spectrum_inital}, we set $A=1.7414\times10^{-2}$, $k_{min}=3$ and $k_{max}=8$  such that the initial kinetic energy is $K_0=0.9241$ and the rms velocity is $u'_0=0.7849$.

In the LBE and DUGKS simulations, we set the domain size $L^3=N^3$, where $N$ is the number of the cells or lattices in each spatial direction. In addition, we must ensure that the local Mach number (Ma) is small enough so that the flow is nearly incompressible, which can be met by choosing a suitable $V_s$. In the simulations, we chose velocity scale $V_s=0.0408$ which leads to the initial kinetic energy $K_0=1.5383\times 10^{-3}$, the corresponding initial rms velocity $u'_0=0.0320$ and maximum velocity magnitude $\|\bm{u}_0\|_{max}=0.1660$ so that the maximum Mach number $ \text{Ma}=\|\bm{u}_0\|_{max}/c_s=0.2875$, here $c_s=\sqrt{RT}, RT=1/3$. The initial velocity field and parameters used in the LBE and DUGKS simulations are identical except the time step size $\Delta t$. In LBE method, the time step size $\Delta t=\Delta x=1$ in LBE units,  while in DUGKS it is solely determined by the CFL condition, i.e., $\Delta t=\gamma \Delta x_{min}/\sqrt{2}c$, where $\gamma$ is the CFL number and $\Delta x_{min}$ is the minimum grid spacing and $\sqrt{2}c$ is the maximum discrete particle speed in D3Q19. In the DUGKS simulations, we set $\gamma=0.7071$ such that the time step $\Delta t=0.5$ for convenient comparison. Moreover, for the MRT-LBE, the specific parameters are set to be $\omega_{\varepsilon}=\omega_{xx}=0 $, $\omega_{\varepsilon j}=-475/63$, $s_2=s_{10}=1.4$,
$s_9=s_{13}={\Delta t}/{(3\nu + 0.5\Delta t)}$, $s_1=1.19, s_4=1.2$, and $s_{16}=1.98$ \cite{d2002multiple}.

\begin{table}[htbp]
\centering
\caption{\label{tablea} Parameters used in the LBE, DUGKS and PS simulations.}
\begin{tabular}[t]
   {c c c c c c c c  c}

   \hline
   \hline

   & method &\hspace{1mm} $L   $ &\hspace{2mm} $N  $ &\hspace{2mm}$  K_0               $  &\hspace{2mm}  $u'_0    $  &\hspace{2mm}  $\nu      $  &\\
   \hline
   & PS128  &\hspace{1mm} $2\pi$ &\hspace{2mm} $128$ &\hspace{2mm}$0.9241              $  &\hspace{2mm}  $0.7849$  &\hspace{2mm}  $1.4933\times{10}^{-2} $&\\
   &LBE128  &\hspace{1mm} $128 $ &\hspace{2mm} $128$ &\hspace{2mm}$1.5383\times{10}^{-3}$  &\hspace{2mm}  $0.0320$  &\hspace{2mm} $1.2395\times{10}^{-2} $&\\
   &DUGKS128&\hspace{1mm} $128 $ &\hspace{2mm} $128$ &\hspace{2mm}$1.5383\times{10}^{-3}$  &\hspace{2mm}  $0.0320$  &\hspace{2mm} $1.2395\times{10}^{-2} $&\\
   & PS256  &\hspace{1mm} $2\pi$ &\hspace{2mm} $256$ &\hspace{2mm}$0.9241              $  &\hspace{2mm}  $0.7849$  &\hspace{2mm}  $1.4933\times{10}^{-2} $  &\\
   &LBE256  &\hspace{1mm} $256 $ &\hspace{2mm} $256$ &\hspace{2mm}$1.5383\times{10}^{-3}$  &\hspace{2mm}  $0.0320$ &\hspace{2mm} $2.4790\times{10}^{-2} $  &\\
   &DUGKS256&\hspace{1mm} $256 $ &\hspace{2mm} $256$ &\hspace{2mm}$1.5383\times{10}^{-3}$  &\hspace{2mm}  $0.0320$ &\hspace{2mm} $2.4790\times{10}^{-2} $  &\\
  \hline
  \hline
\end{tabular}
\end{table}

 Table~\ref{tablea} summarizes the parameters used in the simulations with these three methods.
 Two mesh resolutions are considered in the simulations. In order to fix the initial Taylor microscale Reynolds number $\text{Re}_{\lambda}=26.06$, in the PS simulation we set the kinematic viscosity $\nu=1.4933\times{10}^{-2}$ for both resolutions, while in the LBE and DUGKS simulations, we set the viscosity $\nu=1.2395\times{10}^{-2}$ and $2.4790\times{10}^{-2}$ for the mesh resolutions of $128^3$ and $256^3$, respectively. It should be noted that the flow is over resolved in the PS simulations as the spatial resolution parameter $k_{max}\eta$ is larger than $3.04$ at $128^3$ and $6.16$ at $256^3$. respectively, where $k_{max}$ is the maximum resolved wavenumber \cite{eswaran1988examination}. This implies that the results from the PS simulations at the two grid resolutions would be identical.
The non-dimensional time step size, normalized by the turbulence eddy turnover time $t_0=K_0/\epsilon_0$, is $\Delta t'=\Delta t \epsilon_0/K_0$.

With the initial velocity field $\bm{u}_0$, the initial pressure $p_0$ is obtained by solving the Poisson equation in the spectral space for the PS method. As for the LBE and DUGKS methods,  besides the pressure $p_0$, herein related to the density fluctuation by equation of the state, a consistent initial distribution function including the non-equilibrium part should be specified, which is achieved by using the iterative procedure  described in \cite{mei2006consistent}.
\subsection{Instantaneous velocity and vorticity fields}

 We compare the instantaneous velocity and vorticity magnitude obtained by LBE and DUGKS methods with those from PS simulation on the $xy$ plane at $z=L/2$. The vorticity fields for all three methods are first computed in the spectral space, $\tilde{\omega}=i\bm{k}\times\tilde{\bm u}$, and then  $\tilde{\bm \omega}$ is transferred back to the physical space using inverse fast Fourier translation (FFT).

\begin{figure}[htbp]
\centering
\begin{subfigure}[b]{0.26\textwidth}
\includegraphics[width=\textwidth]{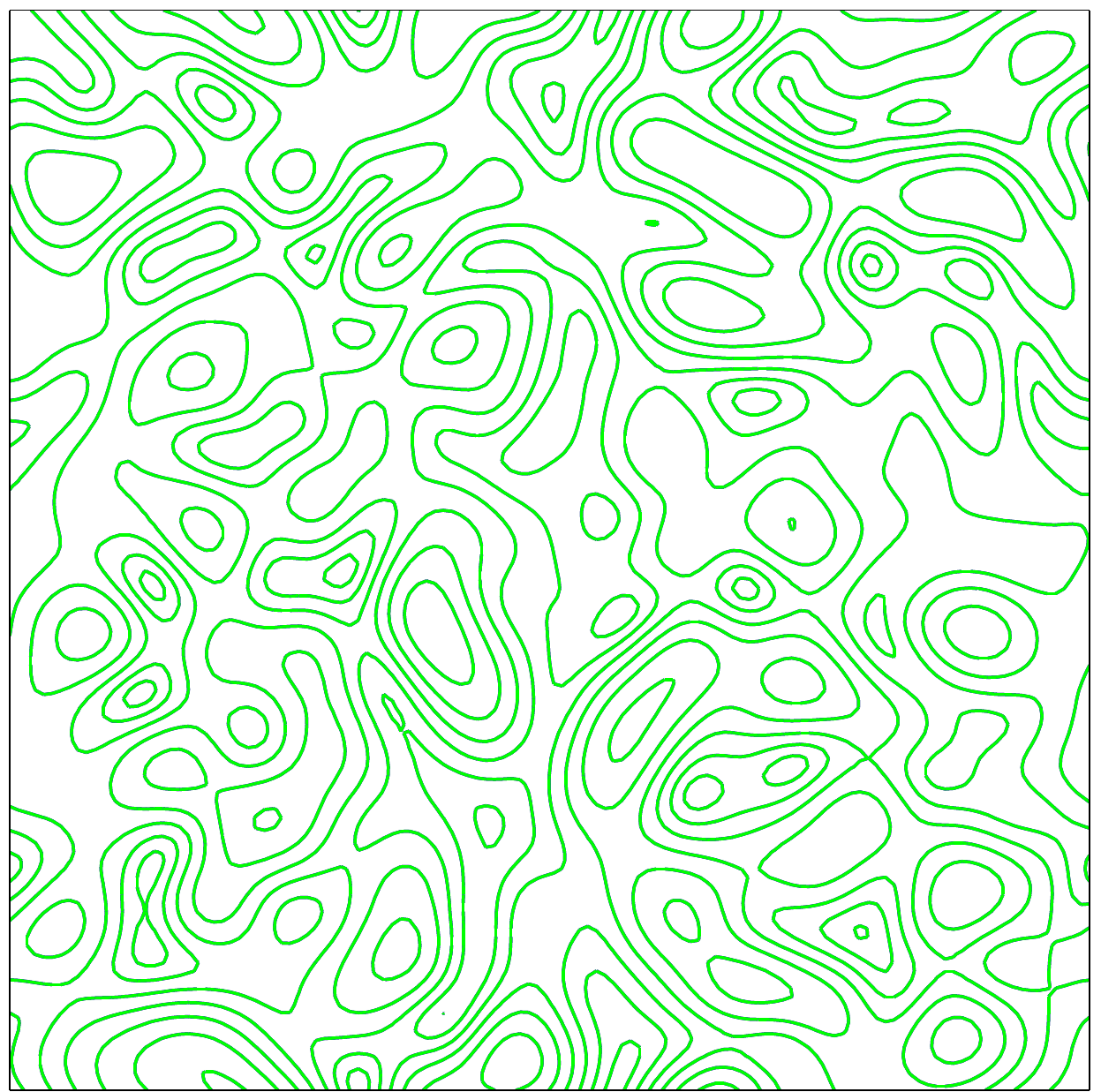}
\caption{$t'=0$}
\label{fig:vel0}
\end{subfigure}~
\begin{subfigure}[b]{0.26\textwidth}
\includegraphics[width=\textwidth]{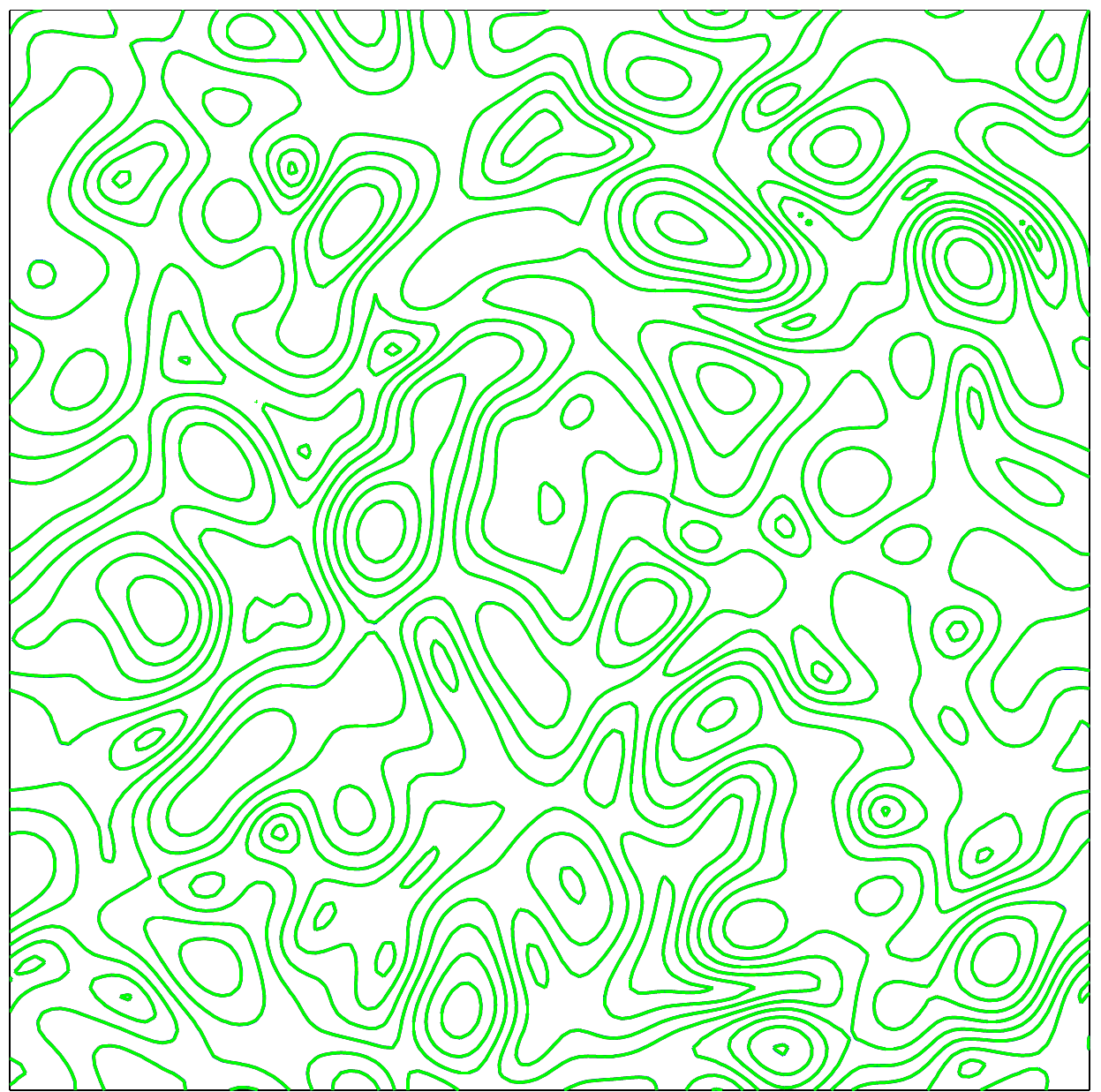}
\caption{$t'=0$}
\label{fig:vor0}
\end{subfigure}\\
\begin{subfigure}[b]{0.26\textwidth}
\includegraphics[width=\textwidth]{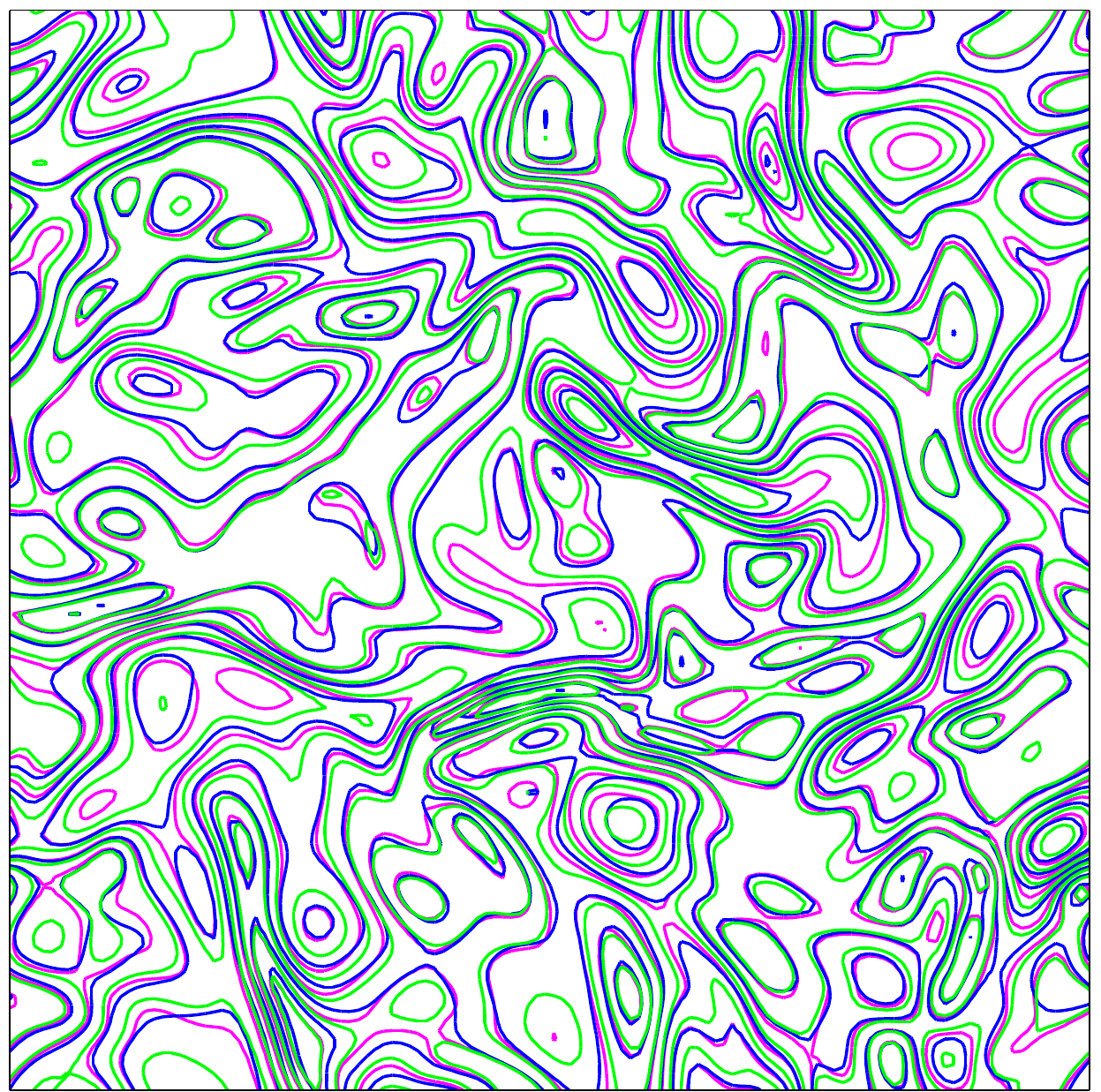}
\caption{$t'=1.21$}
\label{fig:vel02}
\end{subfigure}~
\begin{subfigure}[b]{0.26\textwidth}
\includegraphics[width=\textwidth]{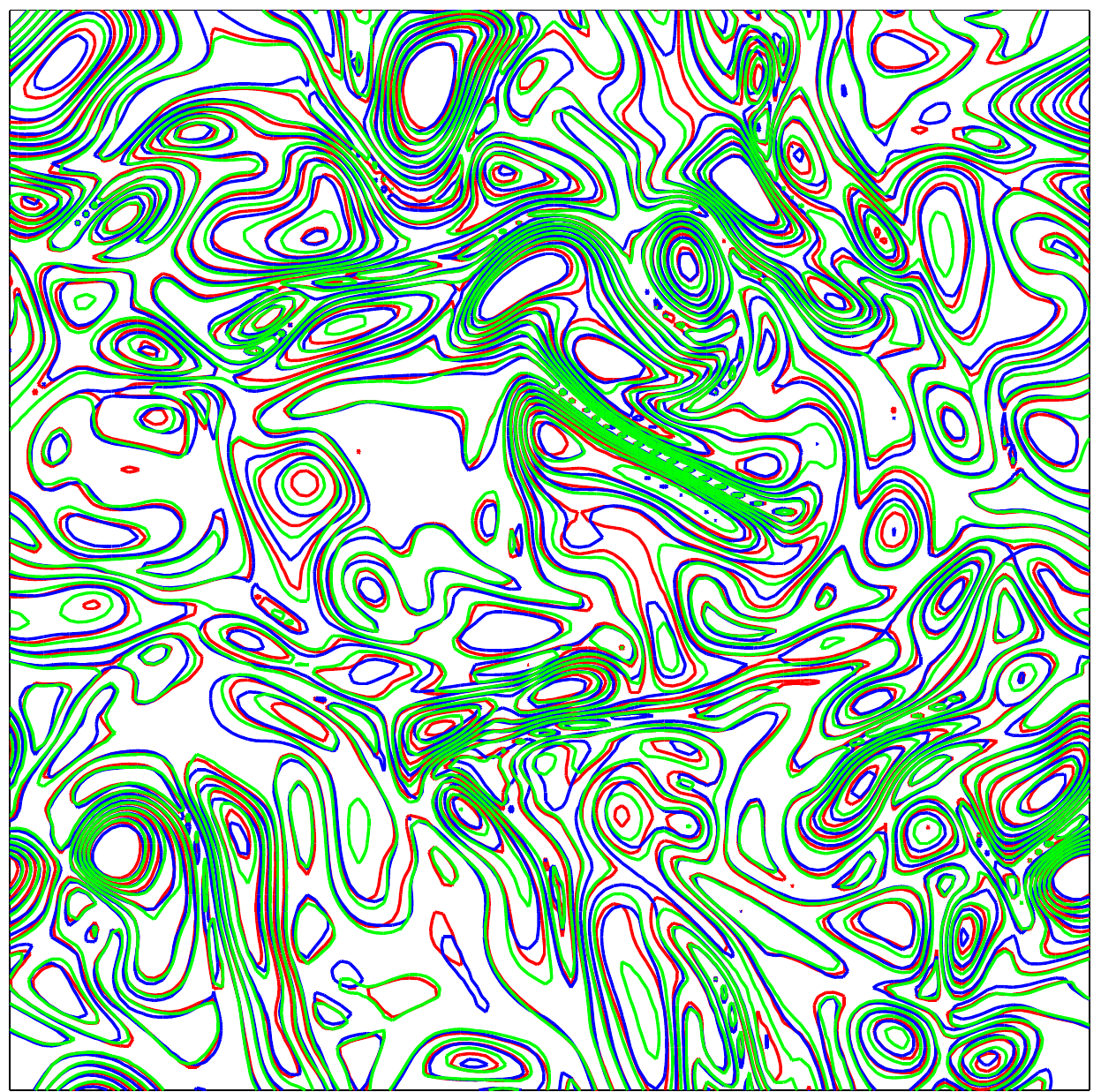}
\caption{$t'=1.21$}
\label{fig:vor02}
\end{subfigure}\\
\begin{subfigure}[b]{0.26\textwidth}
\includegraphics[width=\textwidth]{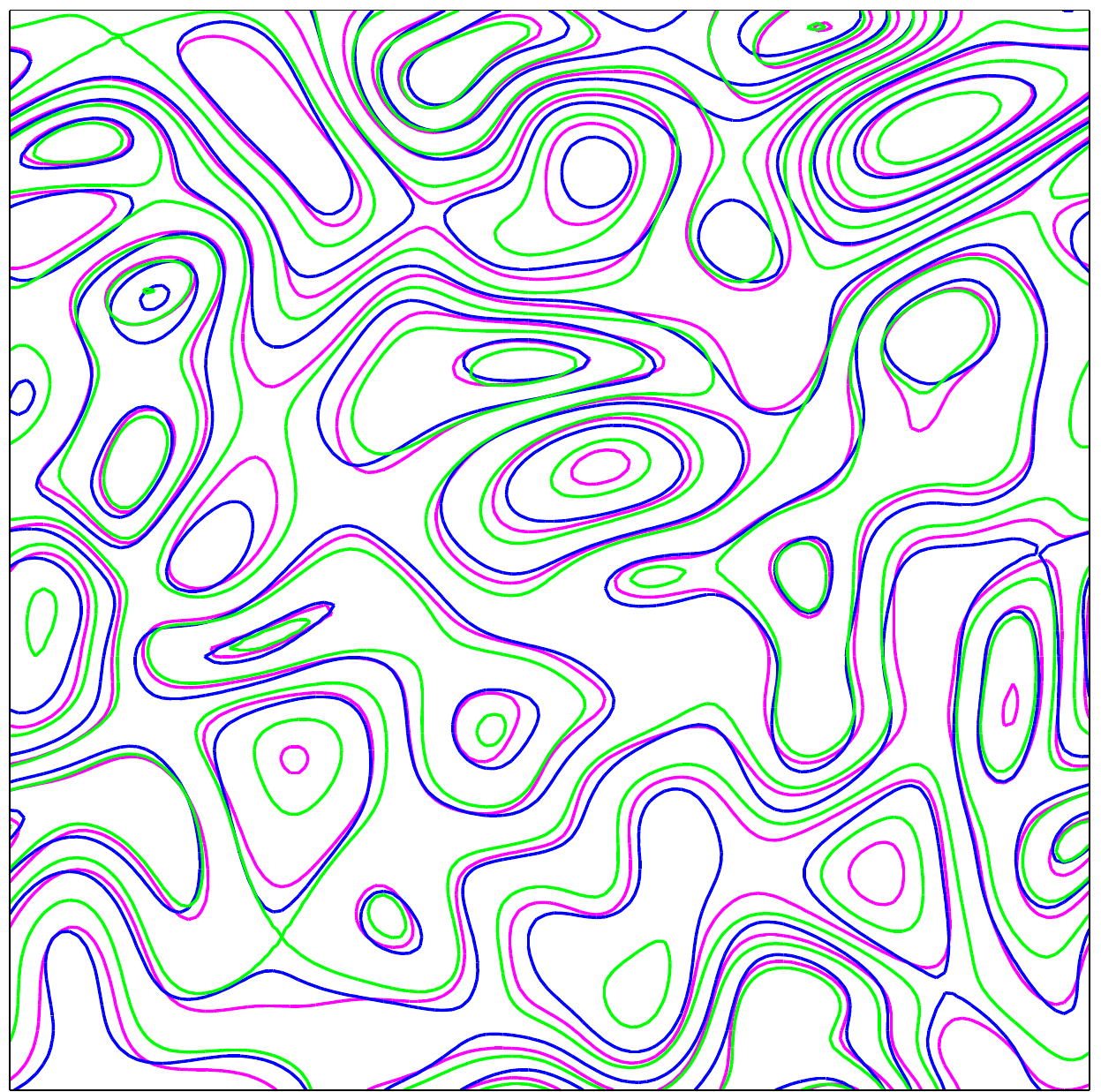}
\caption{$t'=6.08$}
\label{fig:vel10}
\end{subfigure}~
\begin{subfigure}[b]{0.26\textwidth}
\includegraphics[width=\textwidth]{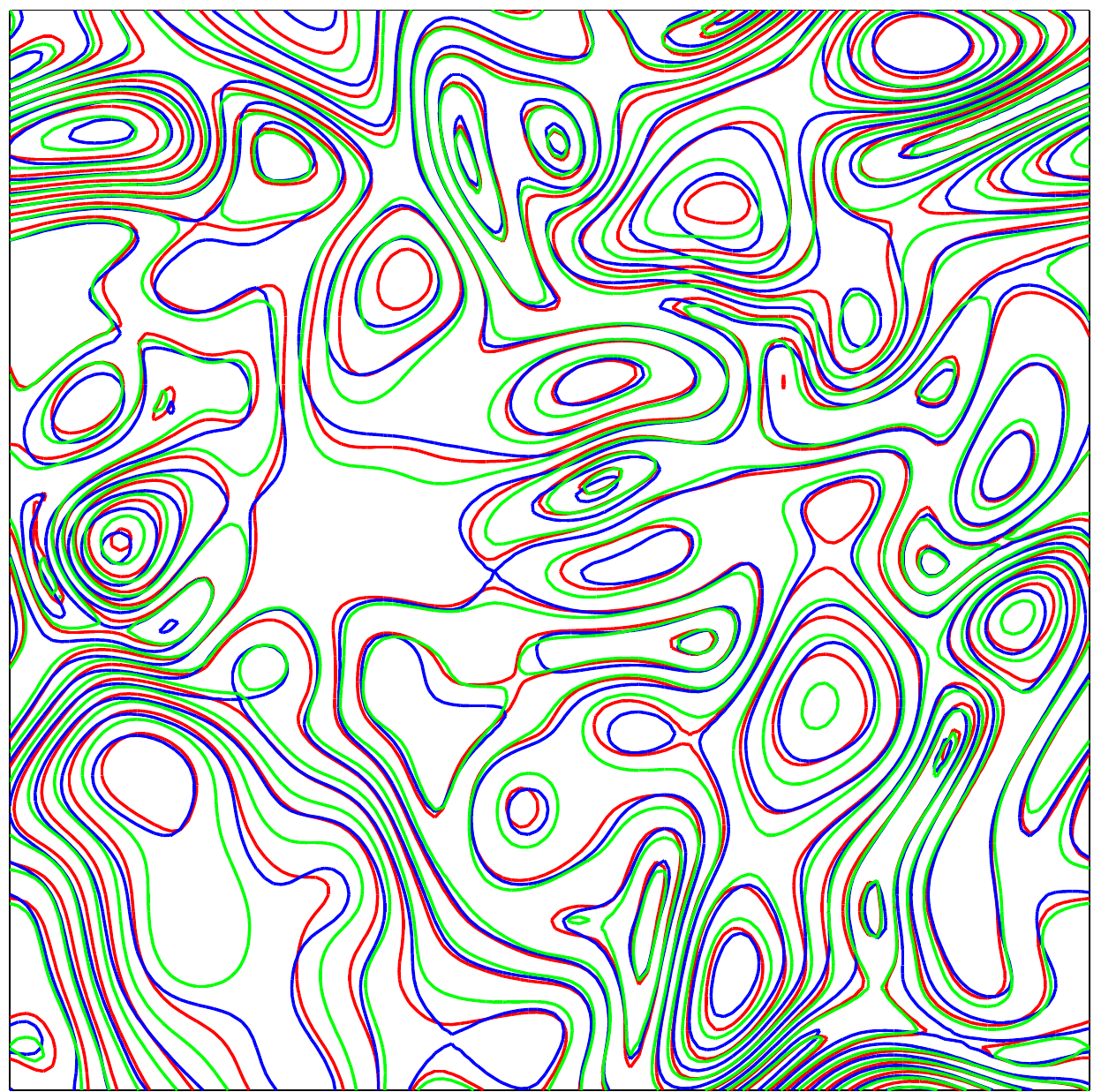}
\caption{$t'=6.08$}
\label{fig:vor10}
\end{subfigure}\\
\begin{subfigure}[b]{0.26\textwidth}
\includegraphics[width=\textwidth]{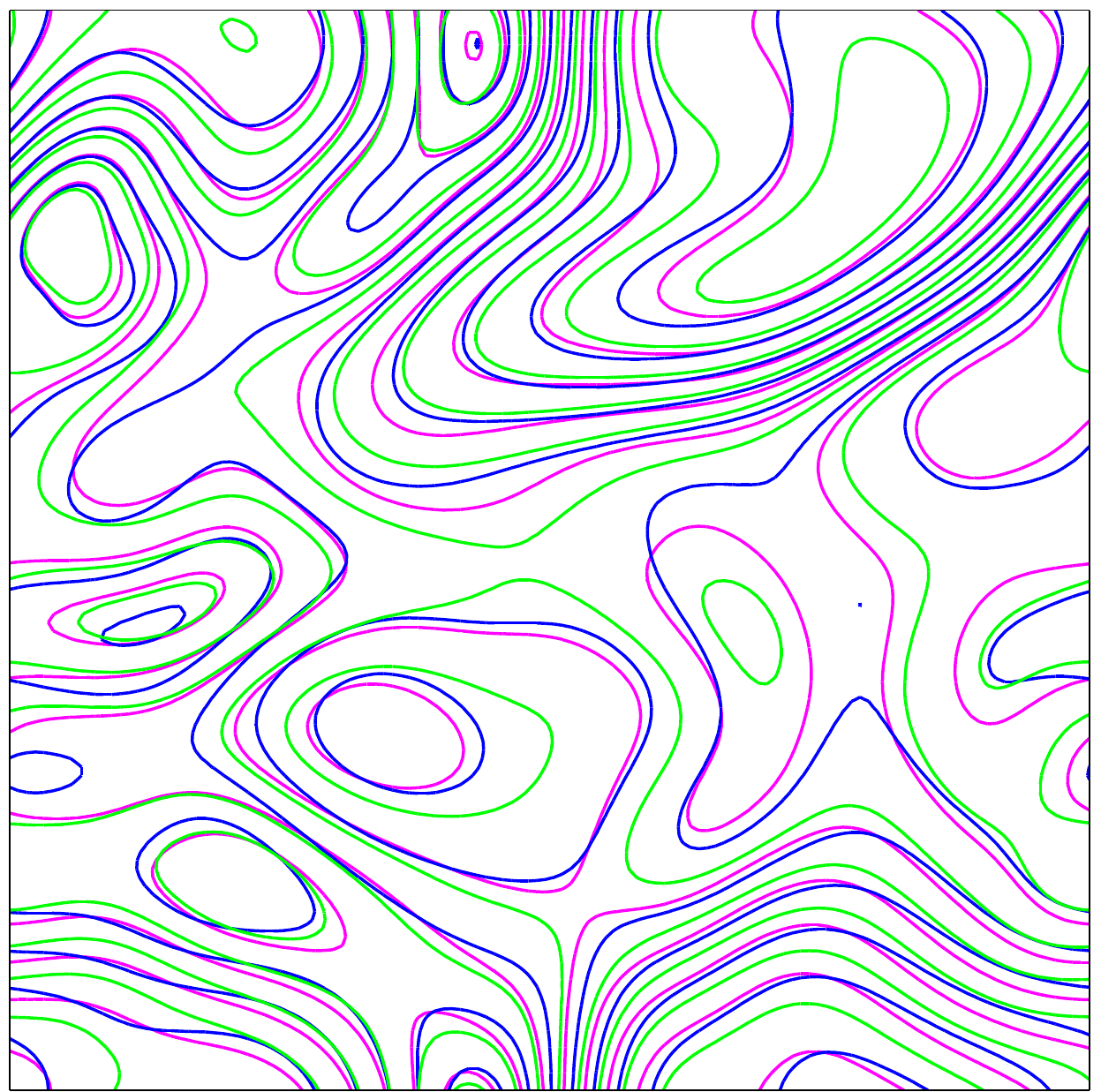}
\caption{$t'$=12.16}
\label{fig:vel20}
\end{subfigure}~
\begin{subfigure}[b]{0.26\textwidth}
\includegraphics[width=\textwidth]{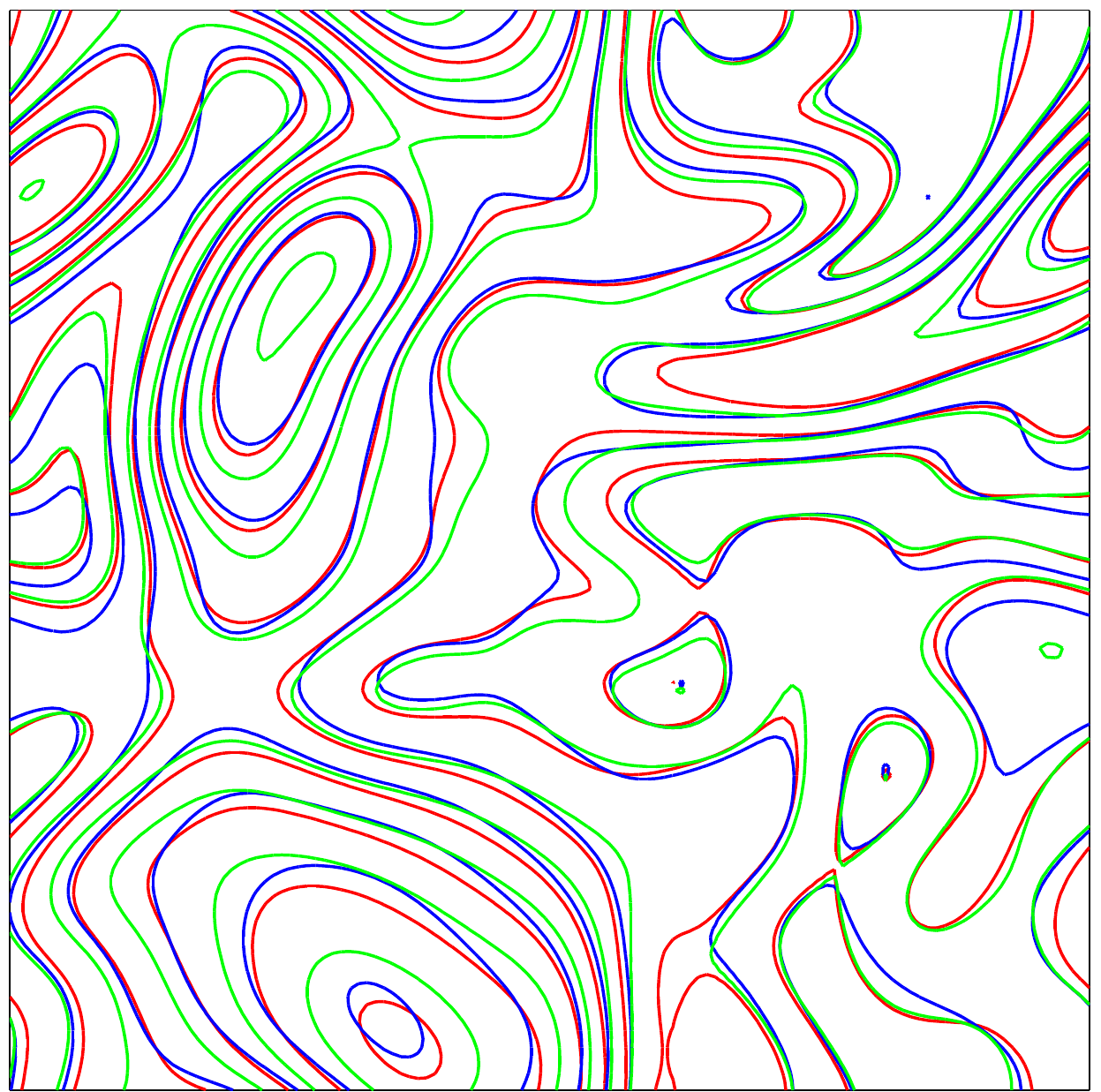}
\caption{$t'$=12.16}
\label{fig:vor20}
\end{subfigure}\\
\caption{
Contours of normalized velocity magnitude $\|\bm{u}\|/u'_0$ (left column) and normalized vorticity magnitude $\|\bm{\omega}\|L/u'_0$ (right column) on the $xy$ plane at $z=L/2$ at time $t'=0, 1.21, 6.08$ and $12.16$ (from top to bottom) with $N^3=128^3$. The solid red, green and blue lines denote results of the PS, LBE and DUGKS, respectively.} \label{fig:vor_vel128}
\end{figure}
 Figure~\ref{fig:vor_vel128} shows the contours of normalized velocity magnitude  $\|\bm{u}\|/u'_0$  and vorticity magnitude $\|\bm{\omega}\|L/u'_0$ at  different non-dimensional times $t'=0, 1.21, 6.08$ and $12.16$  on a mesh of $N^3=128^3$. As shown in Figs.~\ref{fig:vel0} and ~\ref{fig:vor0}, these three methods have the identical initial fields with many large eddies; then small scale eddies are produced by vortex stretching as shown in Figs.~\ref{fig:vel02} and ~\ref{fig:vor02};  in the end, as shown in Figs.~\ref{fig:vel20} and ~\ref{fig:vor20}, the small scale eddies are dissipated by viscous actions. As shown in these figures, although the fields predicted by the LBE and DUGKS methods are similar to each other, and very close to those from the PS simulation in terms of vortex shapes and locations, the discrepancy between the both kinetic methods and the PS method is still visible and increases over time.

\begin{figure}[htbp]
\centering
\begin{subfigure}[b]{0.26\textwidth}
\includegraphics[width=\textwidth]{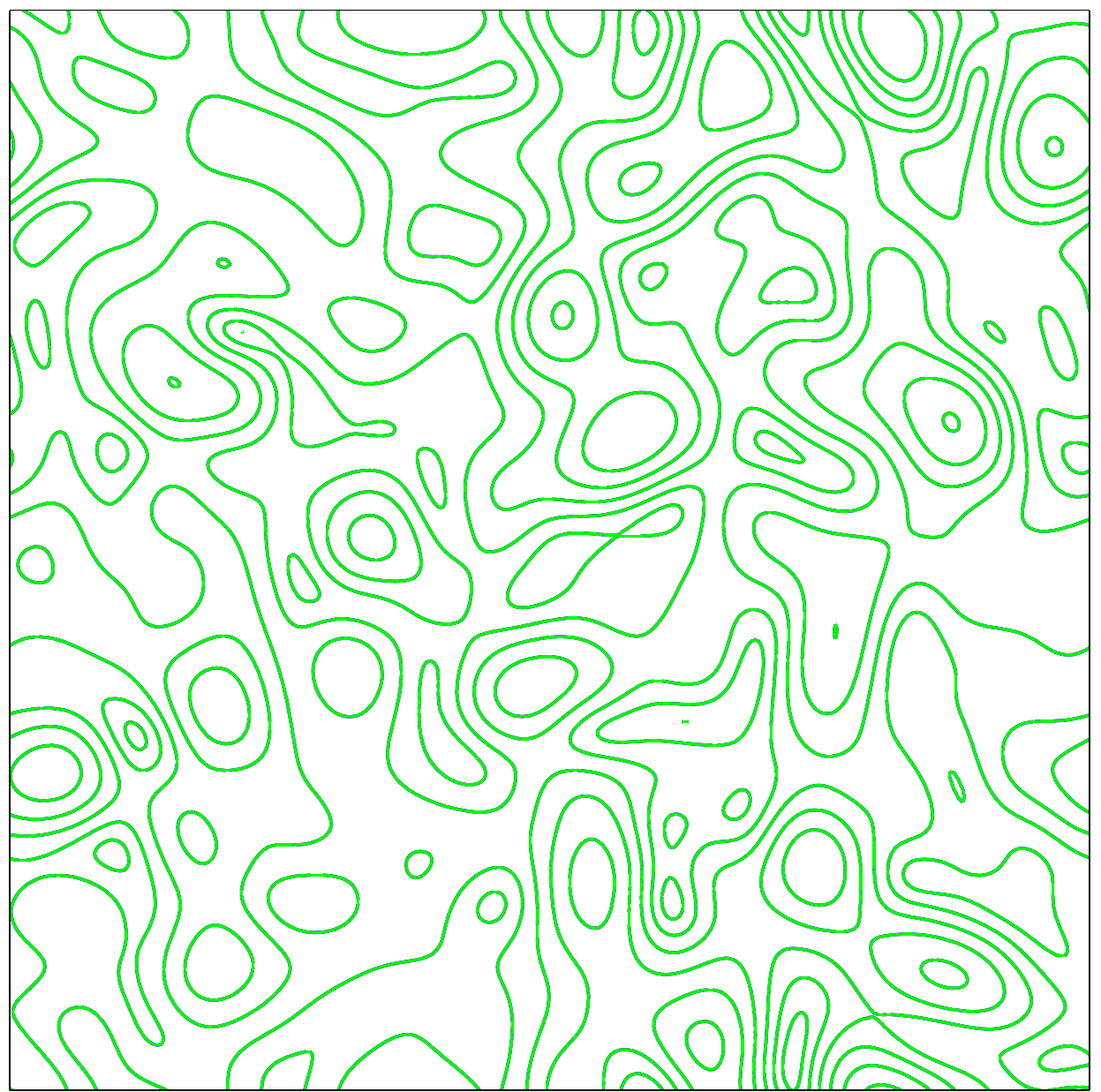}
\caption{$t'=0$}
\label{fig:ed09}
\end{subfigure}~
\begin{subfigure}[b]{0.26\textwidth}
\includegraphics[width=\textwidth]{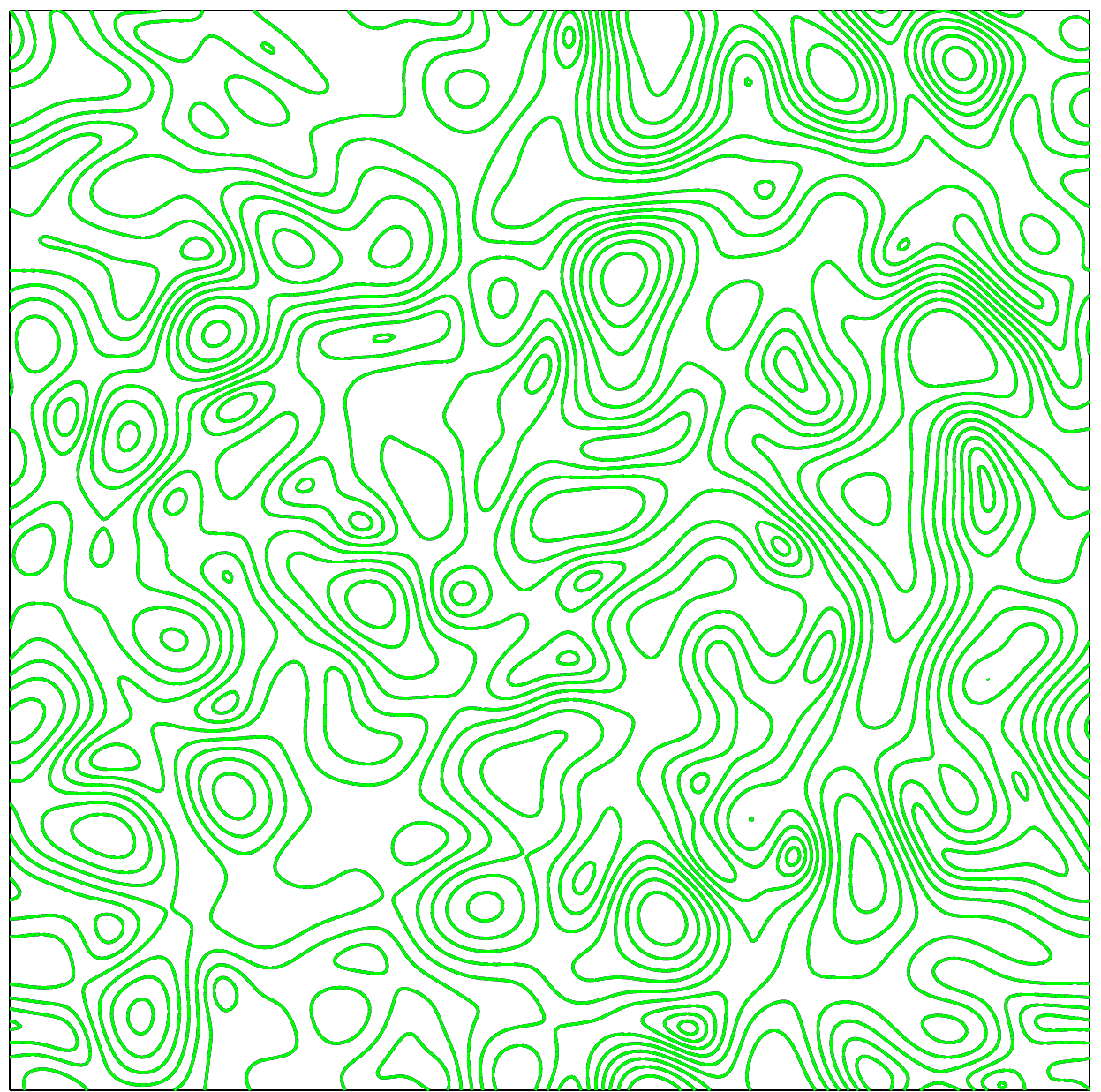}
\caption{$t'=0$}
\label{fig:ed05}
\end{subfigure}\\
\begin{subfigure}[b]{0.26\textwidth}
\includegraphics[width=\textwidth]{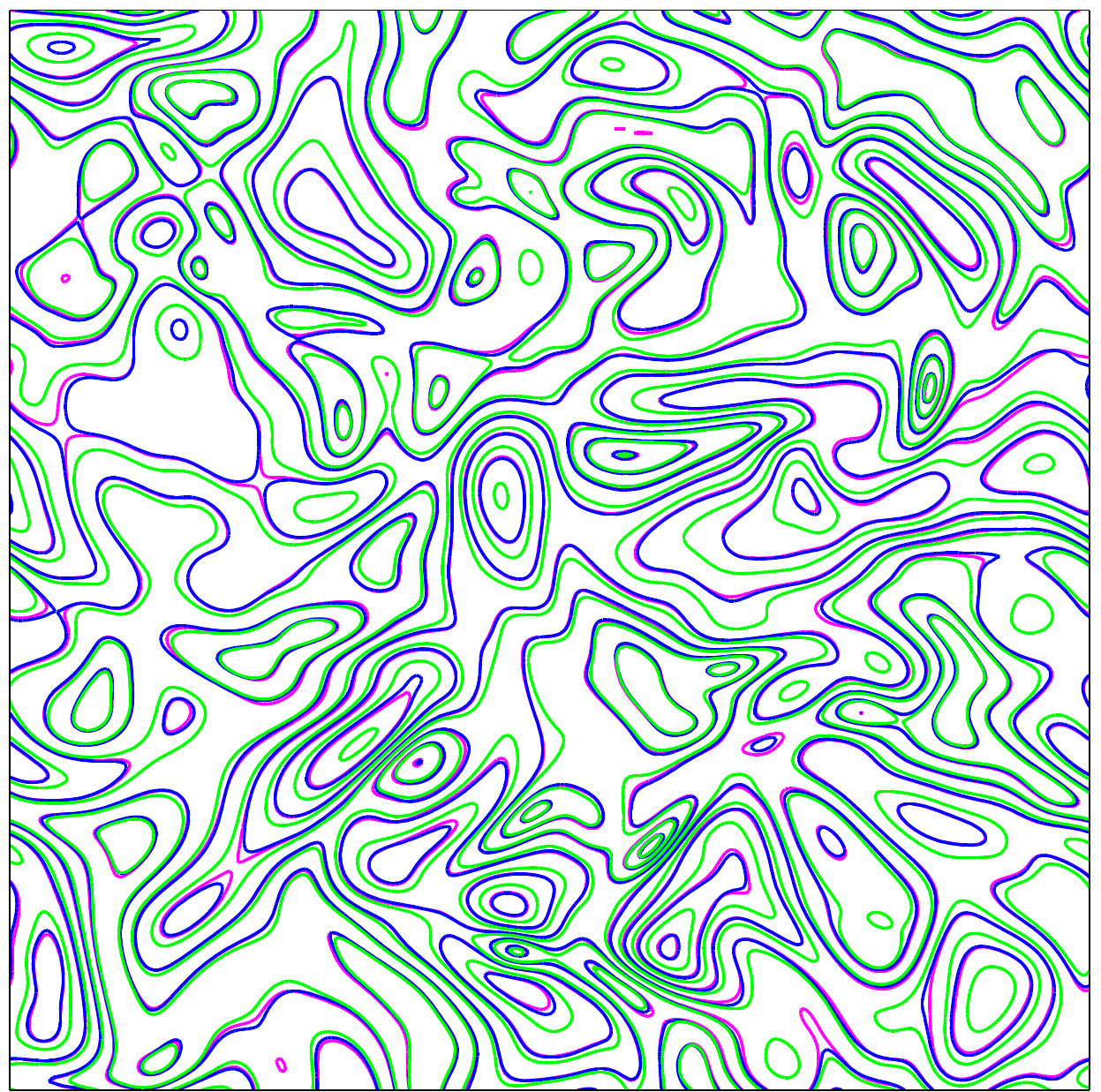}
\caption{$t'=1.21$}
\label{fig:ed09}
\end{subfigure}~
\begin{subfigure}[b]{0.26\textwidth}
\includegraphics[width=\textwidth]{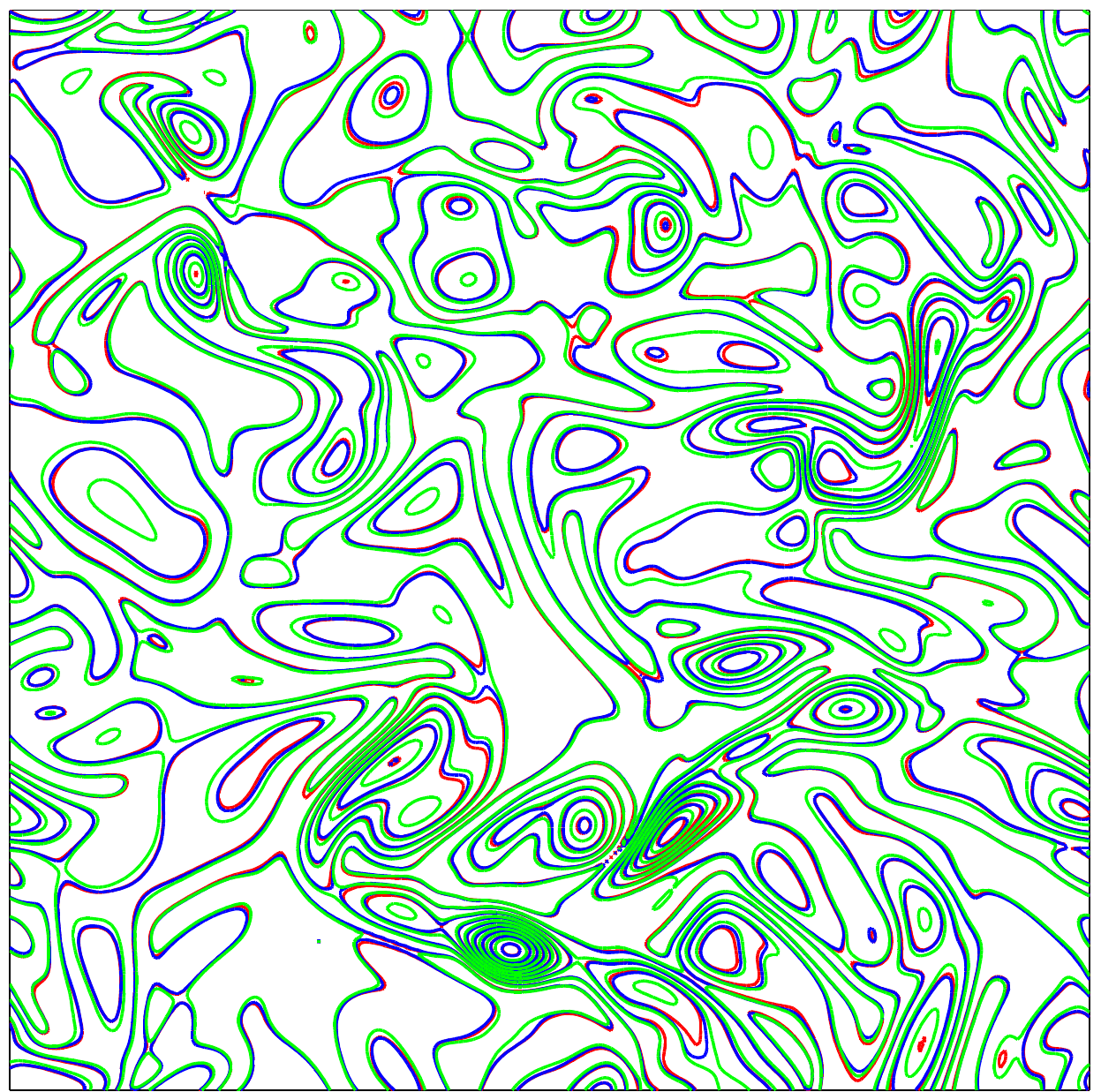}
\caption{$t'=1.21$}
\label{fig:ed05}
\end{subfigure}\\
\begin{subfigure}[b]{0.26\textwidth}
\includegraphics[width=\textwidth]{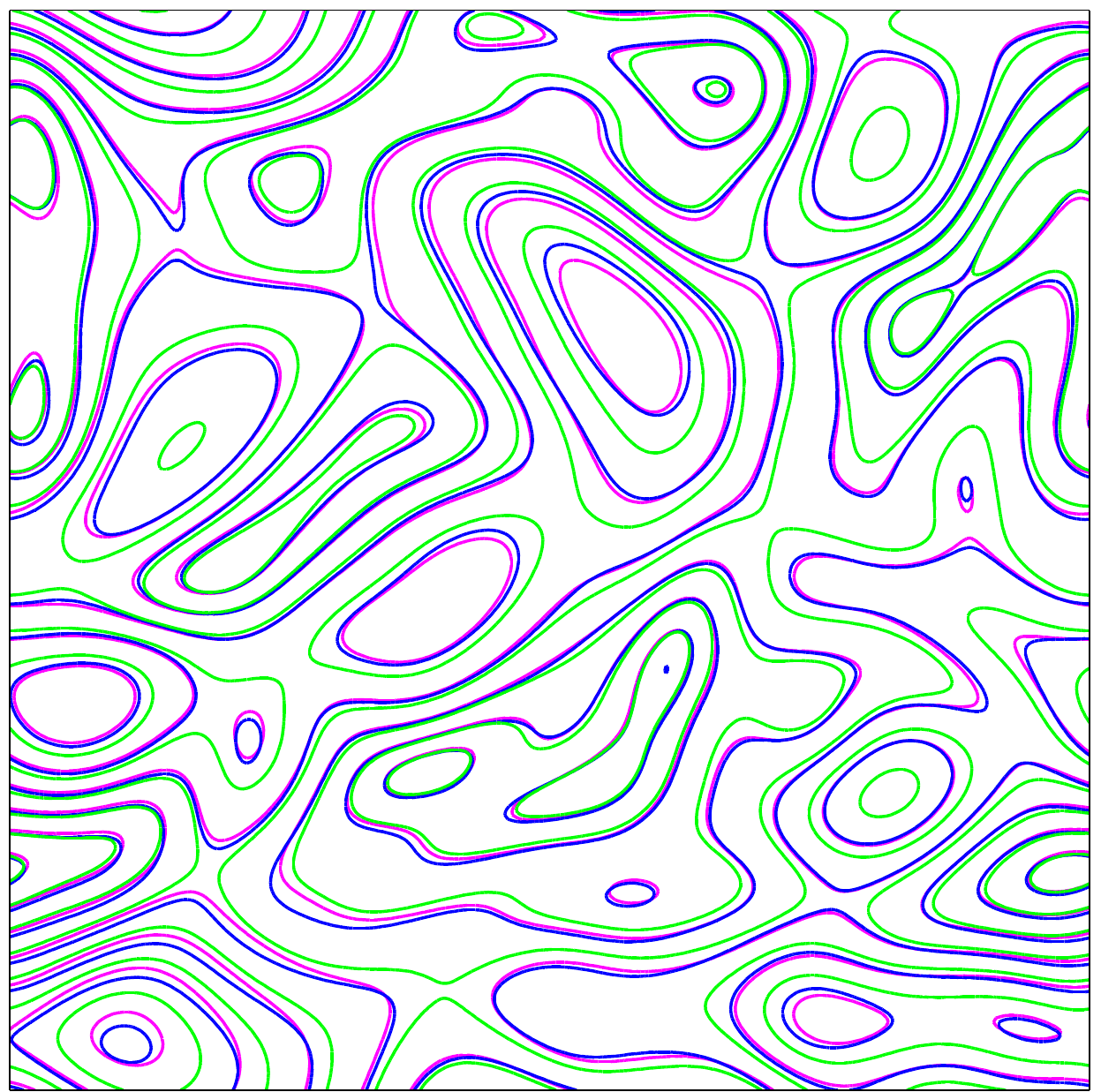}
\caption{$t'=6.08$}
\label{fig:ed09}
\end{subfigure}~
\begin{subfigure}[b]{0.26\textwidth}
\includegraphics[width=\textwidth]{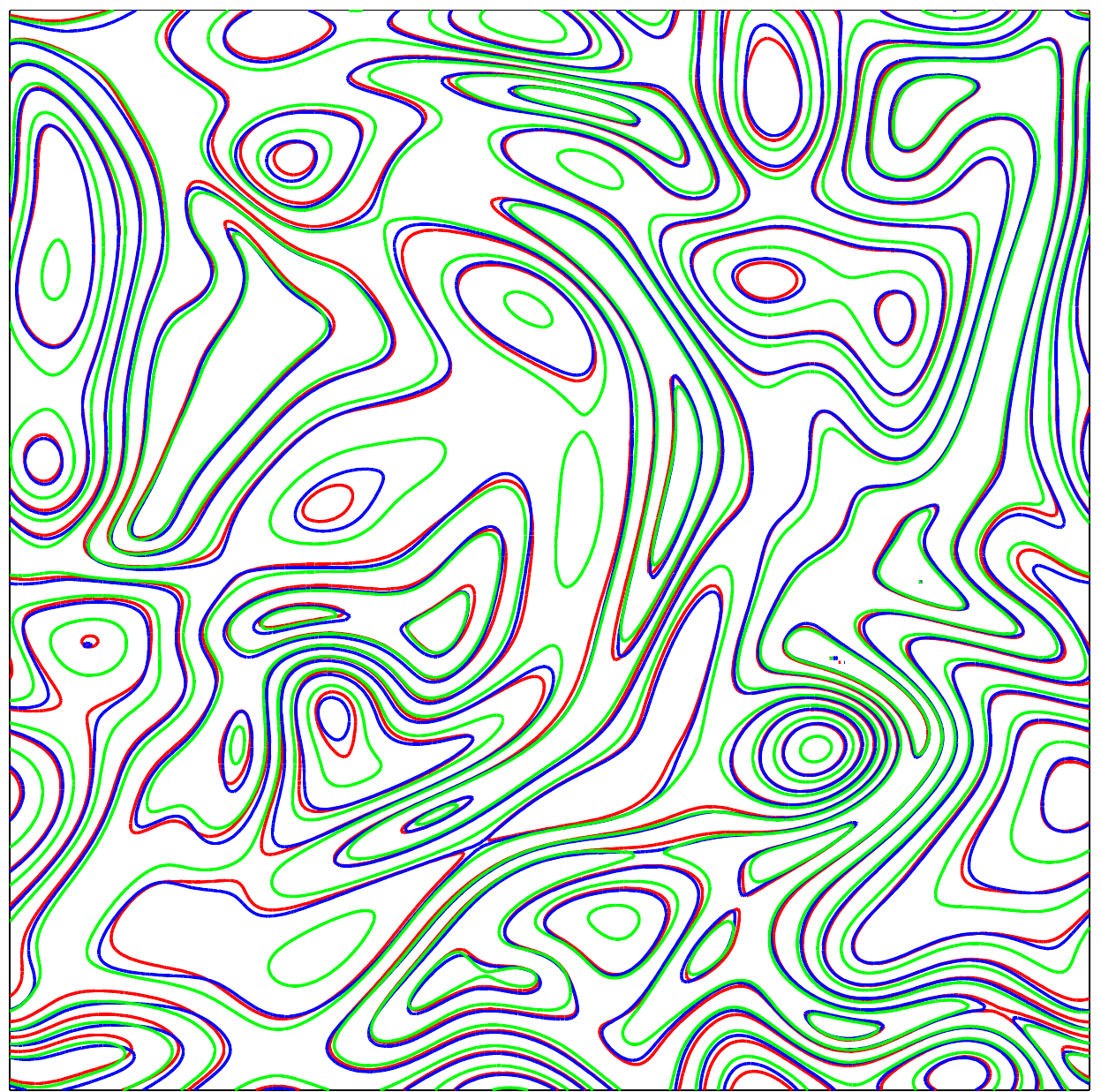}
\caption{$t'=6.08$}
\label{fig:ed05}
\end{subfigure}\\
\begin{subfigure}[b]{0.26\textwidth}
\includegraphics[width=\textwidth]{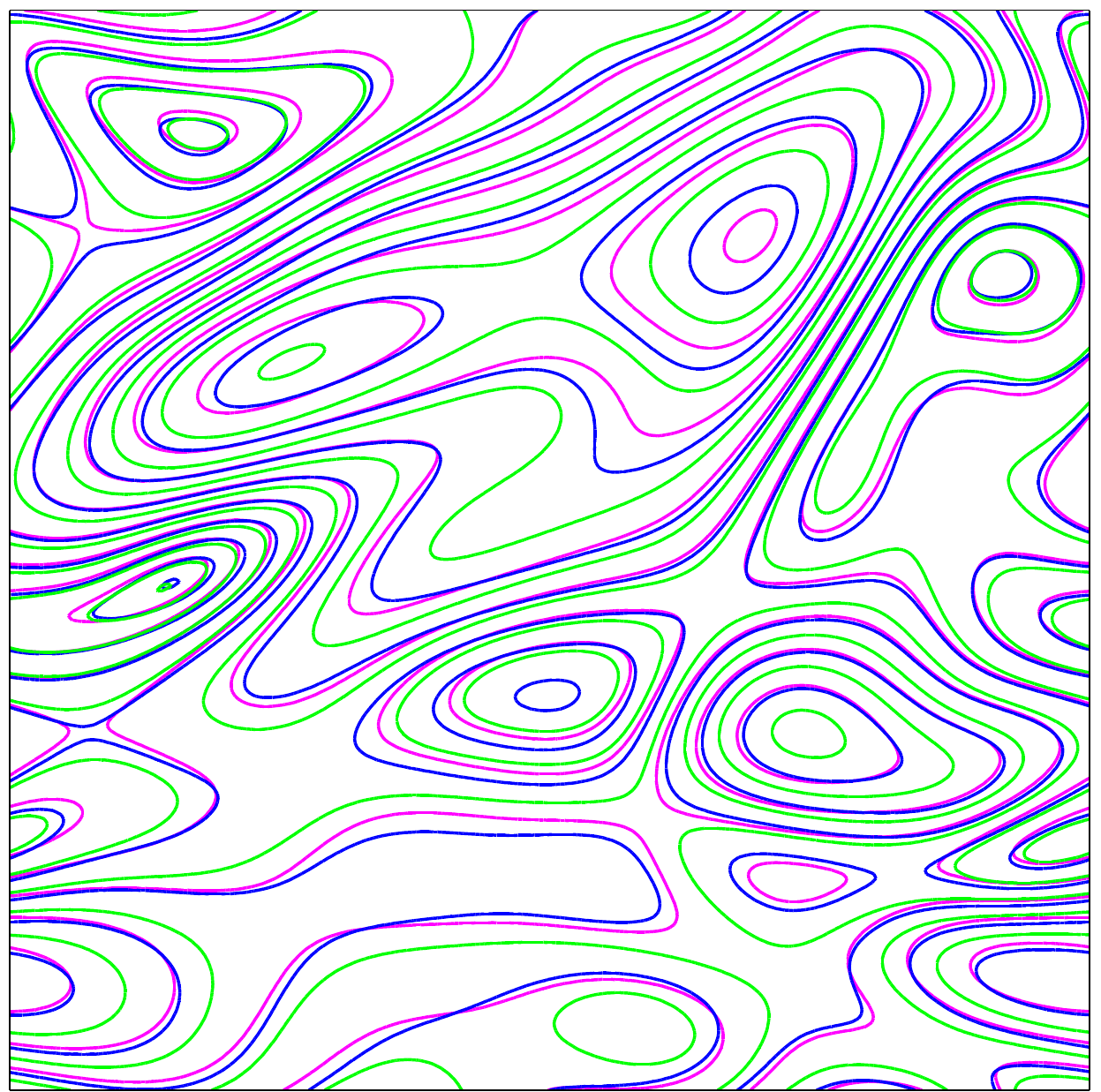}
\caption{$t'$=12.16}
\label{fig:edaa}
\end{subfigure}~
\begin{subfigure}[b]{0.26\textwidth}
\includegraphics[width=\textwidth]{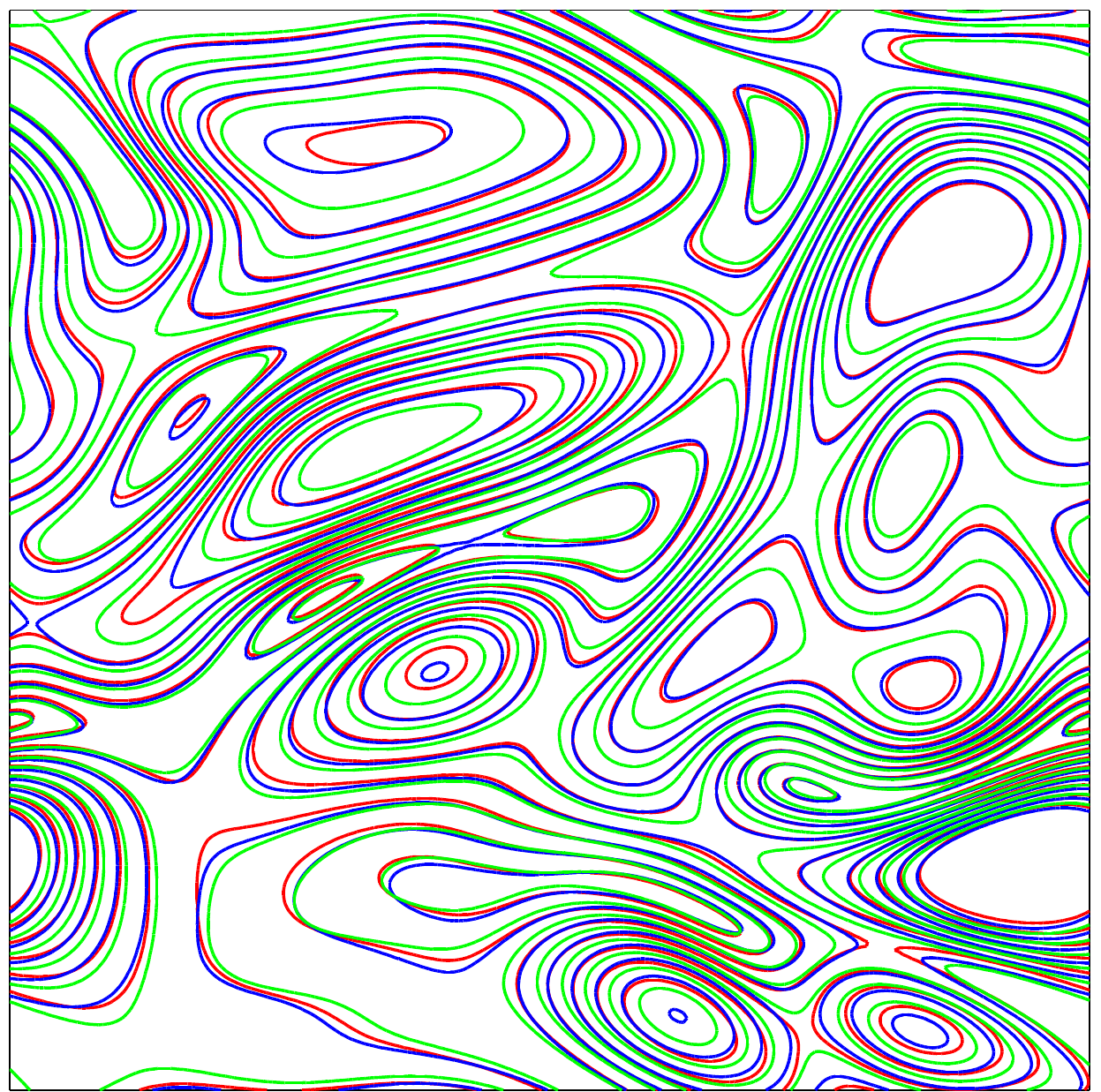}
\caption{$t'$=12.16}
\label{fig:ed05}
\end{subfigure}\\
\caption{
Contours of normalized velocity magnitude $\|\bm{u}\|/u'_0$ (left column) and normalized vorticity magnitude $\|\bm{\omega}\|L/u'_0$ (right column) on the $xy$ plane at $z=L/2$ at time $t'=0, 1.21, 6.08$ and $12.16$ (from top to bottom) with $N^3=256^3$. The solid red, green and blue lines denote results of the PS, LBE and DUGKS, respectively.} \label{fig:vor_vel256}
\end{figure}

 We also conduct the simulations  on a finer mesh of $256^3$ at $\text{Re}_{\lambda}=26.06$. As shown in Fig.~\ref{fig:vor_vel256},
 again the velocity magnitude (left column) and vorticity magnitude (right column) obtained from LBE and DUGKS methods are in good agreement
 with those from PS method. It can be seen that both kinetic methods with the fine resolution give much better prediction than those with the coarse one.

\subsection{Statistical quantities}

In this subsection, we compare some key statistical quantities, including both the low and high order statistical quantities, obtained by the LBE and DUGKS methods with those from the PS method. The simulations of these three methods are performed on both $N^3=128^3$ and $256^3$ mesh resolutions.
\begin{figure}[ht]
\centering
\begin{subfigure}[b]{0.48\textwidth}
\includegraphics[width=\textwidth]{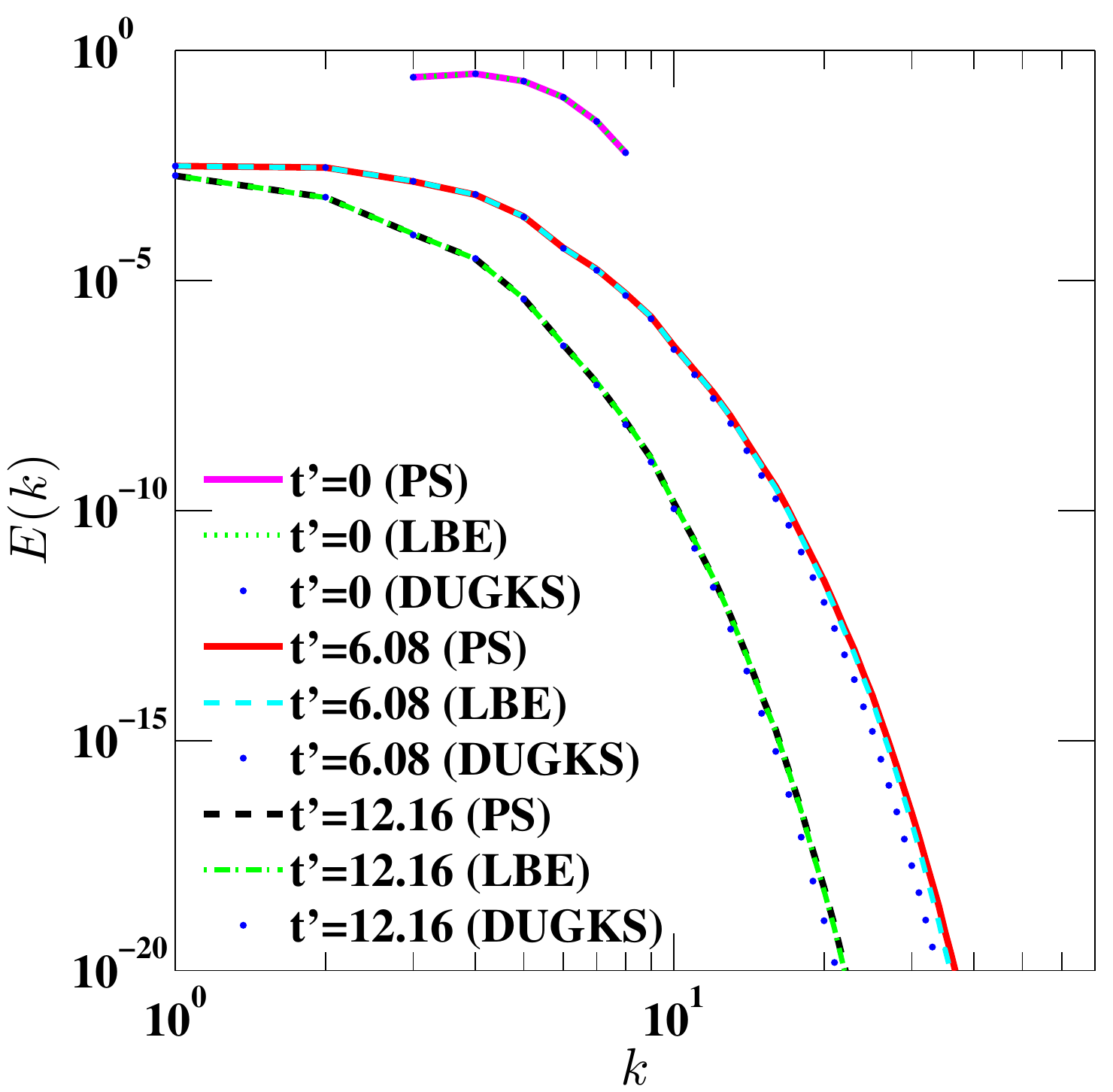}
\caption{$N^3=128^3$}
\label{fig:ed09}
\end{subfigure}~
\begin{subfigure}[b]{0.48\textwidth}
\includegraphics[width=\textwidth]{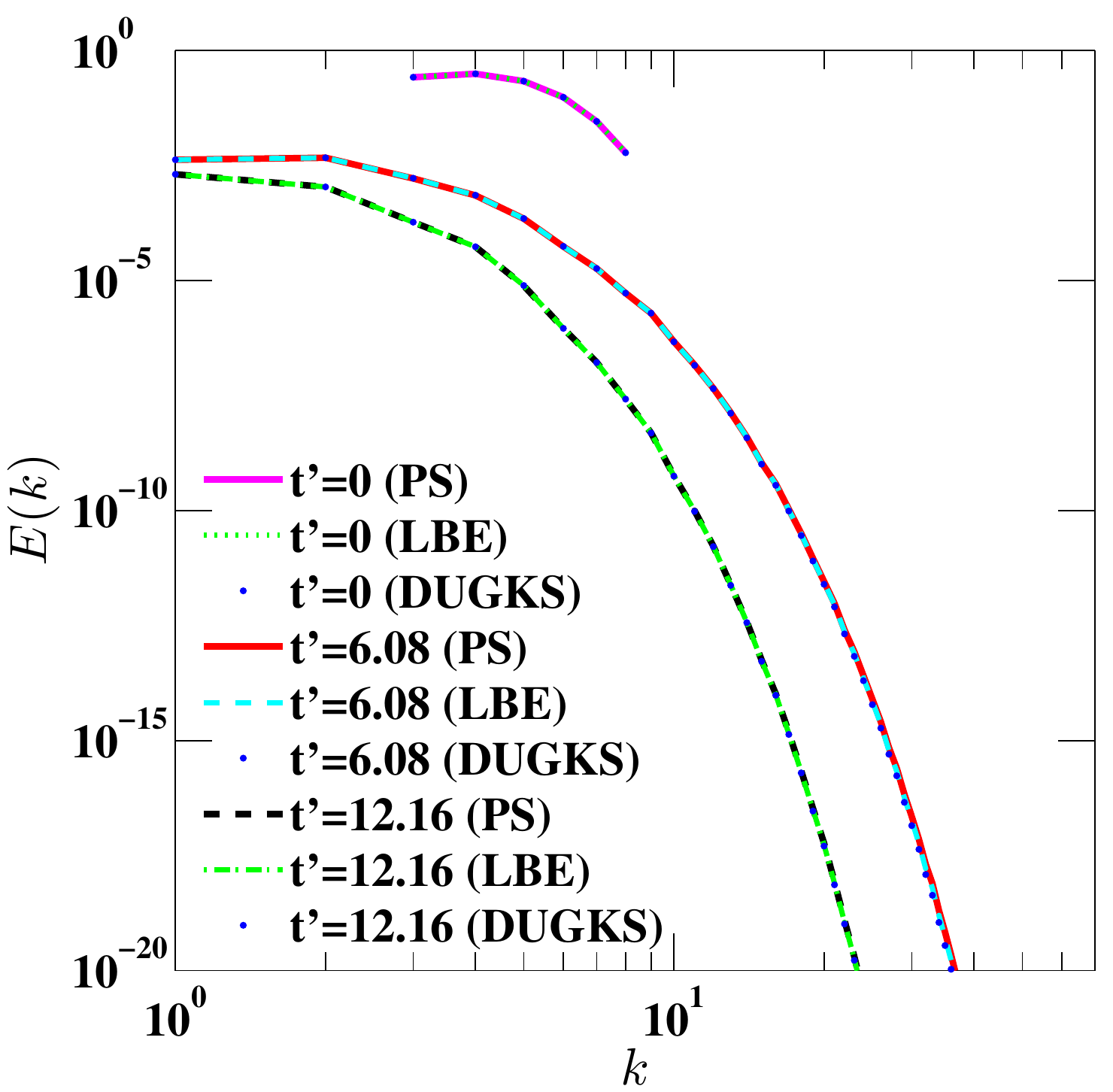}
\caption{$N^3=256^3$}
\label{fig:ed09_mesh256}
\end{subfigure}
\caption{
The energy spectra $E(k,t)$ with different mesh resolutions at $\text{Re}_{\lambda}=26.06$. } \label{fig:espec}
\end{figure}

\begin{figure}[hb]
\centering
\begin{subfigure}[b]{0.48\textwidth}
\includegraphics[width=\textwidth]{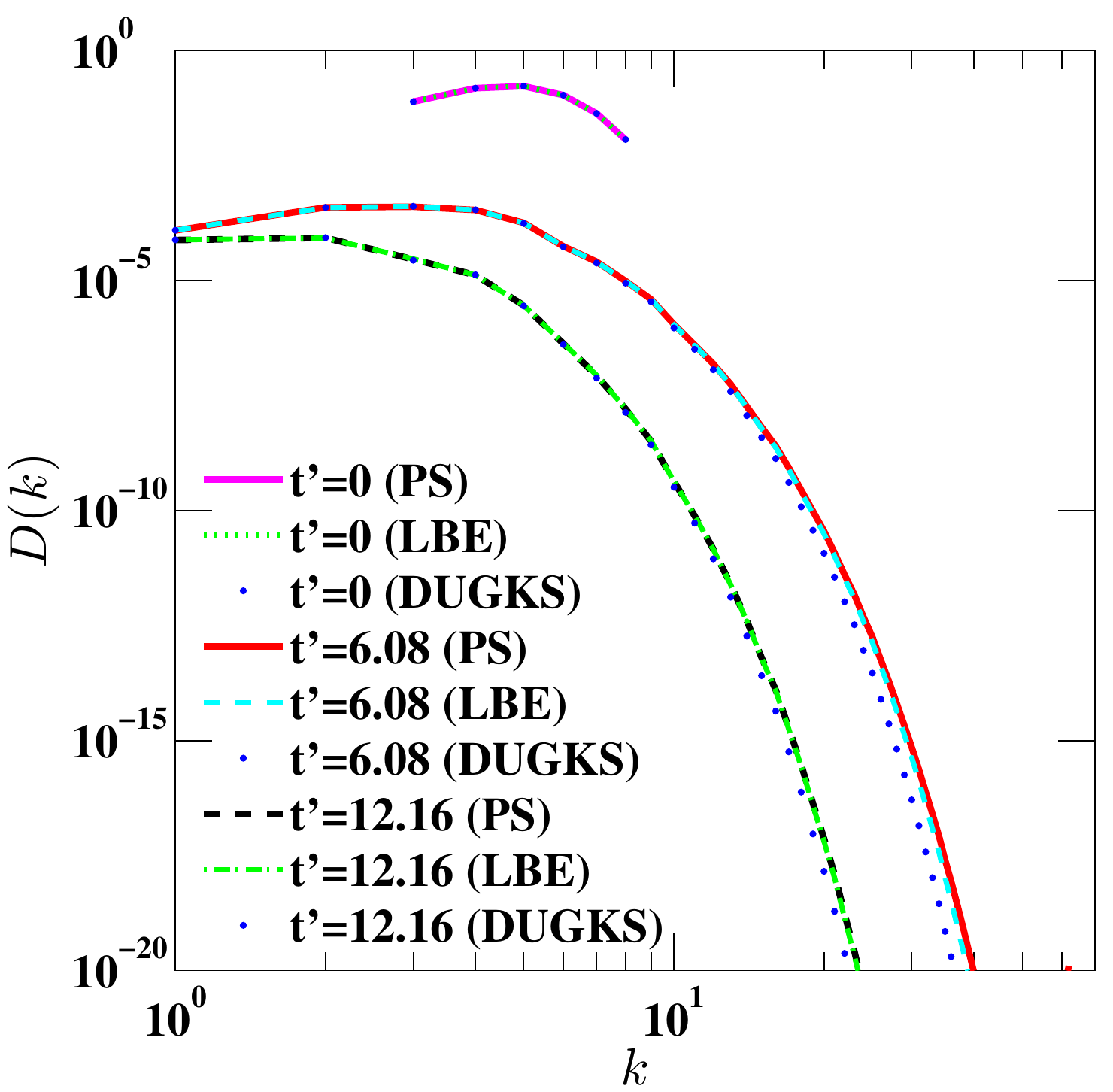}
\caption{$N^3=128^3$}
\label{fig:dis09}
\end{subfigure}~
\begin{subfigure}[b]{0.48\textwidth}
\includegraphics[width=\textwidth]{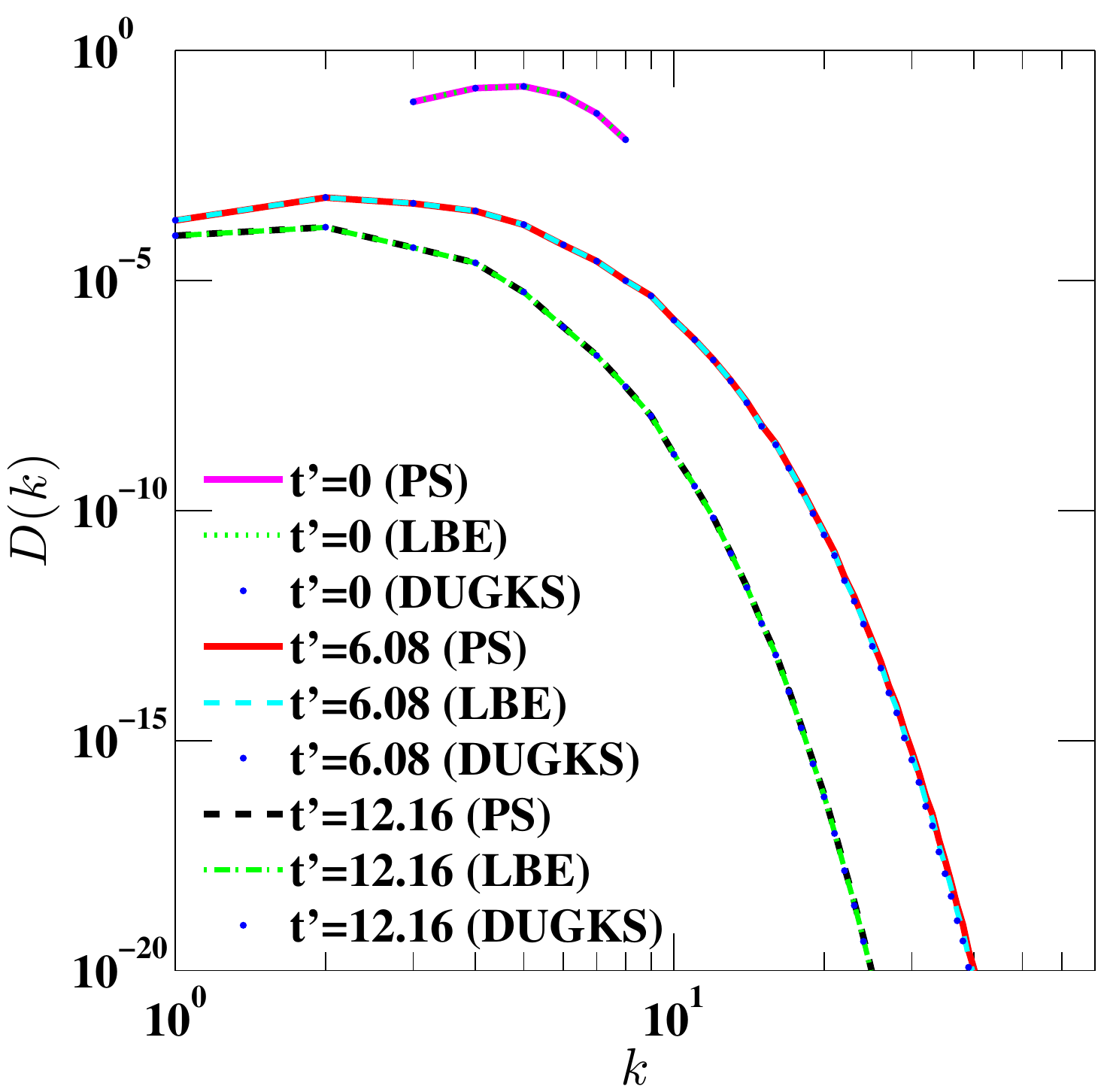}
\caption{$N^3=256^3$}
\label{fig:dis09_mesh256}
\end{subfigure}
\caption{
The dissipation rate spectra $D(k,t)$ with different mesh resolutions at $\text{Re}_{\lambda}=26.06$.} \label{fig:dis}
\end{figure}

\begin{figure}[htbp]
\centering
\begin{subfigure}[b]{0.48\textwidth}
\includegraphics[width=\textwidth]{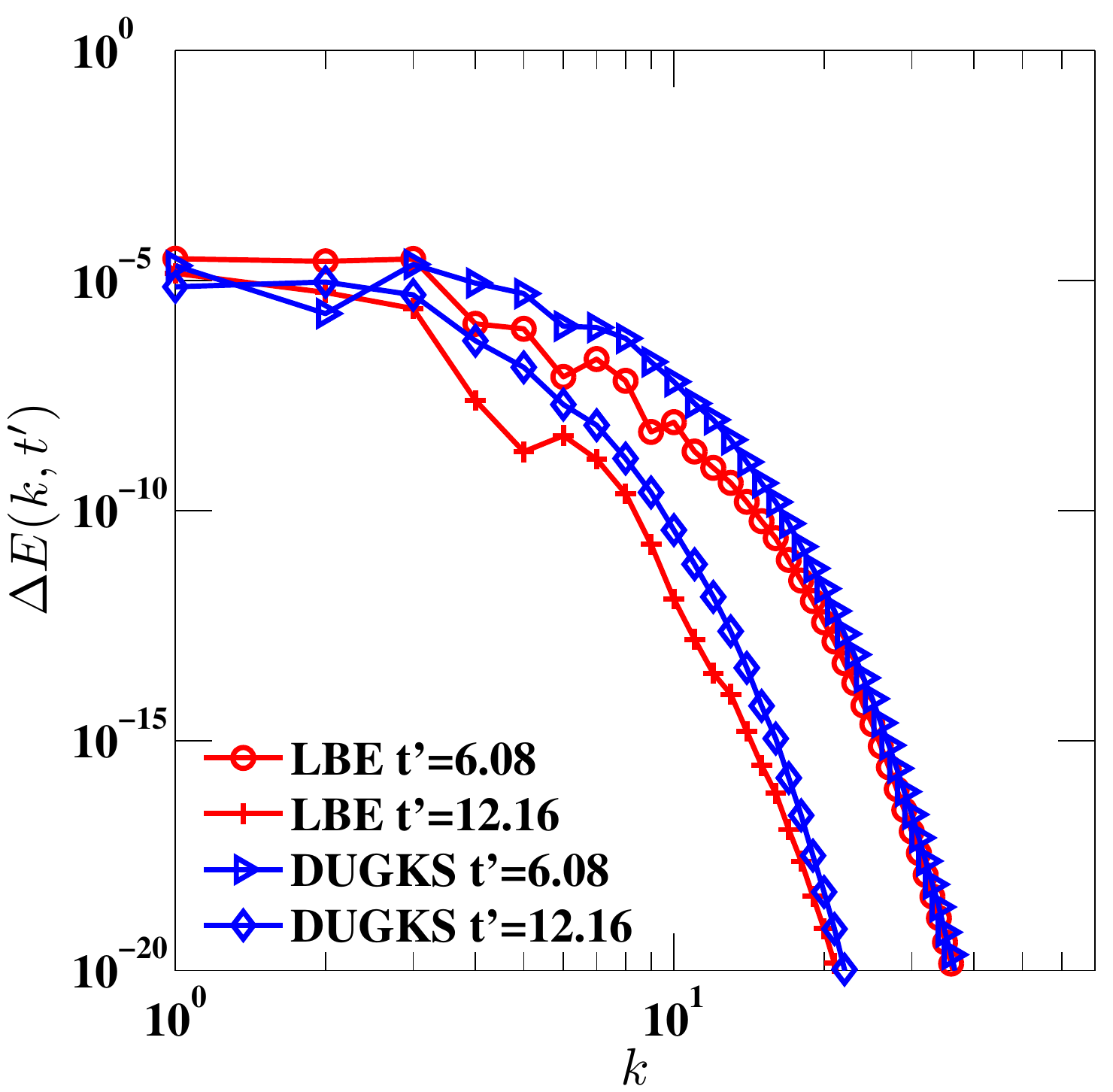}
\caption{$N^3=128^3$}
\end{subfigure}~
\begin{subfigure}[b]{0.48\textwidth}
\includegraphics[width=\textwidth]{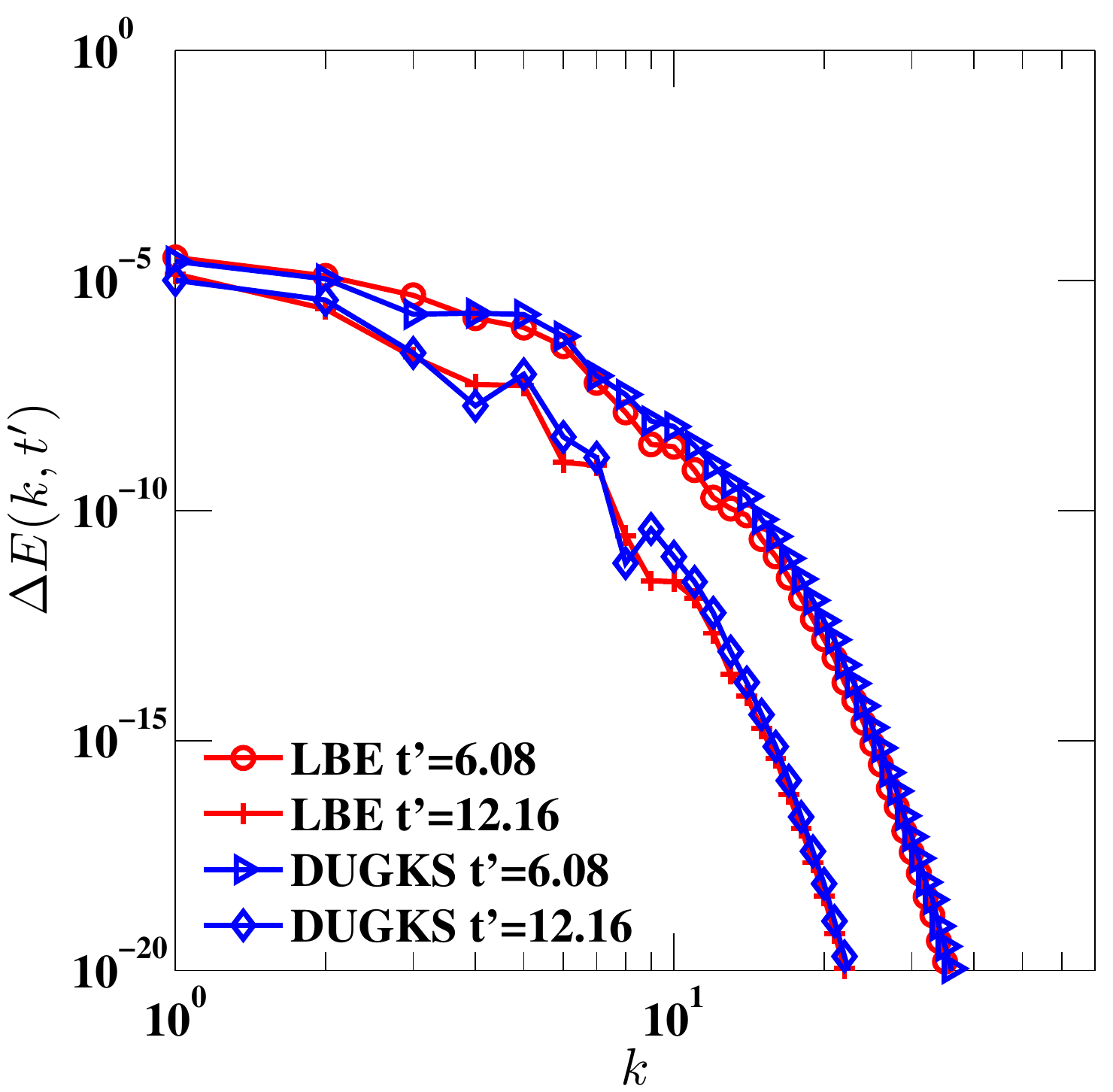}
\caption{$N^3=256^3$}
\end{subfigure}
\caption{
The energy spectra difference $\Delta E(k,t')$ with different mesh resolutions at $\text{Re}_{\lambda}=26.06$.} \label{fig:ED}
\end{figure}

\begin{figure}[htbp]
\centering
\begin{subfigure}[b]{0.48\textwidth}
\includegraphics[width=\textwidth]{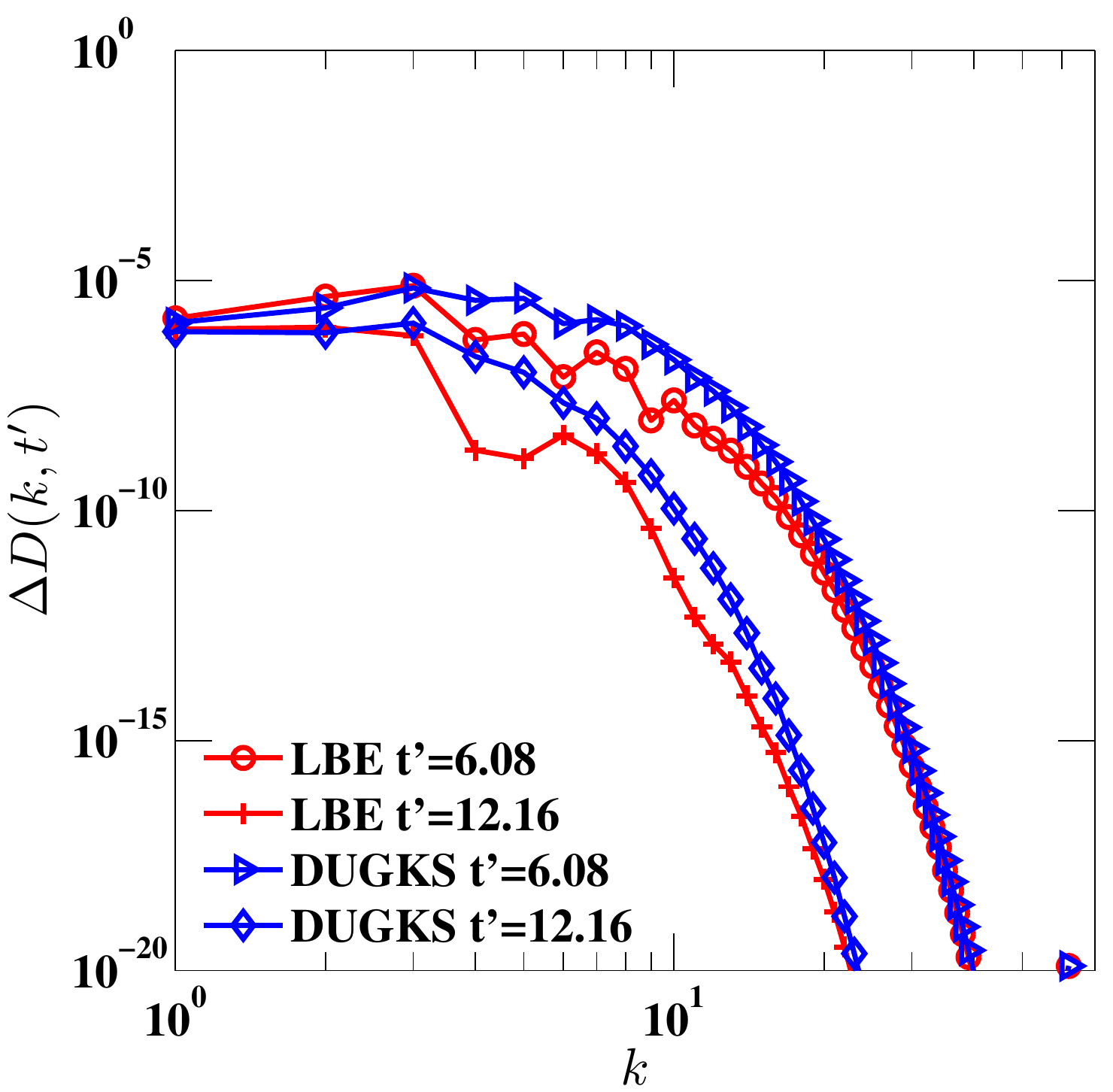}
\caption{$N^3=128^3$}
\end{subfigure}~
\begin{subfigure}[b]{0.48\textwidth}
\includegraphics[width=\textwidth]{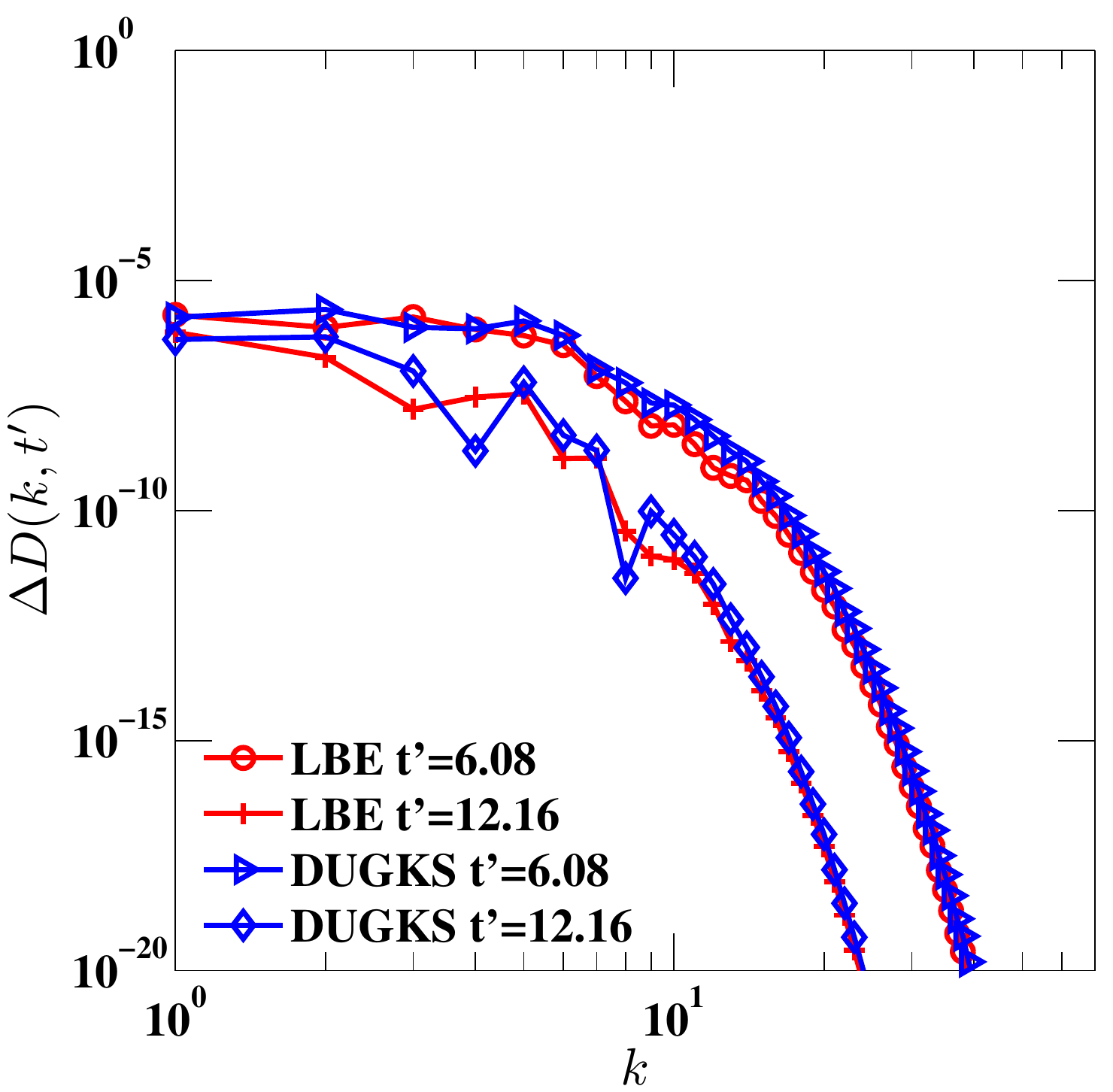}
\caption{$N^3=256^3$}
\end{subfigure}
\caption{
The dissipation rate spectra difference $\Delta D(k,t)$ with different mesh resolutions at $\text{Re}_{\lambda}=26.06$.} \label{fig:DD}
\end{figure}

\begin{figure}[htbp]
\centering
\begin{subfigure}[b]{0.48\textwidth}
\includegraphics[width=\textwidth]{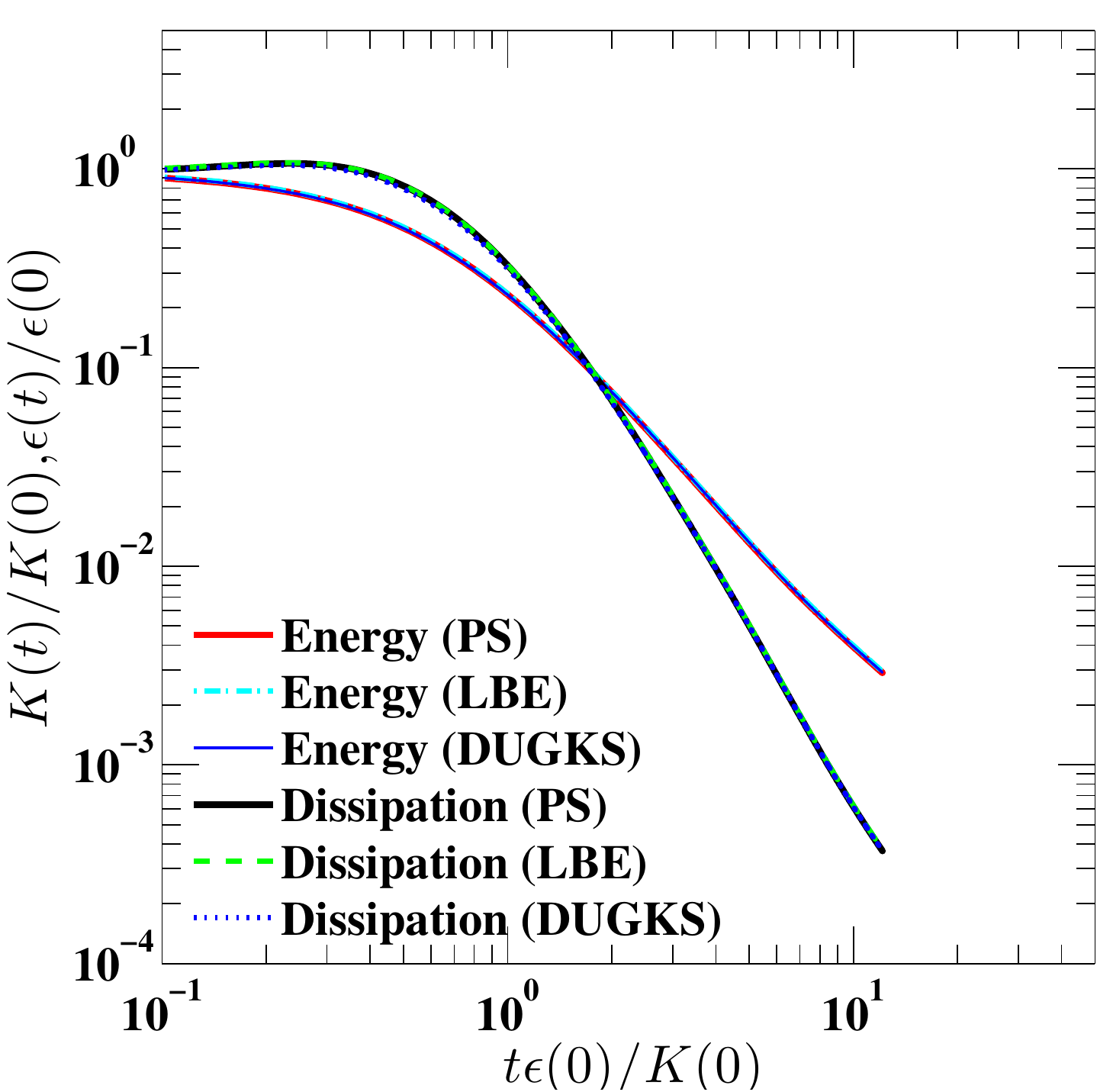}
\caption{$N^3=128^3$}
\label{fig:k09}
\end{subfigure}~
\begin{subfigure}[b]{0.48\textwidth}
\includegraphics[width=\textwidth]{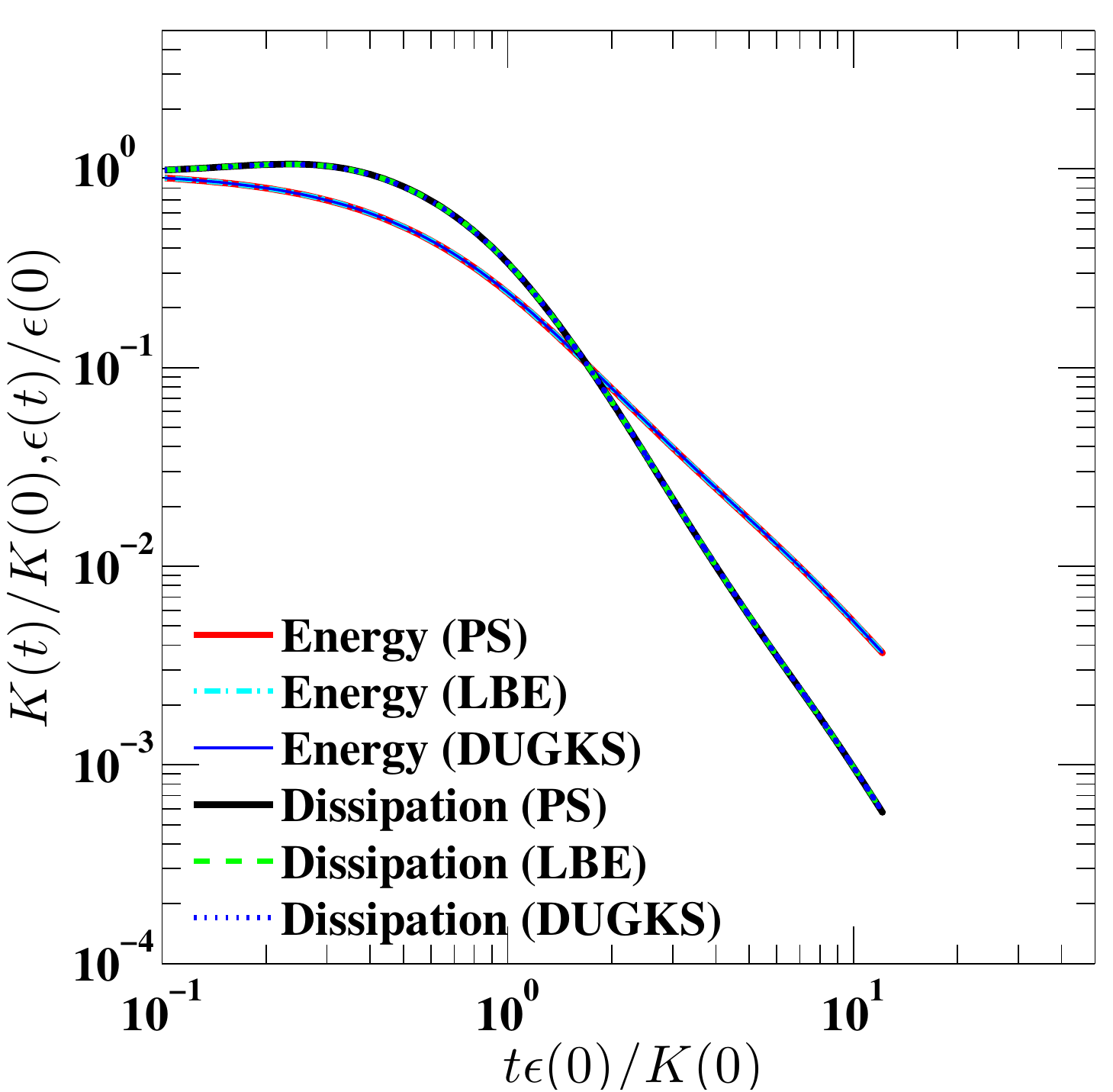}
\caption{$N^3=256^3$}
\label{fig:k09_mesh256}
\end{subfigure}
\caption{
Evolutions of the normalized total kinetic energy $K(t)/K_0$ and the normalized dissipation rate $\epsilon(t)/\epsilon_0$ with different mesh resolutions at $\text{Re}_{\lambda}=26.06$.} \label{fig:ed}
\end{figure}

We first compare the energy spectra $(E(k))$ and dissipation spectra $(D(k))$ at $t'=0, 6.08$ and $12.16$.
As shown in Figs.~\ref{fig:ed09} and \ref{fig:dis09}, the results computed by the LBE and DUGKS agree well with those from the PS counterparts. It should be noted that although there are a little deviations in the high wavenumber region for the results of DUGKS on the mesh of $128^3$, the values of the both spectra have decreased to the $10^{-10}$ magnitude of the maximum initial value, which will not cause significant deviations on the integral quantities, such as the normalized kinetic energy $K/K_0$ and dissipation rate $\epsilon/{\epsilon}_0$ shown in Fig.~\ref{fig:k09}. This discrepancy may be caused by the numerical dissipation, which is proportional to the mesh size. Therefore we can refine the mesh resolution to reduce the numerical dissipation. As expected, as shown in Figs.~\ref{fig:ed09_mesh256} and \ref{fig:dis09_mesh256}, with  mesh resolution of $256^3$ the results of DUGKS show no visible difference with those from the LBE and PS simulations. We also compute the difference of the spectra between both kinetic methods and the PS method, which is defined by
\begin{equation}\label{errs}
\Delta S(k,t')=\| S(k,t')-S_p(k,t') \|,
\end{equation}
where $S$ denotes the results of energy spectra or dissipation rate spectra, and $S_p$ represents the results from the PS simulations. Figures ~\ref{fig:ED} and ~\ref{fig:DD} respectively show the differences of energy spectra $\Delta E(k,t')$ and dissipation rate spectra $\Delta D(k,t')$ on both meshes. We observe that  with the mesh of $128^3$, the results obtained by LBE is slightly better than those from the DUGKS when compared with the PS results, but there is no visible difference when both methods used the fine mesh of $256^3$. These results indicate that the dissipation of the DUGKS is slightly larger than the LBE method, though both LBE and DUGKS methods have low numerical dissipation.

Secondly, we compare the evolutions of normalized kinetic energy $K(t)/K_0$ and dissipation rate $\epsilon(t)/{\epsilon}_0$ . As shown in Fig.~\ref{fig:ed}, both $K(t)/K_0$ and $\epsilon(t)/{\epsilon}_0$ calculated by LBE and DUGKS methods are in excellent agreement with those from PS simulation on both meshes. Quantitatively, we compare the the maximum errors of $K(t)$ and $\epsilon(t)$  relative to the PS results, which is defined by
\begin{equation}
R_m(s)=\left\|\frac{s-s_p}{s_p}\right\|_{max},
\end{equation}
where $s=K(t)$ or $\epsilon(t)$, and $s_p$ is the corresponding quantity from the PS method.
As shown in Table~\ref{tabled}, the maximum relative errors are less than $1\%$ except that of $\epsilon(t)$ from the DUGKS with $N^3=128^3$ which reaches to $3.9\%$ around the peak value, and it decreases to $0.49\%$ in the $256^3$ simulation.
We also observe that the normalized energy dissipation rate attains a peak value at $t'=0.23$ due to the energy cascade, before decreasing with increasing time due to the viscous dissipation.
\begin{table}[!htb]
\centering
\caption{\label{tabled} The maximum errors of $K(t)/K_0$ and $\epsilon(t)/{\epsilon}_0$ relative to PS results. }
\begin{tabular}[t]
   {c c c c c c c}
   \hline
    \hline
  & $\text{Case}$& \hspace{2mm}  $\text{LBE128}$  &\hspace{2mm} $\text{LBE256}$ & \hspace{2mm} $\text{DUGKS128}$ & \hspace{2mm} $\text{DUGKS256}$ &\\
\hline
   & $R_m(K)$    & \hspace{2mm}  $0.42\%       $  &\hspace{2mm} $0.34\%$        & \hspace{2mm} $0.84\%$          & \hspace{2mm}  $0.49\%$         &\\
& $R_m(\epsilon)$& \hspace{2mm}  $0.83\%       $  &\hspace{2mm}  $0.31\%$       & \hspace{2mm} $3.90\%$          & \hspace{2mm}  $0.49\%$         &\\
  \hline
   \hline
\end{tabular}
\end{table}

\begin{figure}[htbp]
\centering
\begin{subfigure}[b]{0.48\textwidth}
\includegraphics[width=\textwidth]{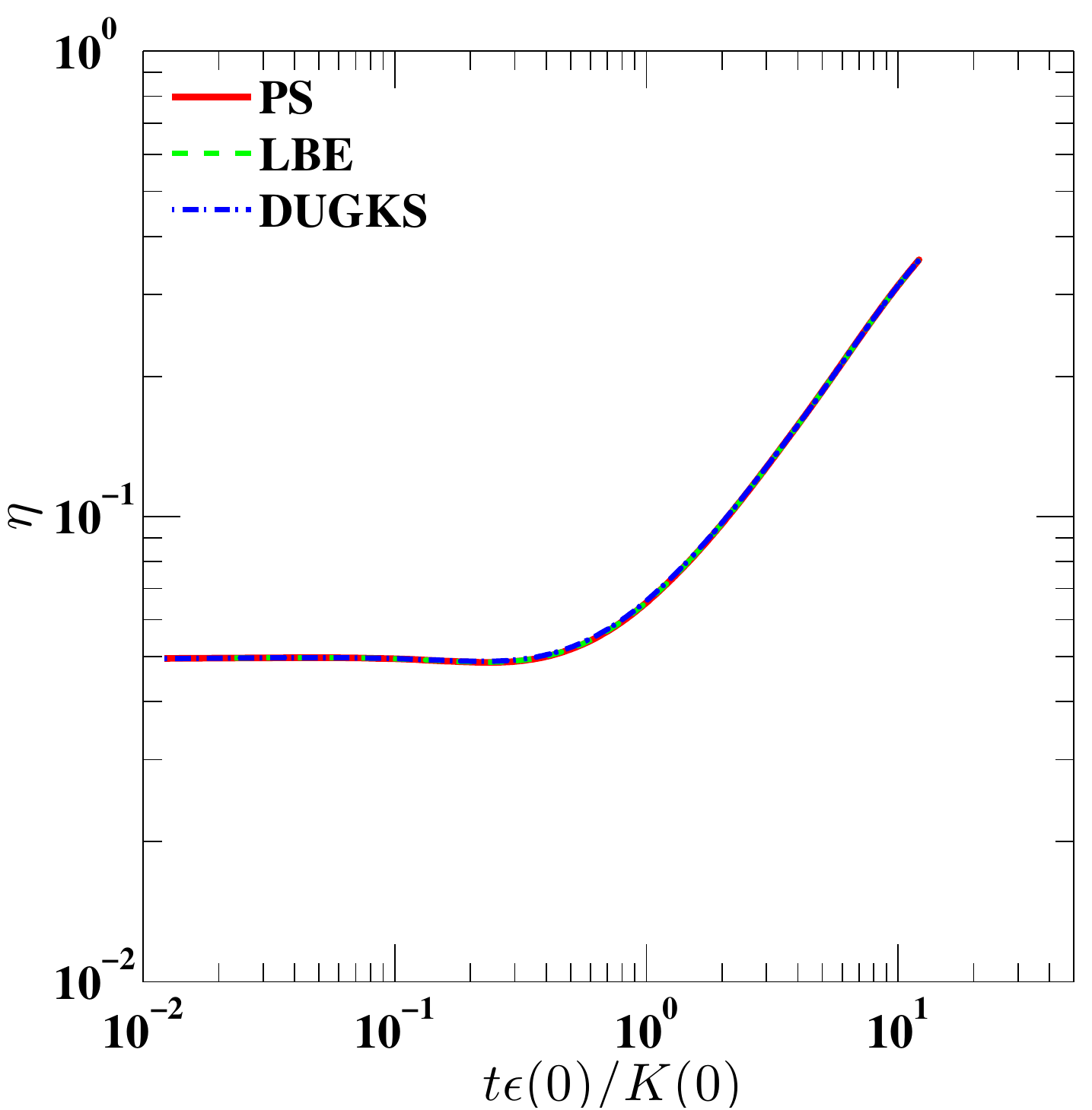}
\caption{$N^3=128^3$}
\label{fig:komo09}
\end{subfigure}~
\begin{subfigure}[b]{0.48\textwidth}
\includegraphics[width=\textwidth]{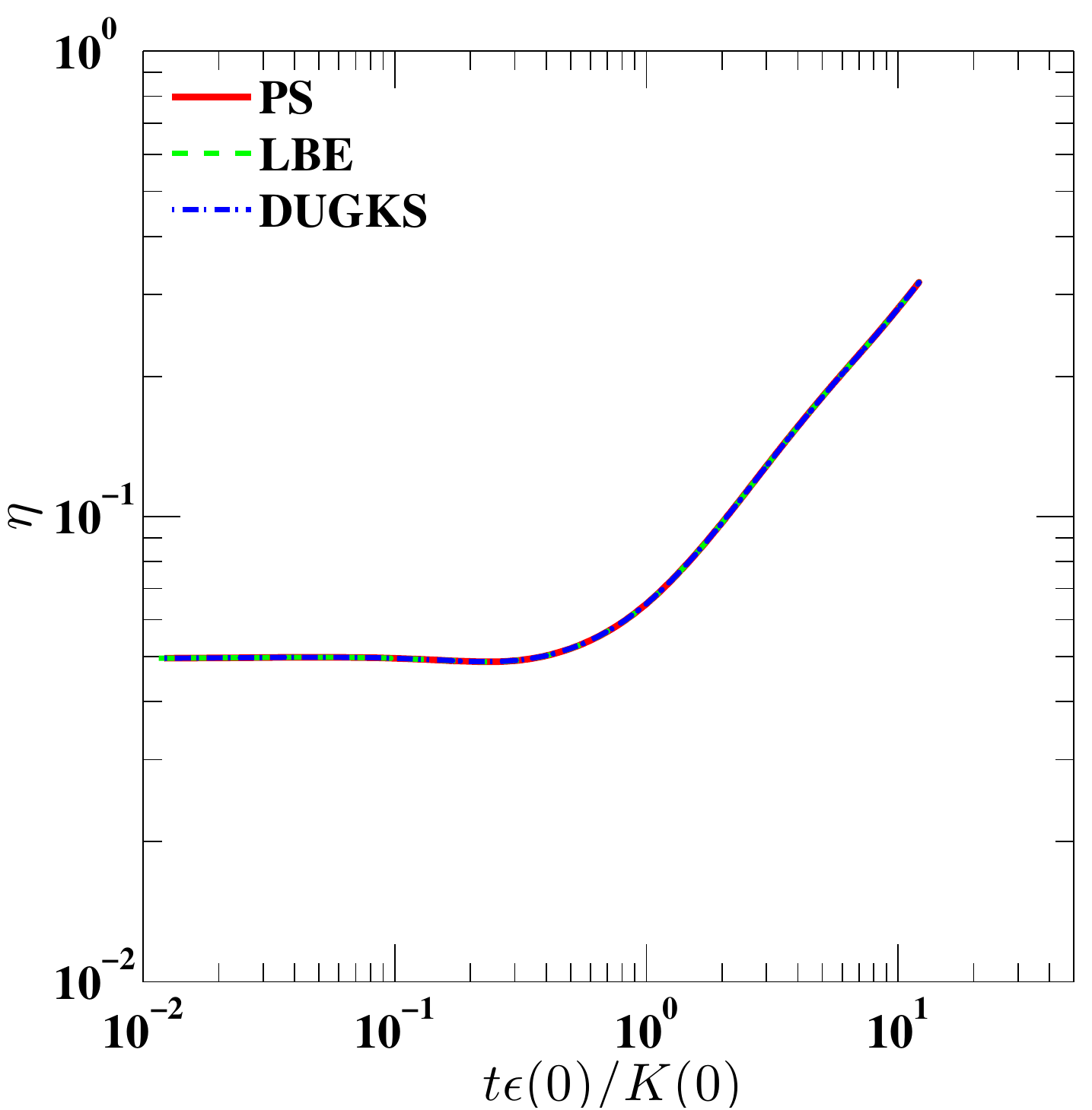}
\caption{$N^3=256^3$}
\label{fig:komo09_mesh256}
\end{subfigure}
\caption{Evolutions of the Kolmogorov length $\eta$ with different mesh resolutions  at $\text{Re}_{\lambda}=26.06$.} \label{fig:komo}
\end{figure}

\begin{figure}[htbp]
\centering
\begin{subfigure}[b]{0.48\textwidth}
\includegraphics[width=\textwidth]{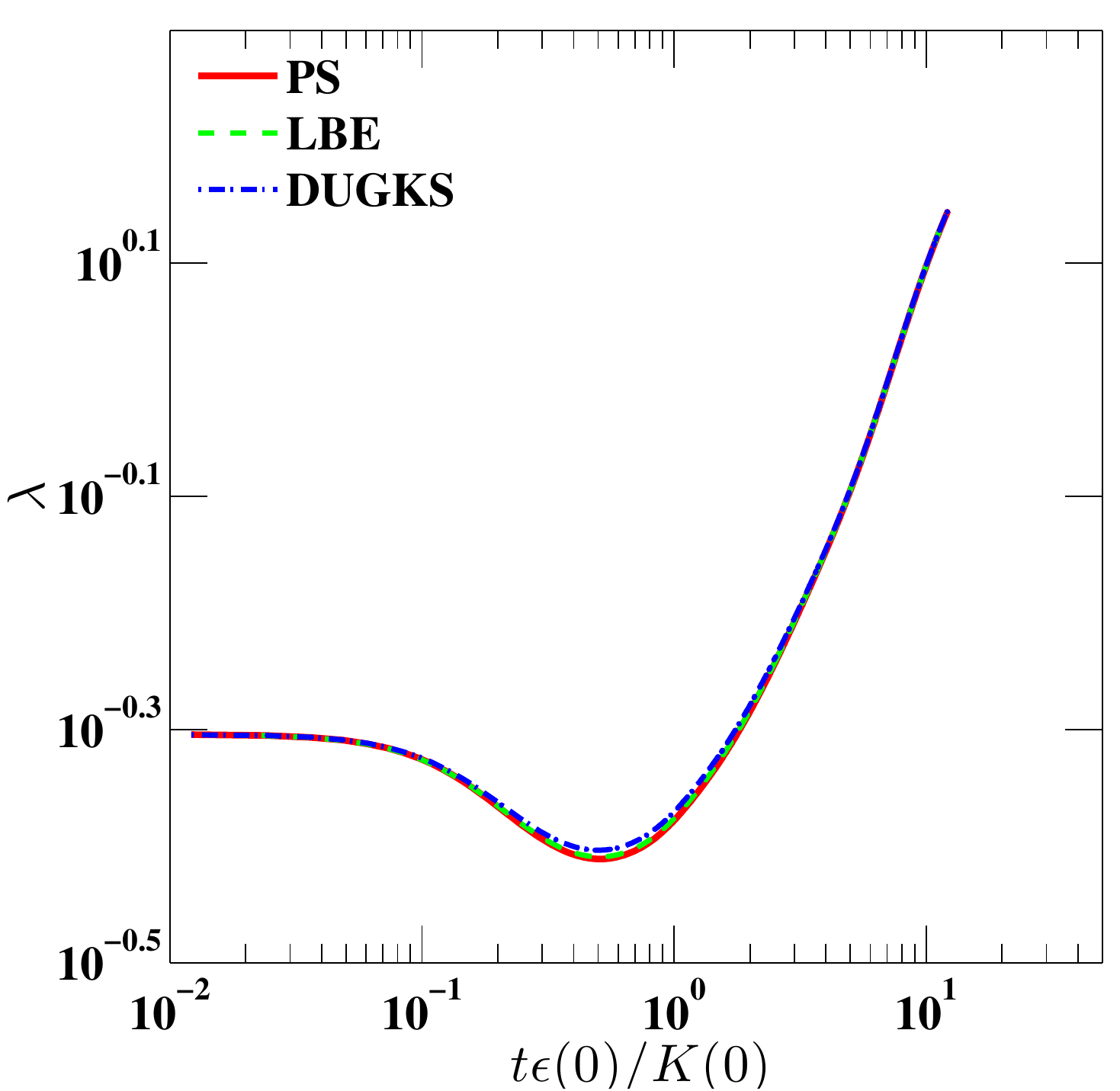}
\caption{$N^3=128^3$}
\label{fig:taylor09}
\end{subfigure}~
\begin{subfigure}[b]{0.48\textwidth}
\includegraphics[width=\textwidth]{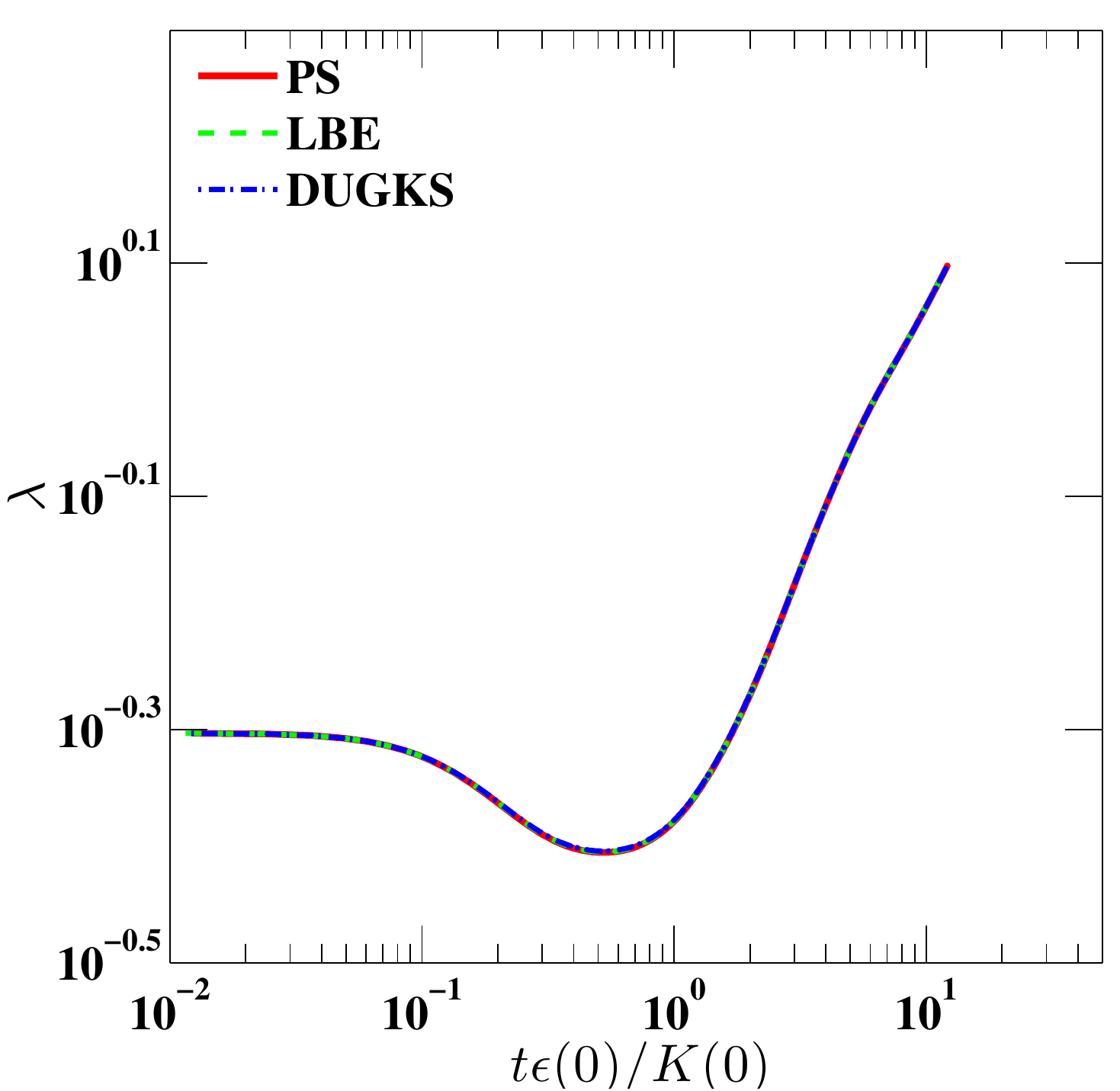}
\caption{$N^3=256^3$}
\label{fig:taylor09_mesh256}
\end{subfigure}
\caption{Evolutions of the Taylor microscale length $\lambda$ with different mesh resolutions at $\text{Re}_{\lambda}=26.06$.} \label{fig:taylor}
\end{figure}

Thirdly, we compare the evolutions of the Kolmogorov length $\lambda$  and the Taylor microscale length $\eta$. The Kolmogorov length is the smallest scale in turbulence flow, at which the viscous effect dominates and the turbulence kinetic energy is converted irreversibly into heat. The Taylor microscale is the intermediate scale between the largest and the smallest scales at which fluid viscosity significantly affects the dynamics of turbulent eddies in the flow. Figures~\ref{fig:komo} and \ref{fig:taylor} show the evolutions of the Kolmogorov length scale and Taylor microscale length. It is found that results of both scales from the DUGKS and the LBE methods agree well with those from the PS method. We also note that there are slightly differences around the minimum of $\lambda$ obtained by the DUGKS with the mesh resolution of $128^3$, but as shown in Table~\ref{tablee}, the maximum relative errors of $\lambda$ and $\eta$ are all less than $2\%$, and reduce to $0.27\%$ as the resolution increases to ${256}^3$.

\begin{table}[!htb]
\centering
\caption{\label{tablee} The maximum errors of $\lambda$ and $\eta$ relative to PS results. }
\begin{tabular}[t]
   {c c c c c c c}
   \hline
    \hline
  & $\text{Case}$& \hspace{2mm}  $\text{LBE128}$  &\hspace{2mm} $\text{LBE256}$ & \hspace{2mm} $\text{DUGKS128}$ & \hspace{2mm} $\text{DUGKS256}$ &\\
\hline
& $R_m(\lambda)$ & \hspace{2mm}  $0.21\%       $  &\hspace{2mm} $0.08\%$        & \hspace{2mm} $1.00\%$          & \hspace{2mm}  $0.12\%$         &\\
& $R_m(\eta)$    & \hspace{2mm}  $0.44\%       $  &\hspace{2mm} $0.12\%$        & \hspace{2mm} $1.84\%$          & \hspace{2mm}  $0.27\%$         &\\
  \hline
   \hline
\end{tabular}
\end{table}

\begin{figure}[hbp]
\centering
\begin{subfigure}[b]{0.48\textwidth}
\includegraphics[width=\textwidth]{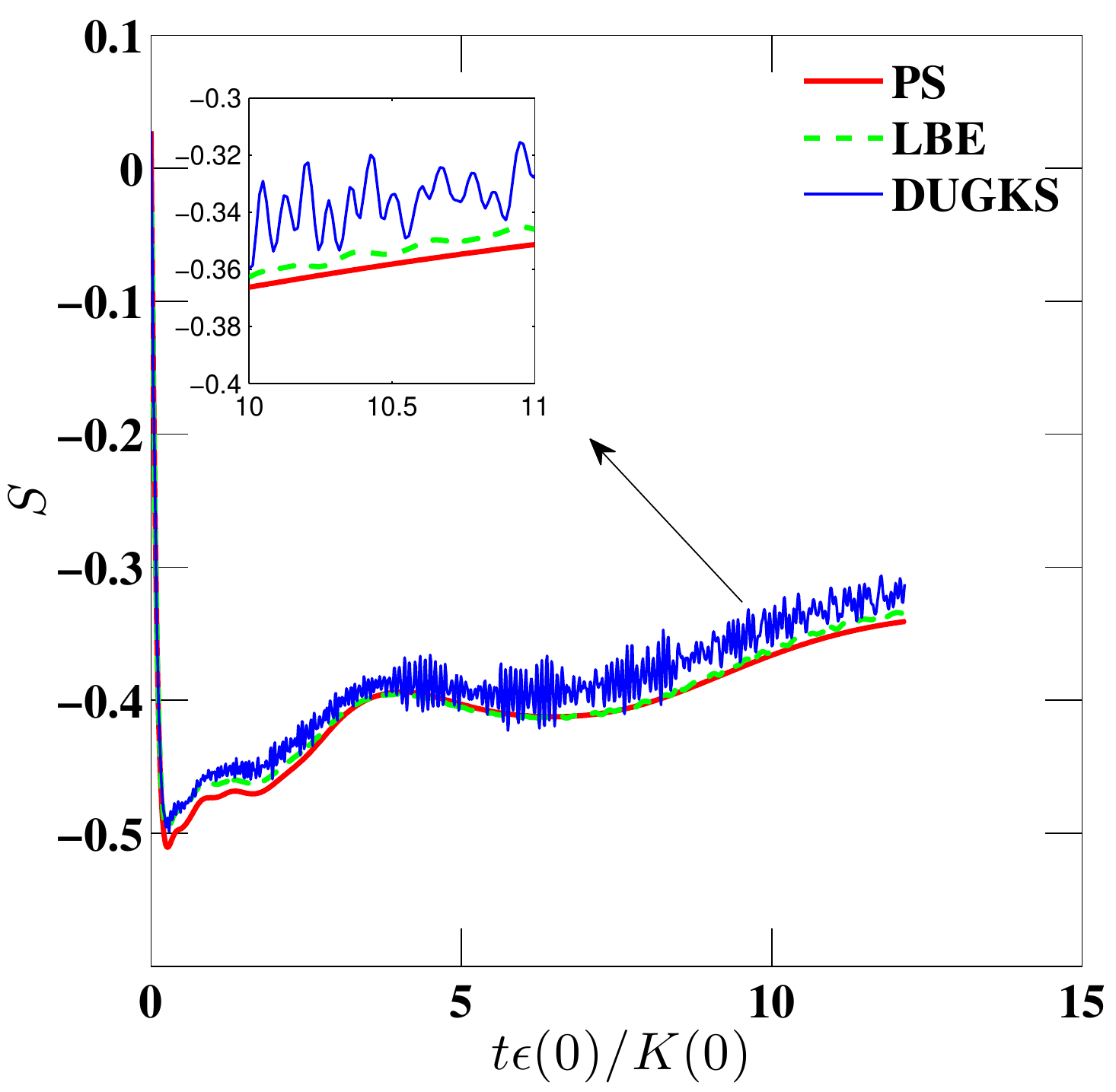}
\caption{$N^3=128^3$}
\label{fig:ske09}
\end{subfigure}~
\begin{subfigure}[b]{0.56\textwidth}
\includegraphics[width=\textwidth]{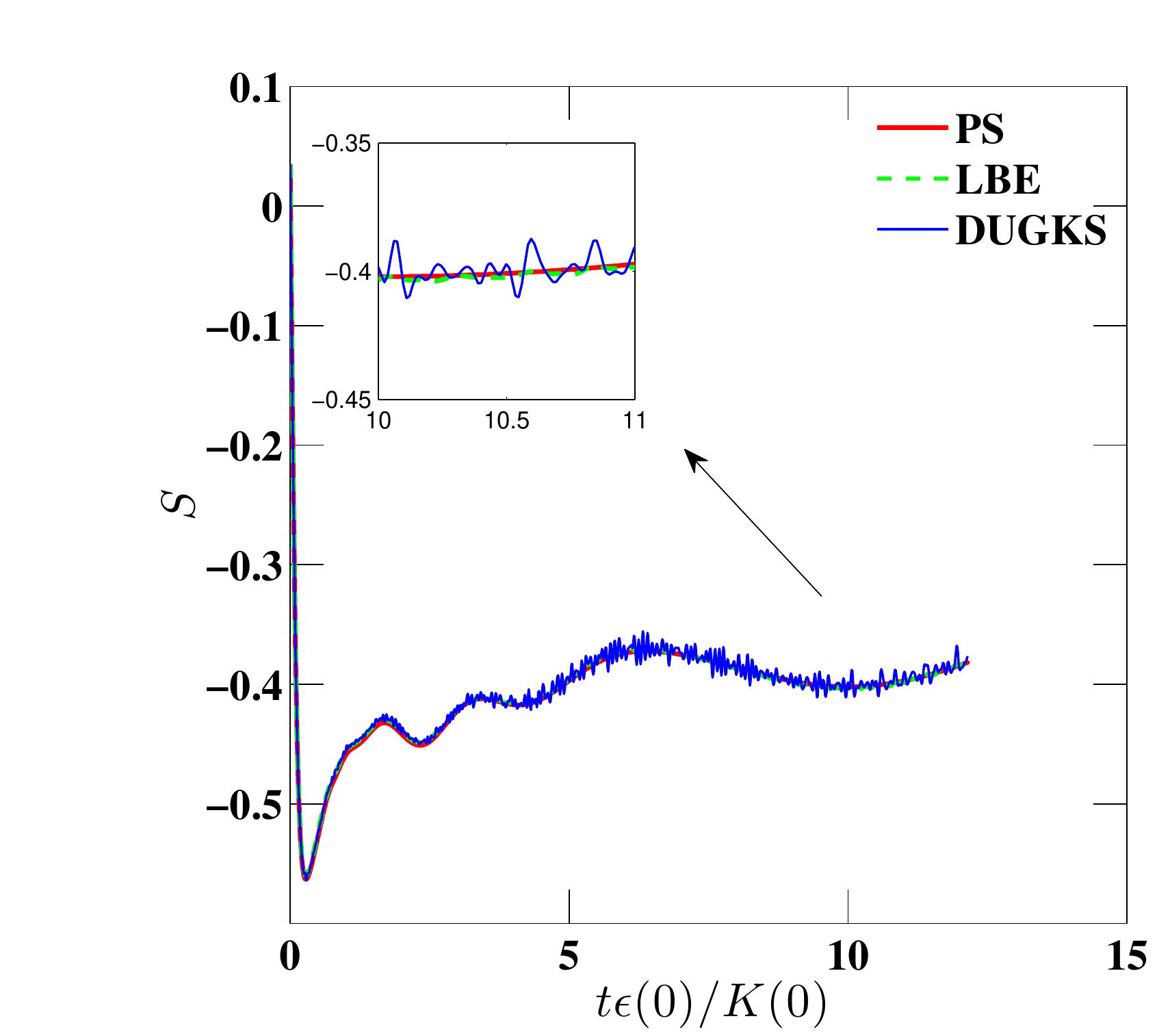}
\caption{$N^3=256^3$}
\label{fig:ske09_mesh256}
\end{subfigure}
\caption{Evolutions of velocity-derivative skewness $S$ with different mesh resolutions at $\text{Re}_{\lambda}=26.06$.} \label{fig:skewness}
\end{figure}

\begin{figure}[hbp]
\centering
\begin{subfigure}[b]{0.48\textwidth}
\includegraphics[width=\textwidth]{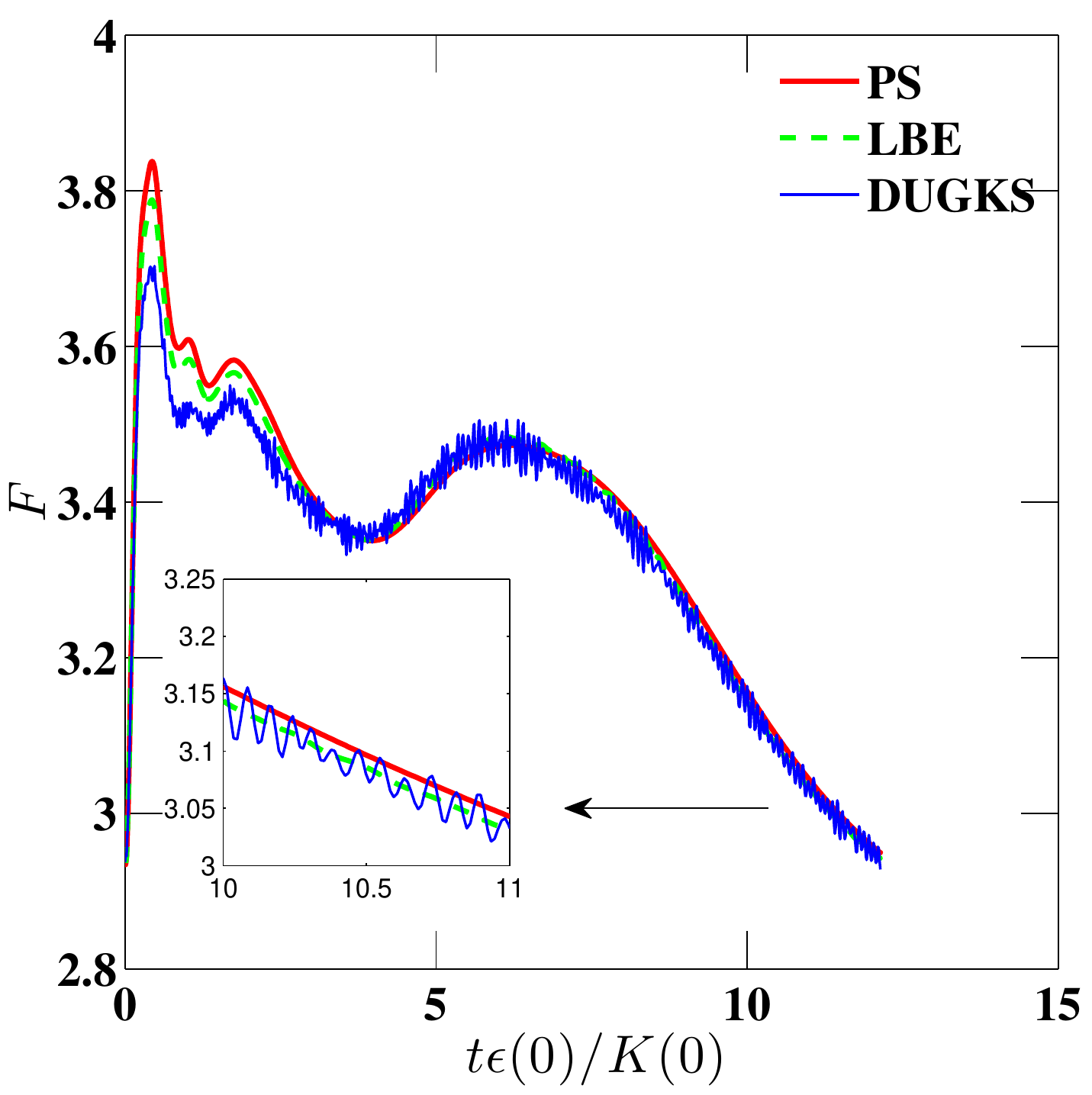}
\caption{$N^3=128^3$}
\label{fig:fla09}
\end{subfigure}~
\begin{subfigure}[b]{0.57\textwidth}
\includegraphics[width=\textwidth]{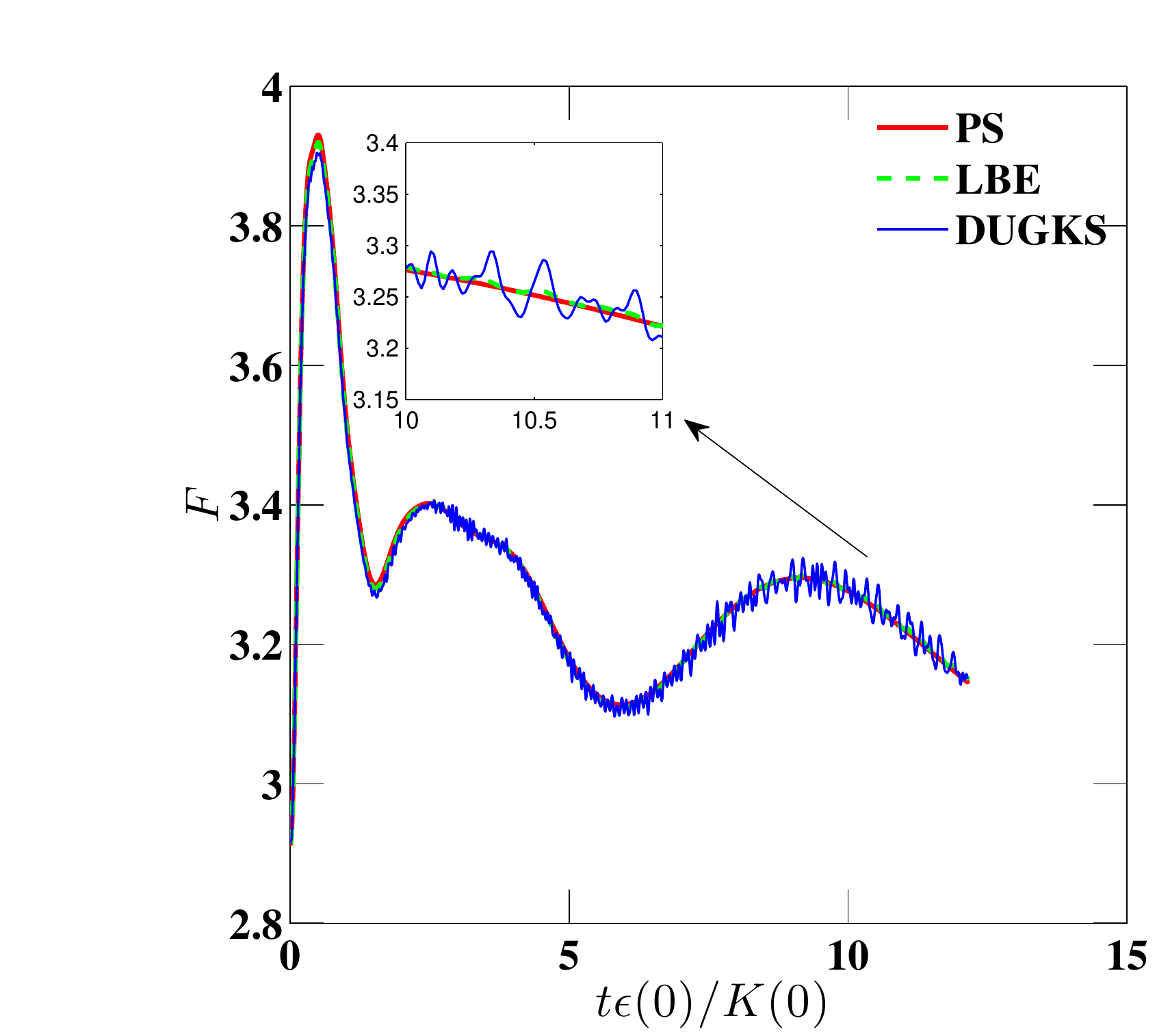}
\caption{$N^3=256^3$}
\label{fig:fla09_mesh256}
\end{subfigure}
\caption{Evolutions of velocity-derivative flatness $F$ with different mesh resolutions at $\text{Re}_{\lambda}=26.06$.} \label{fig:flatness}
\end{figure}

The time evolutions of the averaged velocity-derivative skewness and flatness predicted by these three methods are shown in Figs.~\ref{fig:skewness} and ~\ref{fig:flatness}, respectively. It can be seen that  the results of LBE with the mesh of $128^3$ are in good agreement with the PS solutions, while those of the DUGKS show some high frequency oscillations, although the tendency agrees reasonably with the results from the PS simulation. The oscillations can be attributed to the acoustic waves in the system.
The remarkable discrepancy between the LBE and DUGKS results may be caused by the following reasons: firstly, in the MRT-LBE model, the bulk viscosity can be adjusted by tuning the relaxation
time $s_1$ to absorb the acoustic waves, whereas the BGK based DUGKS does't have such a dissipation mechanism due to the single relaxation time in the BGK equation.  Actually, the results of MRT-LBE with small bulk viscosity also have high frequency oscillations shown in Ref.~\cite{peng2010comparison}, where the bulk viscosity $\zeta$ = $0.0273$ compared to $\zeta=0.1134$ in the present simulation;
secondly, since the velocity-derivative skewness and flatness are the third order and four order moments of $\nabla \bm{u}$, respectively, it is a significant challenge for a second-order method to compute such high-order quantities that are governed by small scales. As demonstrated, both the LBE and DUGKS methods have small numerical dissipation so that both methods can accurately compute the low-order statistic quantities that are governed by large scales. But the numerical dissipation of DUGKS is slightly larger than the LBE method, and yet, the absent acoustic-wave dissipation mechanism enlarges the discrepancy as the velocity field decays and consequently results in errors in high-order quantities.  The high-order errors, however, seem to have little impact on the kinetic energy and dissipation rate.

\begin{figure}[htbp]
\centering
\begin{subfigure}[b]{0.48\textwidth}
\includegraphics[width=\textwidth]{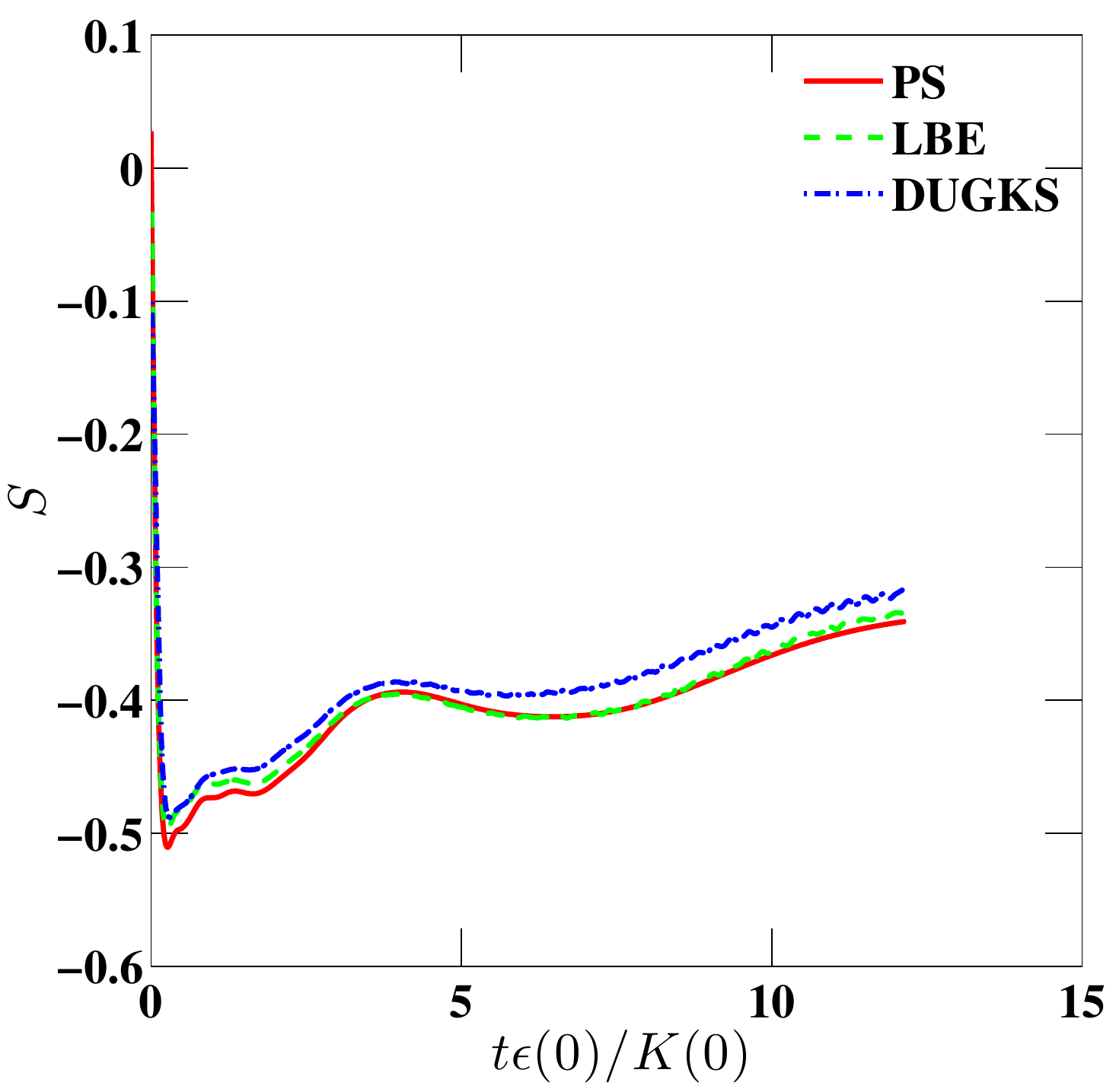}
\caption{$N^3=128^3$}
\label{fig:ske09s}
\end{subfigure}~
\begin{subfigure}[b]{0.48\textwidth}
\includegraphics[width=\textwidth]{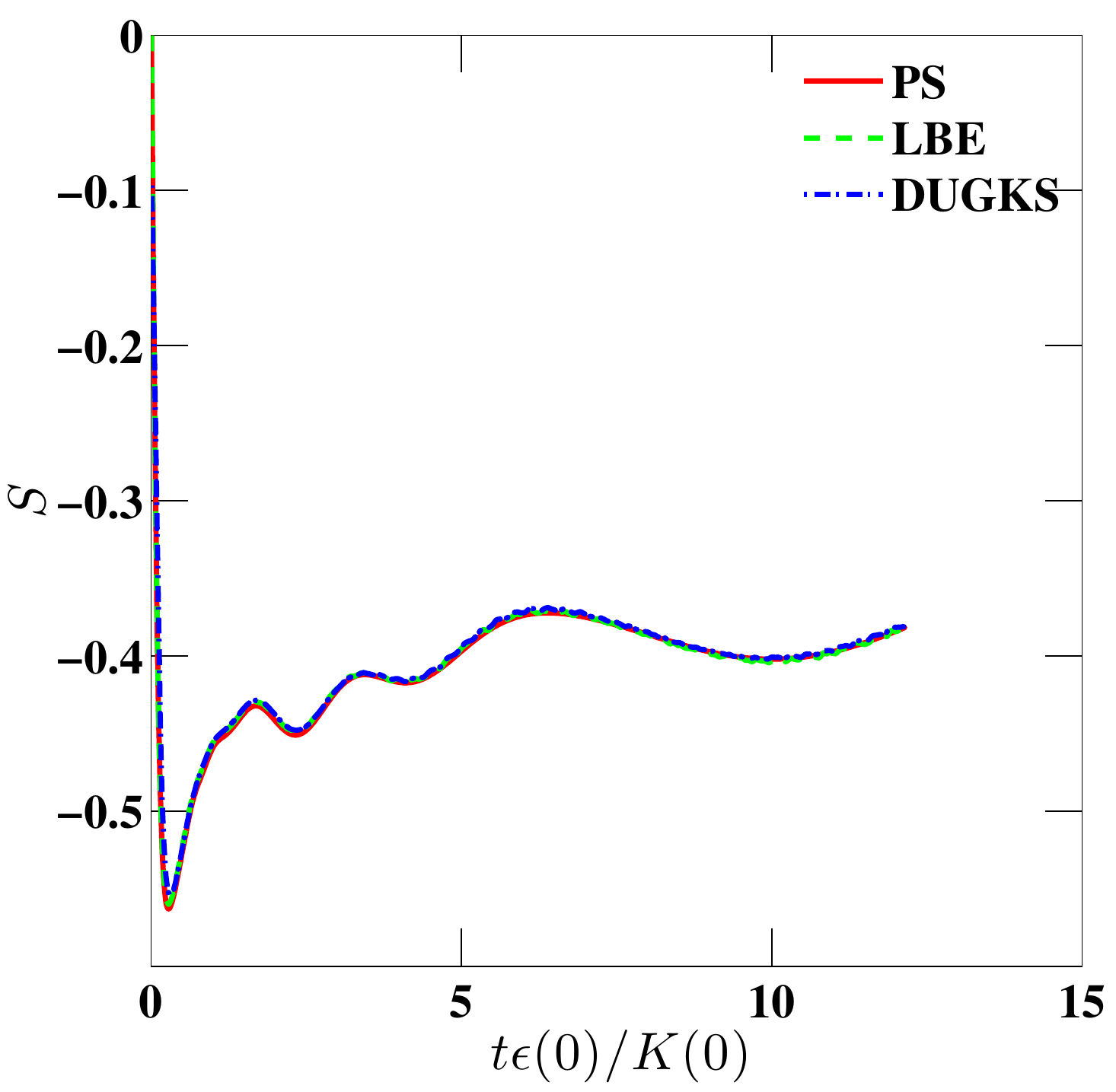}
\caption{$N^3=256^3$}
\label{fig:ske09s_mesh256}
\end{subfigure}
\caption{Evolutions of smoothed velocity-derivative skewness $S$ with different mesh resolutions at $\text{Re}_{\lambda}=26.06$.} \label{fig:skewness_smooth}
\end{figure}

\begin{figure}[htbp]
\centering
\begin{subfigure}[b]{0.48\textwidth}
\includegraphics[width=\textwidth]{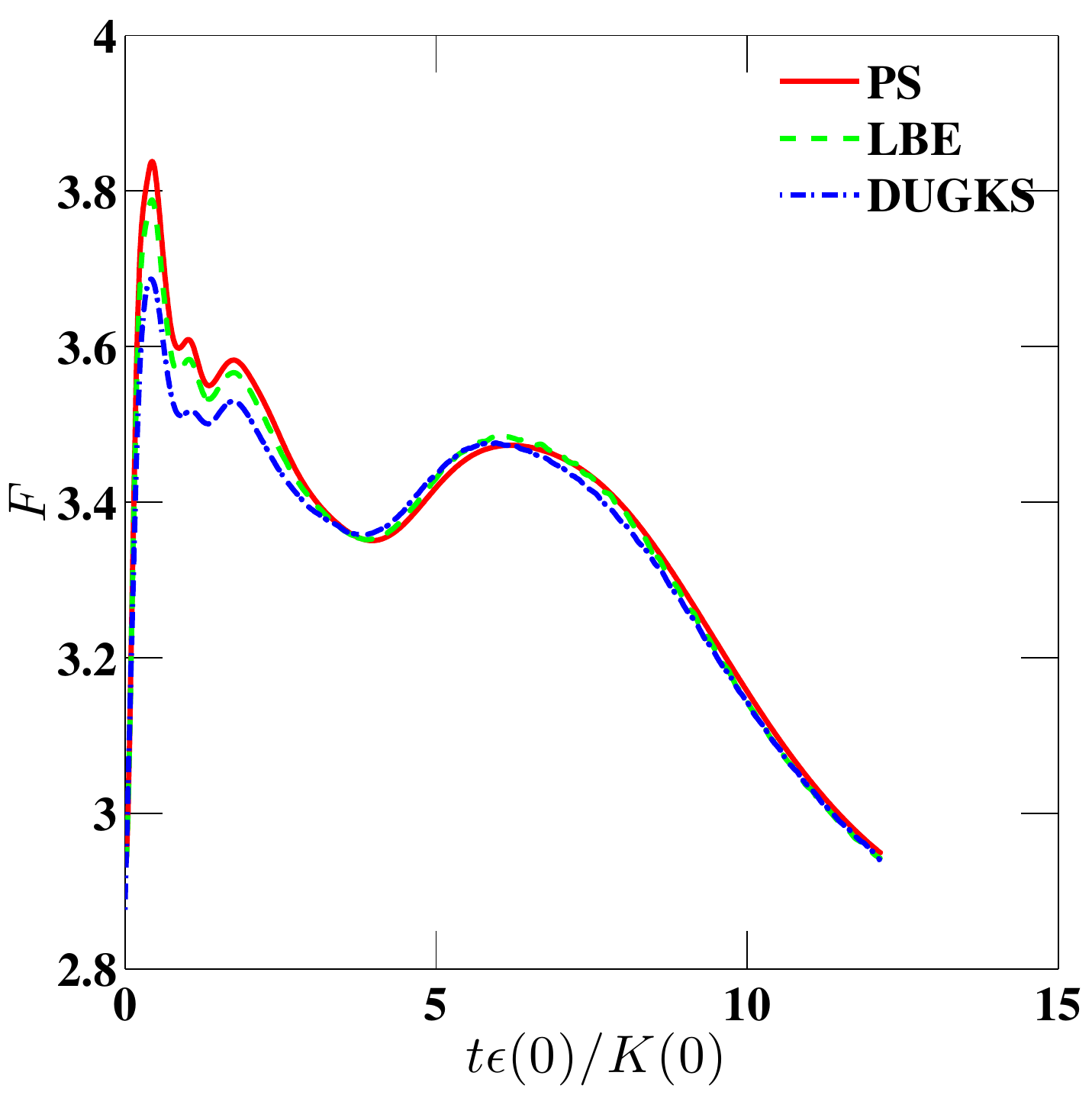}
\caption{$N^3=128^3$}
\label{fig:fla09s}
\end{subfigure}~
\begin{subfigure}[b]{0.48\textwidth}
\includegraphics[width=\textwidth]{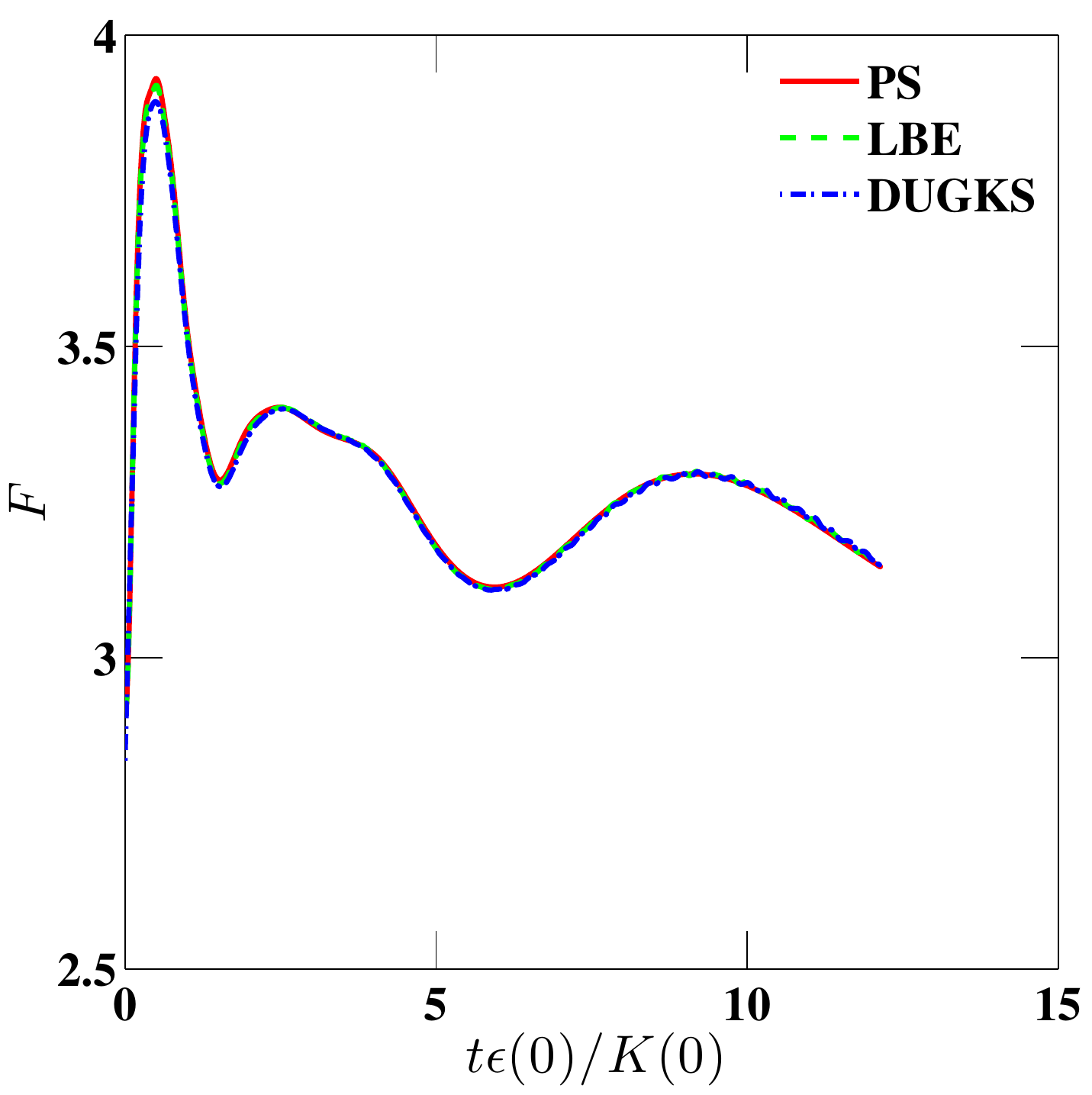}
\caption{$N^3=256^3$}
\label{fig:fla09s_mesh256}
\end{subfigure}
\caption{Evolutions of smoothed velocity-derivative flatness $F$  with different mesh resolutions at $\text{Re}_{\lambda}=26.06$.} \label{fig:flatness_smooth}
\end{figure}

As previous noted, increasing the mesh resolution can reduce the numerical dissipation, thus better results should be obtained in $256^3$ simulations. As expected, as shown in Figs.~\ref{fig:ske09_mesh256} and ~\ref{fig:fla09_mesh256}, the results of DUGKS with $N^3=256^3$ are better than those with the coarse mesh, and the magnitudes of the oscillations are also reduced.

\begin{table}[!htb]
\centering
\caption{\label{tablef} The maximum errors of $S$ and $F$ relative to PS results. }
\begin{tabular}[t]
   {c c c c c c c}
   \hline
    \hline
  & $\text{Case}$& \hspace{2mm}  $\text{LBE128}$  &\hspace{2mm} $\text{LBE256}$ & \hspace{2mm} $\text{DUGKS128}$ & \hspace{2mm} $\text{DUGKS256}$ &\\
\hline
& $R_m(S)$       & \hspace{2mm}  $3.35\%       $  &\hspace{2mm} $4.52\%$        & \hspace{2mm} $11.97\%$          & \hspace{2mm}  $4.98\%$         &\\
& $R_m(F)$       & \hspace{2mm}  $1.30\%       $  &\hspace{2mm} $0.35\%$        & \hspace{2mm} $3.97\%$          & \hspace{2mm}   $1.11\%$         &\\
  \hline
   \hline
\end{tabular}
\end{table}

For convenient comparison, the results of the DUGKS can be filtered out by simple smoothing through averaging (using the smooth function in the matlab), as suggested in Ref.~\cite{peng2010comparison}. The smoothed skewness and flatness results are shown in Figs.~\ref{fig:skewness_smooth} and ~\ref{fig:flatness_smooth}, respectively. It is found that both LBE and DUGKS results indeed agree well with the PS results.
Quantitatively, as given in Table~\ref{tablef}, the maximum relative error of $S$ predicted by the DUGKS with the mesh of $128^3$ is $11.97\%$, while for the LBE, this value is $3.35\%$. As the resolution increases to $256^3$, the maximum relative error of $S$ computed by the DUGKS reduces to $4.98\%$.

\subsection{Effects of the Reynolds number}
 In the above subsections, we have made some detailed comparisons  between the LBE and DUGKS methods with the initial $\text{Re}_{\lambda}=26.06$, at which the initial flow fields can be well-resolved by both methods. In order to further compare the performance of the LBE and DUGKS methods at higher $\text{Re}_{\lambda}$, we conduct the DNS of the DHIT at $\text{Re}_{\lambda}=52.12$ and $104.24$ with a fixed mesh of $128^3$. Accordingly, the spatial resolution parameters $k_{max}\eta$ is $2.18$ for $\text{Re}_{\lambda}=52.12$ and $1.54$ for $\text{Re}_{\lambda}=104.24$, suggesting that the PS method can adequately resolve the initial flow field \cite{eswaran1988examination}. However, it is not clear whether this resolution is sufficient for the LBE and DUGKS methods at these $\text{Re}_{\lambda}$. Herein we compare some key statistic quantities obtained by both kinetic approaches at these $\text{Re}_{\lambda}$ with those from the PS simulations.

\begin{figure}[htbp]
\centering
\begin{subfigure}[b]{0.48\textwidth}
\includegraphics[width=\textwidth]{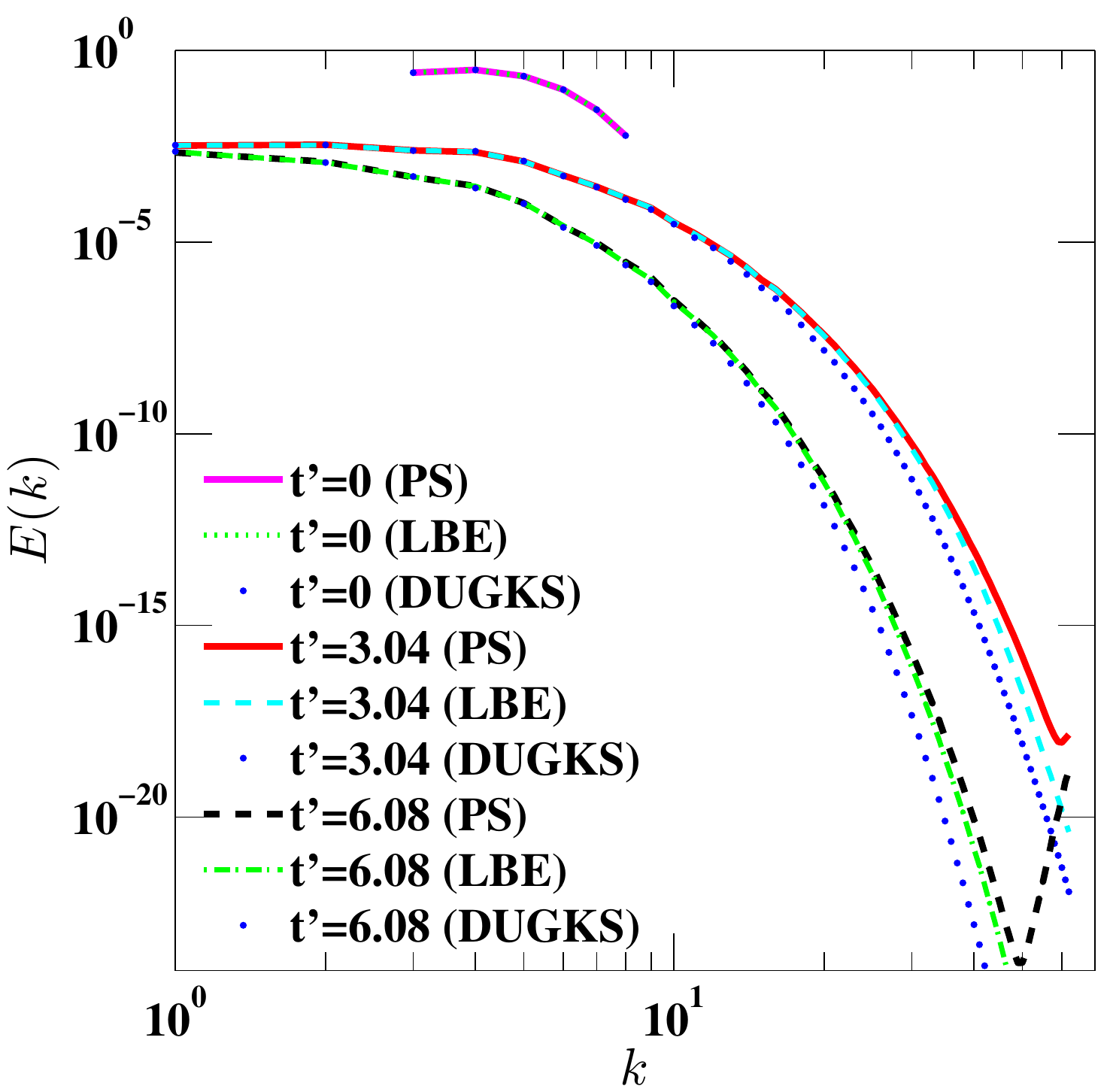}
\caption{$\text{Re}_{\lambda}=52.12,  N^3=128^3$}
\end{subfigure}~
\begin{subfigure}[b]{0.48\textwidth}
\includegraphics[width=\textwidth]{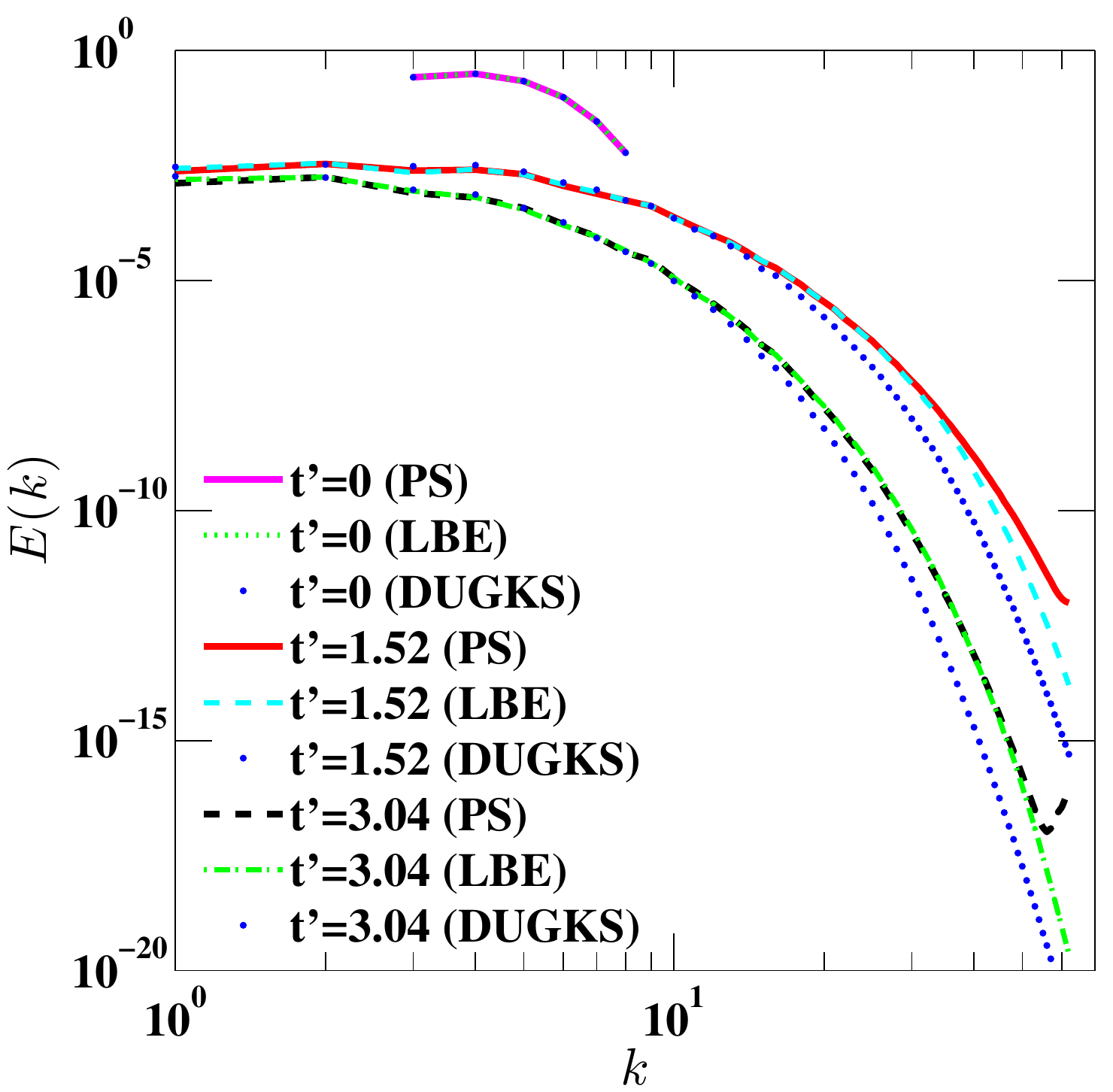}
\caption{$\text{Re}_{\lambda}=104.24,  N^3=128^3$}
\end{subfigure}
\caption{
The energy spectra $E(k,t)$ at different $\text{Re}_{\lambda}$. } \label{fig:especs}
\end{figure}

\begin{figure}[htbp]
\centering
\begin{subfigure}[b]{0.48\textwidth}
\includegraphics[width=\textwidth]{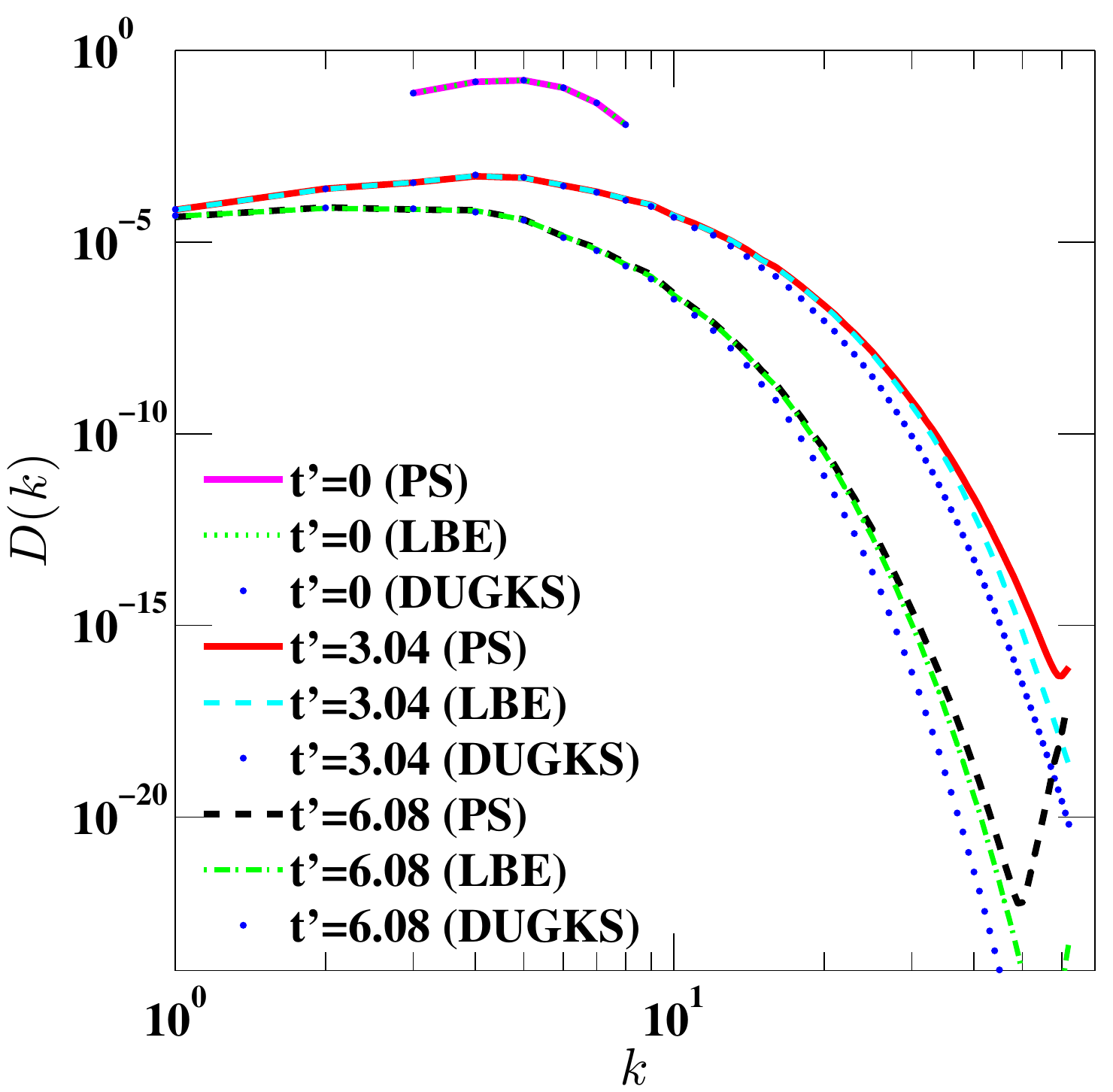}
\caption{$\text{Re}_{\lambda}=52.12,  N^3=128^3$}
\end{subfigure}~
\begin{subfigure}[b]{0.48\textwidth}
\includegraphics[width=\textwidth]{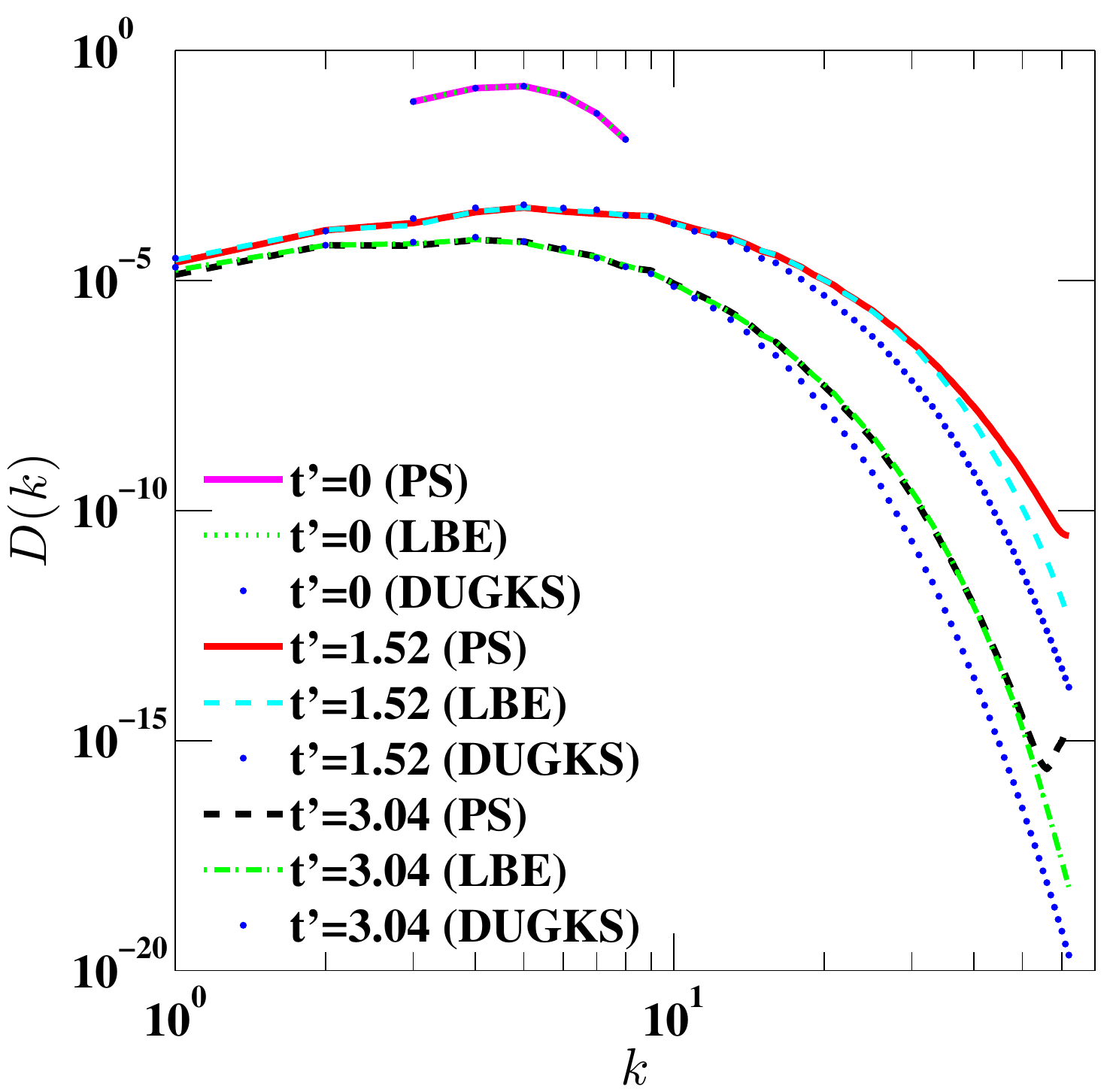}
\caption{$\text{Re}_{\lambda}=104.24, N^3=128^3$}
\end{subfigure}
\caption{
The dissipation rate spectra $D(k,t)$ at different $\text{Re}_{\lambda}$.} \label{fig:diss}
\end{figure}

\begin{figure}[htbp]
\centering
\begin{subfigure}[b]{0.48\textwidth}
\includegraphics[width=\textwidth]{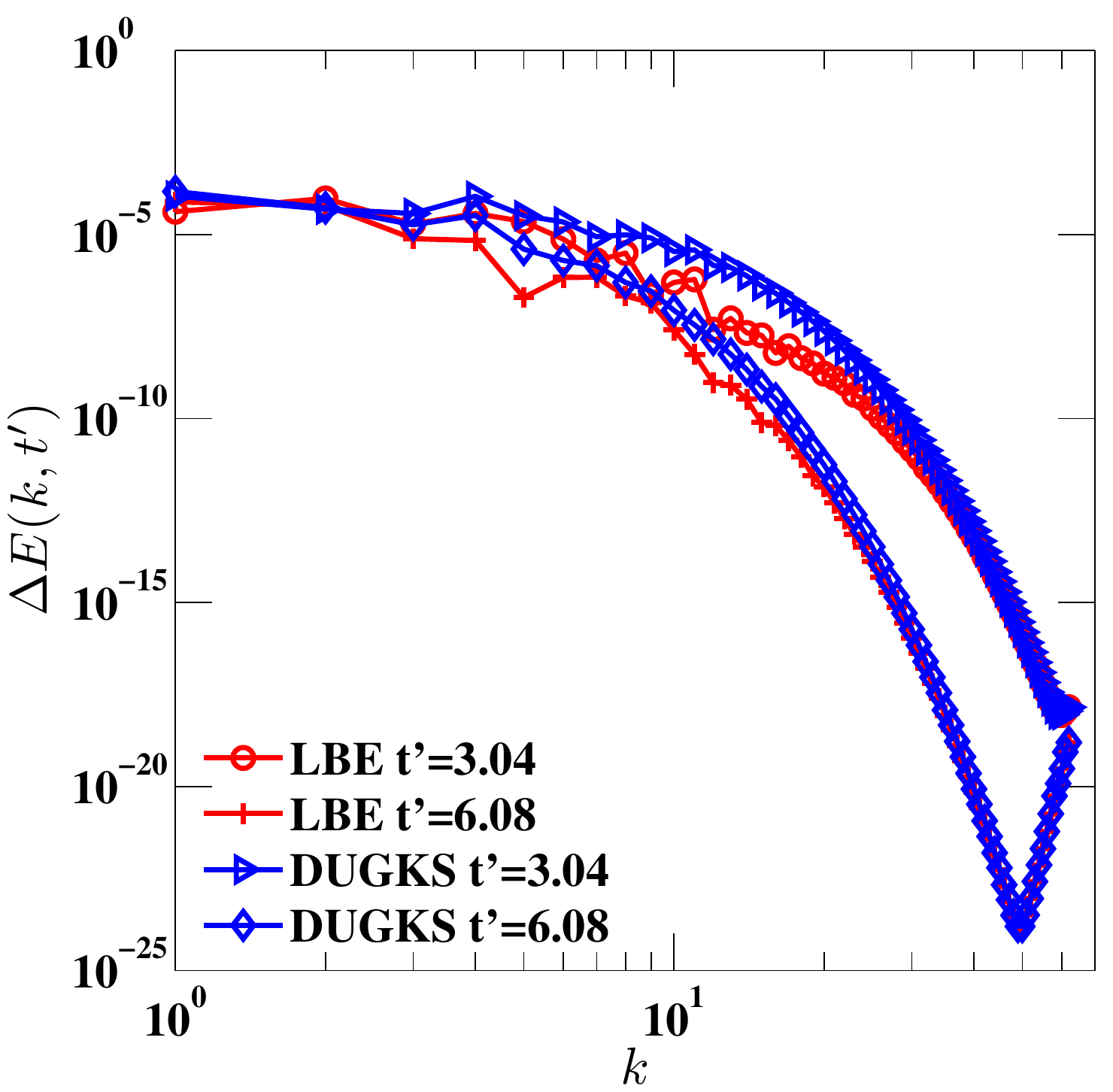}
\caption{$\text{Re}_{\lambda}=52.12,  N^3=128^3$}
\end{subfigure}~
\begin{subfigure}[b]{0.48\textwidth}
\includegraphics[width=\textwidth]{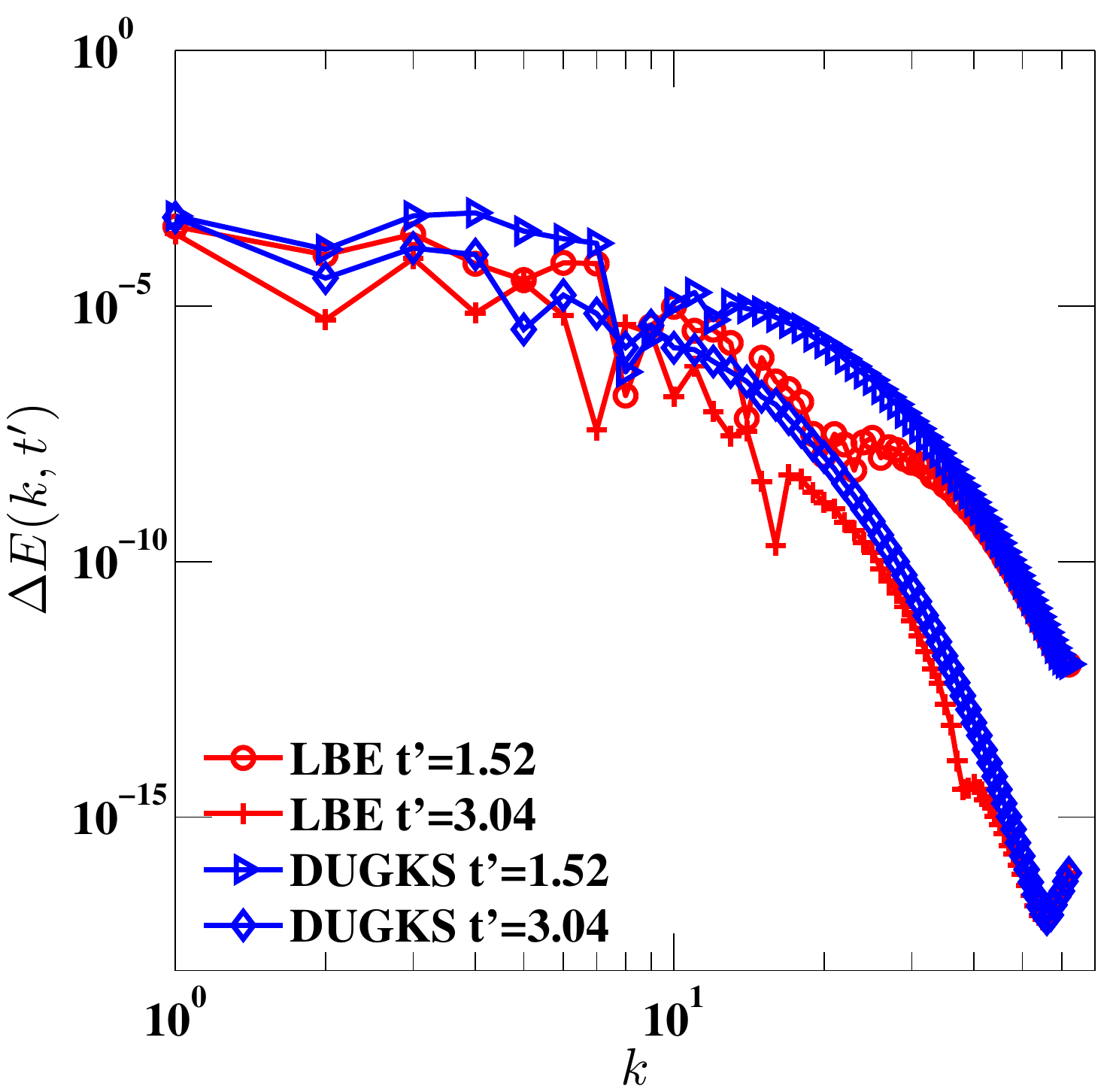}
\caption{$\text{Re}_{\lambda}=104.24, N^3=128^3$}
\end{subfigure}
\caption{
The energy spectra difference $\Delta E(k,t')$ at different $\text{Re}_{\lambda}$.} \label{fig:EDs}
\end{figure}

\begin{figure}[htbp]
\centering
\begin{subfigure}[b]{0.48\textwidth}
\includegraphics[width=\textwidth]{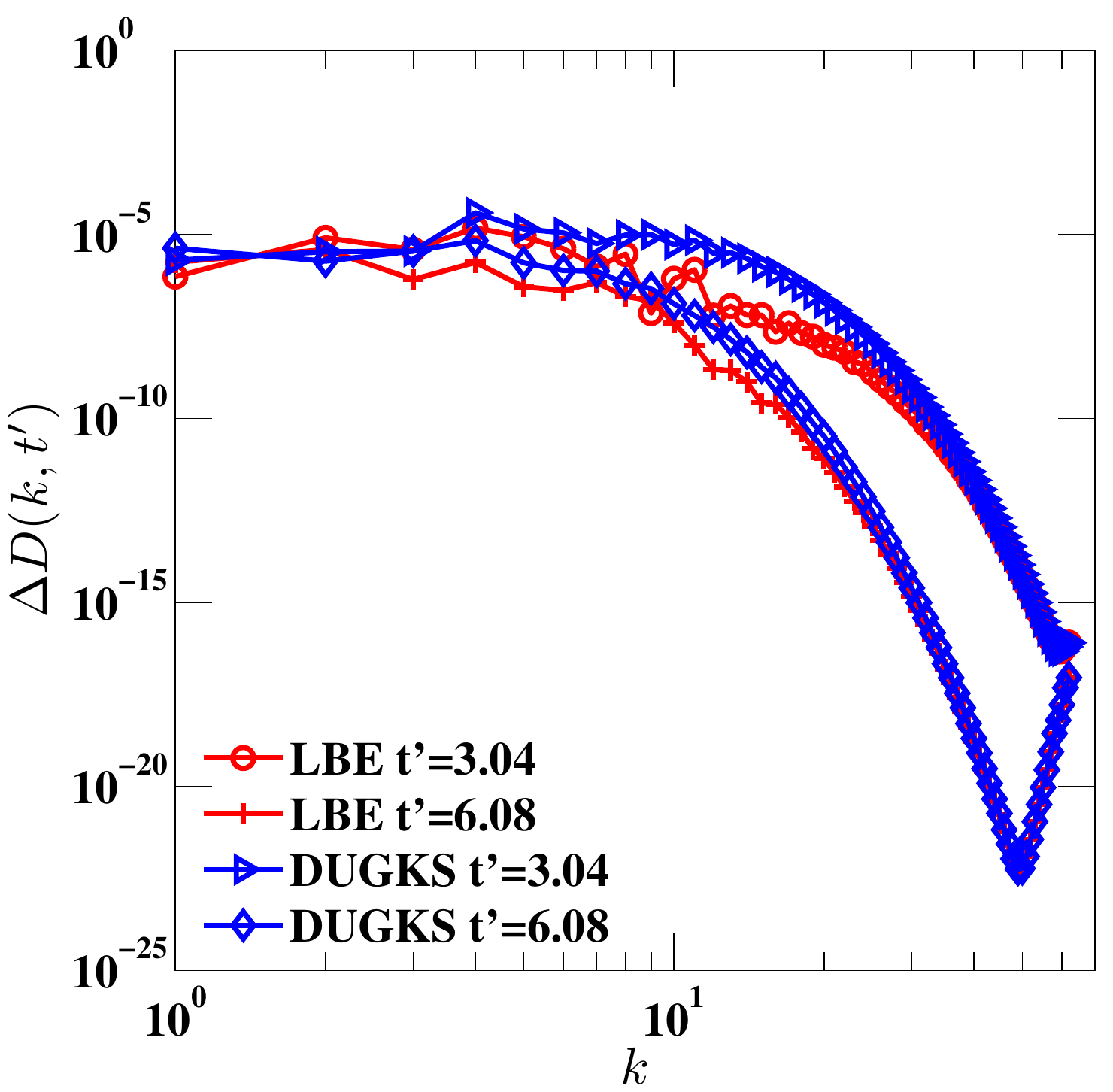}
\caption{$\text{Re}_{\lambda}=52.12,  N^3=128^3$}
\end{subfigure}~
\begin{subfigure}[b]{0.48\textwidth}
\includegraphics[width=\textwidth]{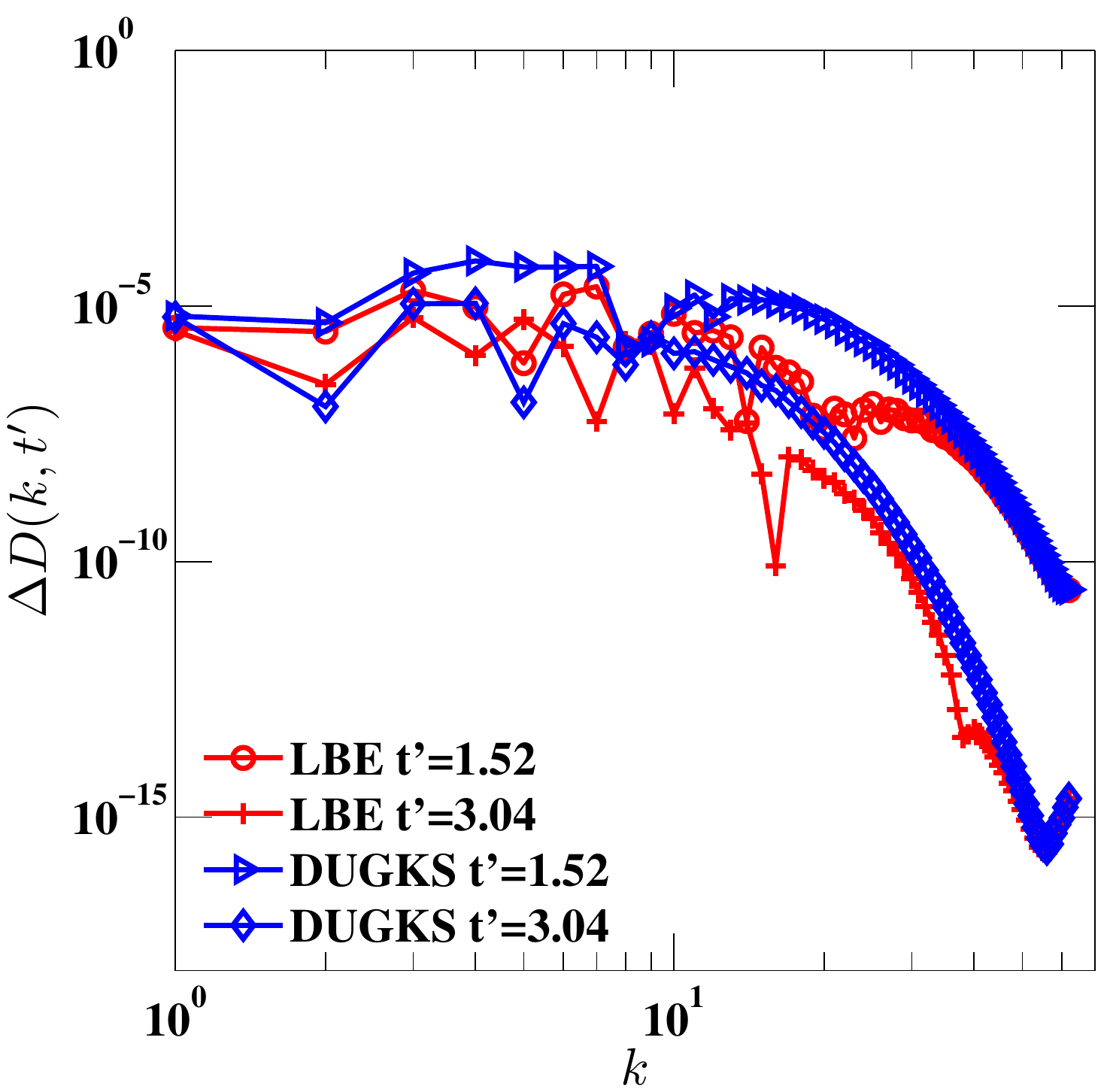}
\caption{$\text{Re}_{\lambda}=104.24, N^3=128^3$}
\end{subfigure}
\caption{
The dissipation rate spectra difference $\Delta D(k,t)$ at different $\text{Re}_{\lambda}$.} \label{fig:DDs}
\end{figure}

Figures~\ref{fig:especs} and ~\ref{fig:diss} show the energy spectra $E(k,t)$ and the dissipation rate spectra $D(k,t)$ at different times. It is observed that,  $E(k,t)$ and $D(k,t)$ obtained by the LBE method
are still in good agreement with those from the PS method, while the results from the DUGKS clearly deviate from the PS results in the high wavenumber region and the discrepancies increase with $\text{Re}_{\lambda}$.
The differences of the spectra between both kinetic methods and the PS method, as defined by Eq.~\eqref{errs}, are shown in Figs.~\ref{fig:EDs} and \ref{fig:DDs}. It can be clearly seen that the LBE method yields better predictions than the DUGKS.

\begin{figure}[htbp]
\centering
\begin{subfigure}[b]{0.48\textwidth}
\includegraphics[width=\textwidth]{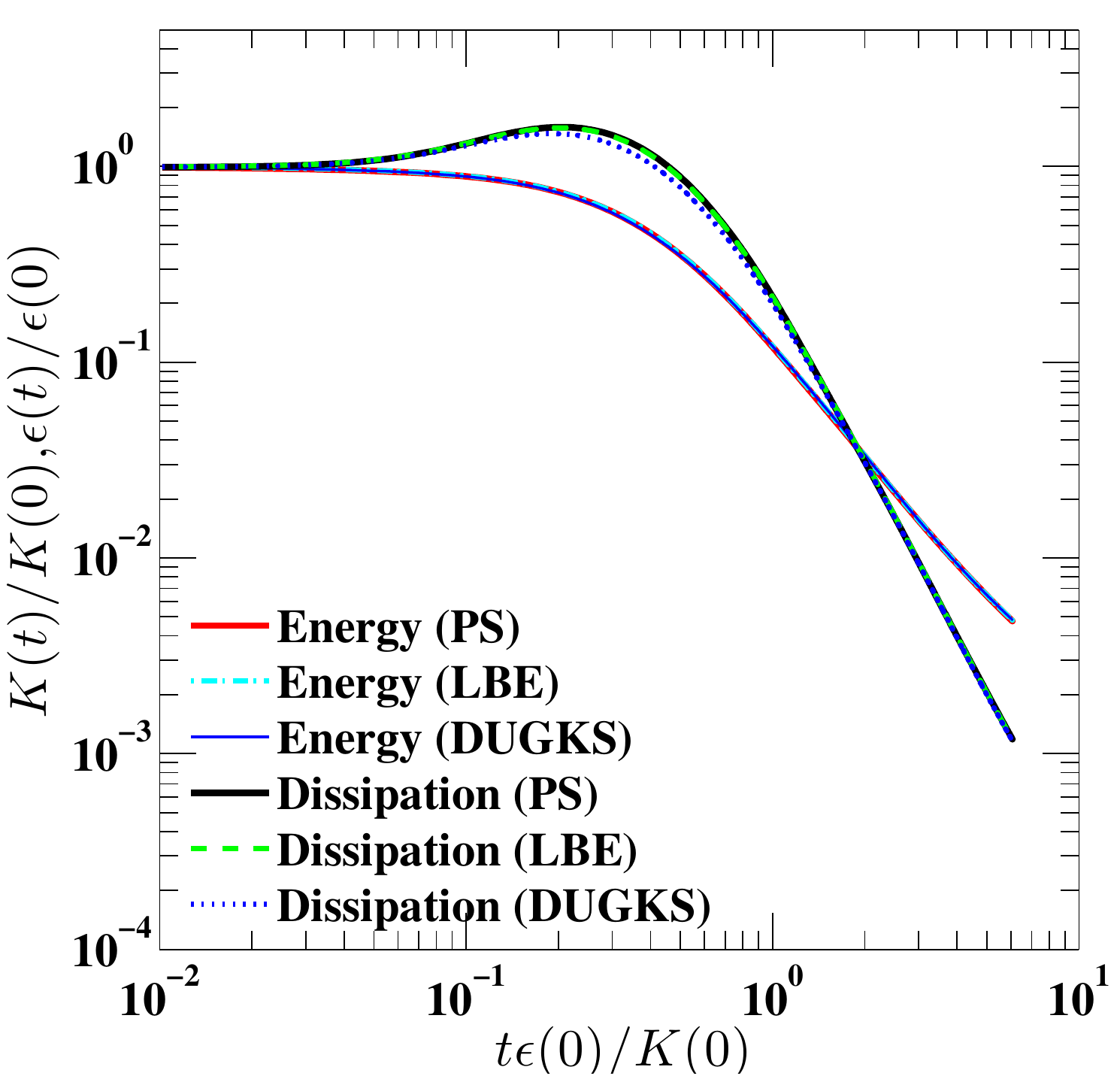}
\caption{$\text{Re}_{\lambda}=52.12, N^3=128^3$}
\end{subfigure}~
\begin{subfigure}[b]{0.48\textwidth}
\includegraphics[width=\textwidth]{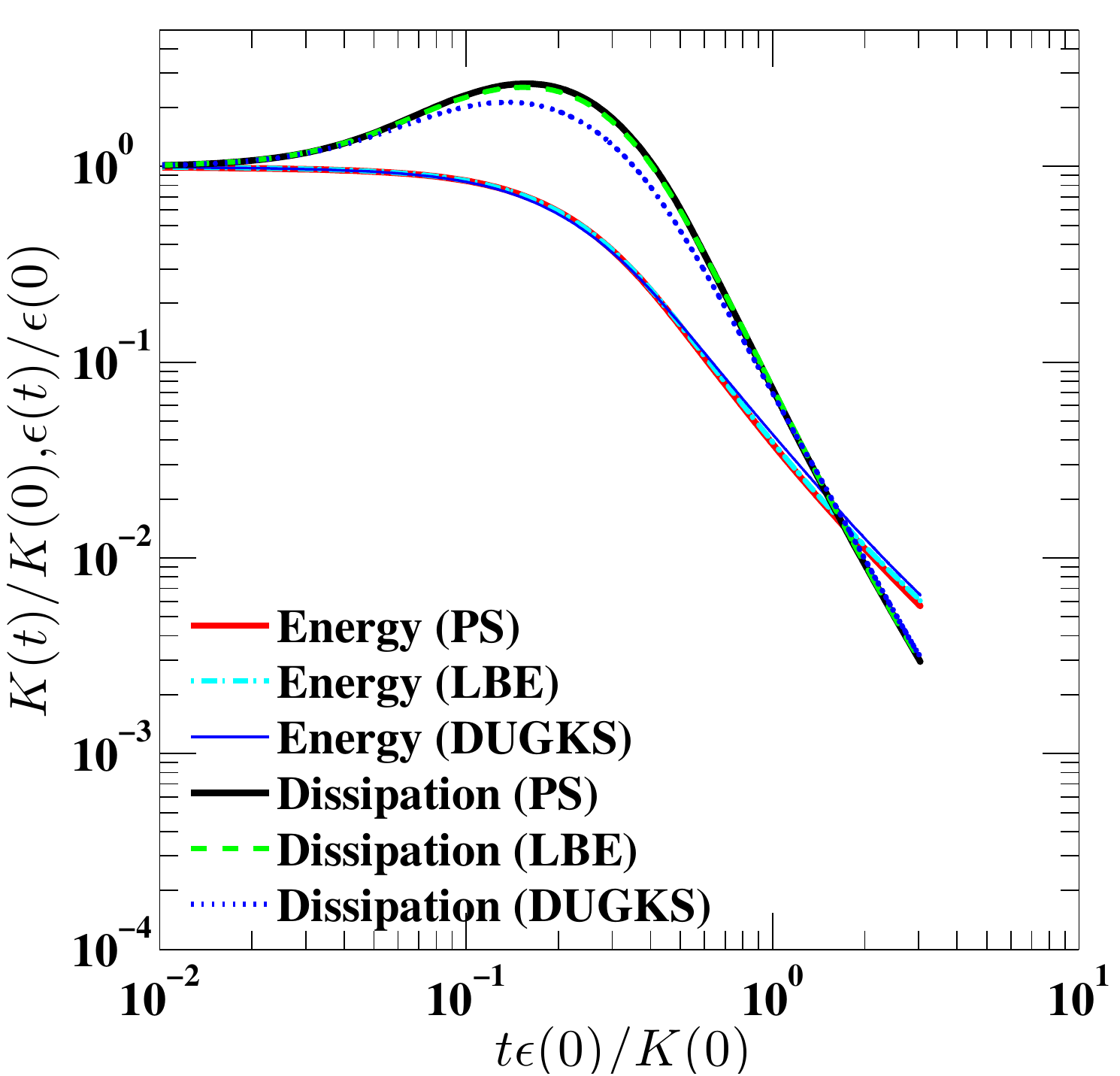}
\caption{$\text{Re}_{\lambda}=104.24, N^3=128^3$}
\end{subfigure}
\caption{
Evolutions of the normalized total kinetic energy $K(t)/K_0$ and the normalized dissipation rate $\epsilon(t)/\epsilon_0$ at different $\text{Re}_{\lambda}$.} \label{fig:eds}
\end{figure}

We also compare the evolutions of the normalized kinetic energy and the dissipation rate. As shown in Fig.~\ref{fig:eds}, $K(t)/K_0$ obtained by LBE and DUGKS methods are in good agreements with the PS results.
However, the differences are visible around the peak values of $\epsilon(t)/\epsilon_0$ computed by both methods, and it can be clearly seen that the LBE method gives a better prediction than the DUGKS. For example, for the case of $\text{Re}_{\lambda}=104.24$, the maximum relative error of $\epsilon(t)/\epsilon_0$ predicted by the DUGKS is $26.7\%$, while for the LBE that is $4.48\%$. This indicates that at $\text{Re}_{\lambda}=104.24$, the mesh resolution for the DUGKS is insufficient to resolve the flow field at different times.

\begin{figure}[htbp]
\centering
\begin{subfigure}[b]{0.48\textwidth}
\includegraphics[width=\textwidth]{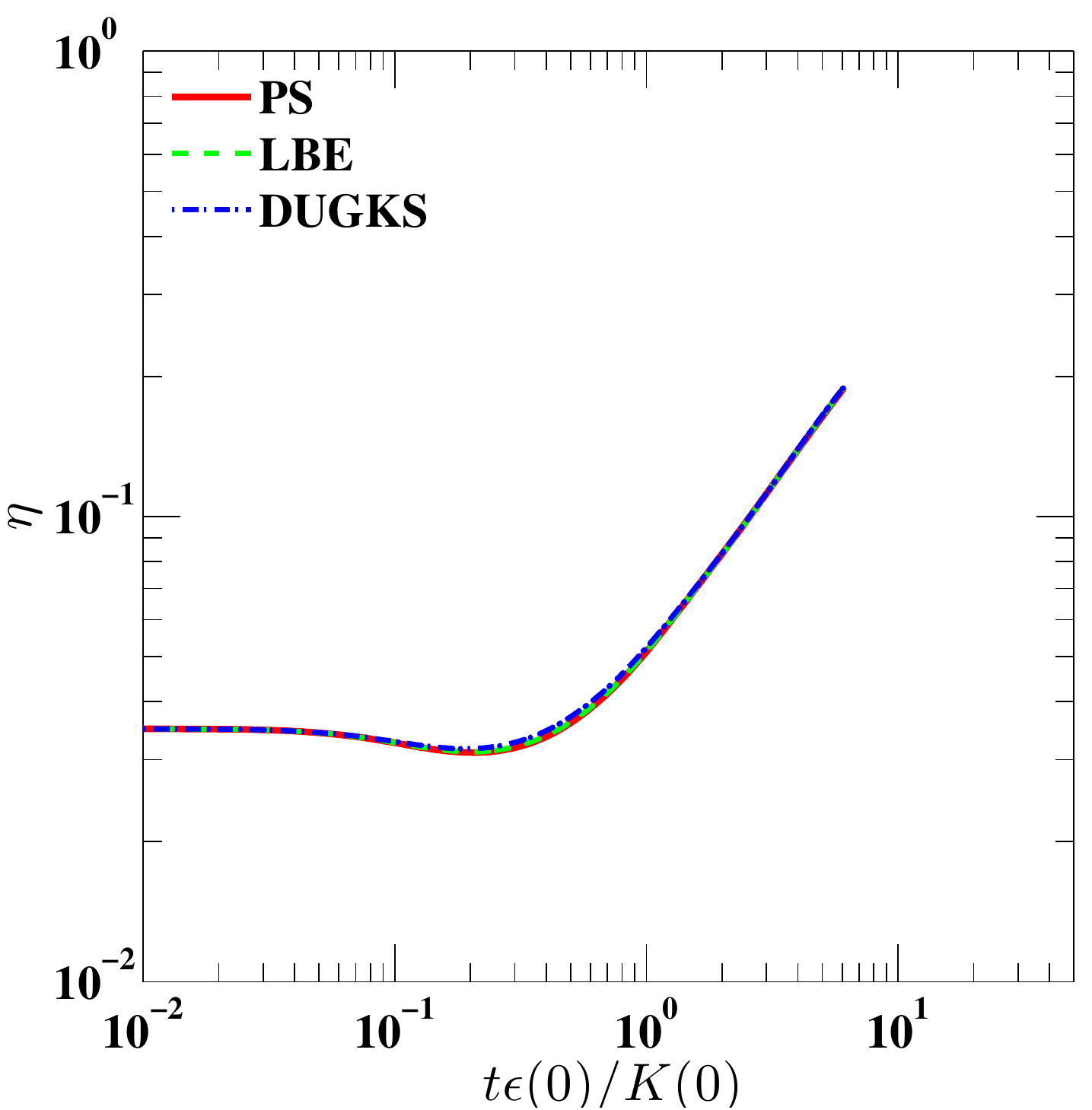}
\caption{$\text{Re}_{\lambda}=52.12, N^3=128^3$}
\end{subfigure}~
\begin{subfigure}[b]{0.48\textwidth}
\includegraphics[width=\textwidth]{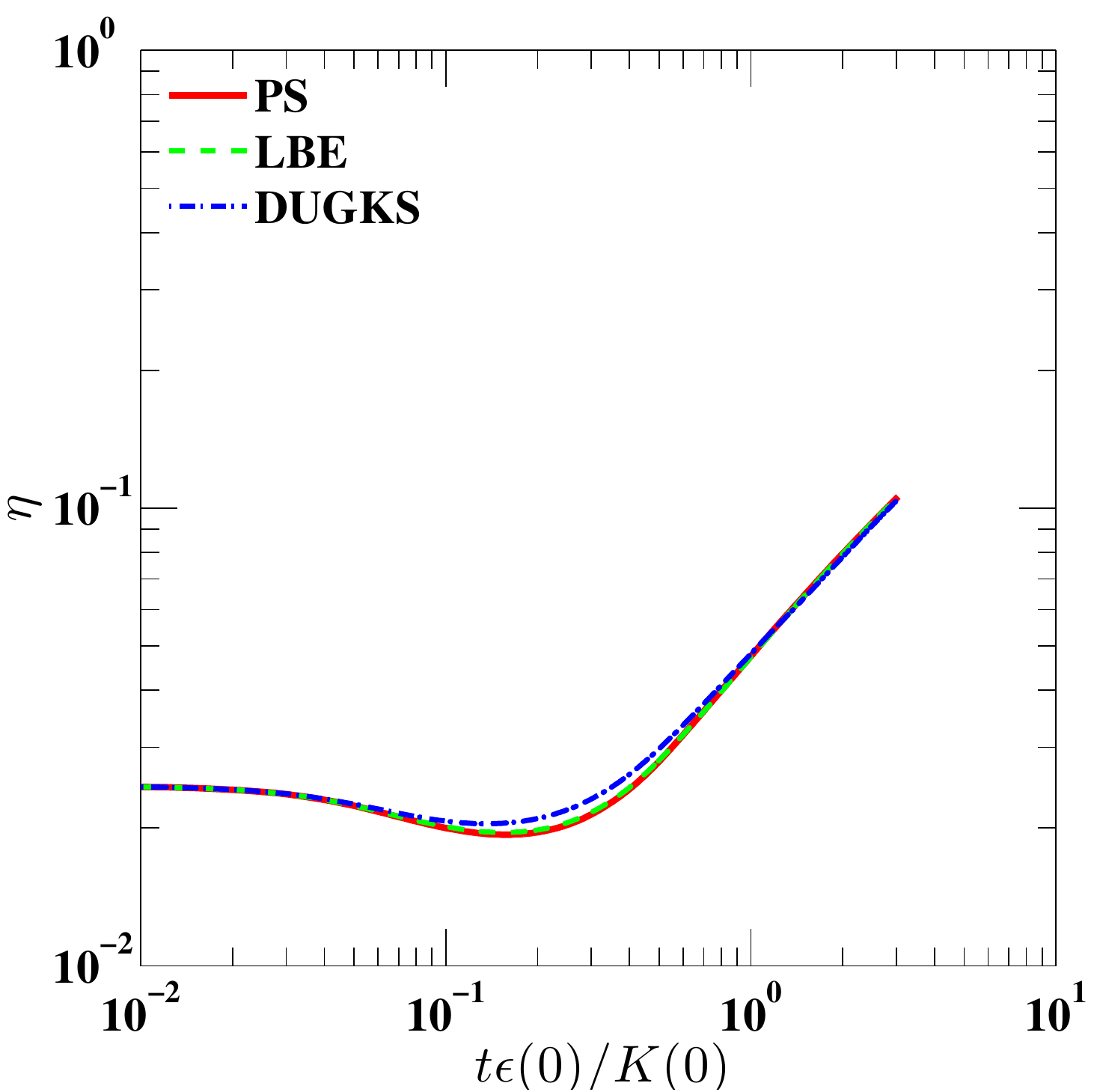}
\caption{$\text{Re}_{\lambda}=104.24, N^3=128^3$}
\end{subfigure}
\caption{Evolutions of the Kolmogorov length $\eta$ at different $\text{Re}_{\lambda}$.} \label{fig:komos}
\end{figure}

\begin{figure}[htbp]
\centering
\begin{subfigure}[b]{0.48\textwidth}
\includegraphics[width=\textwidth]{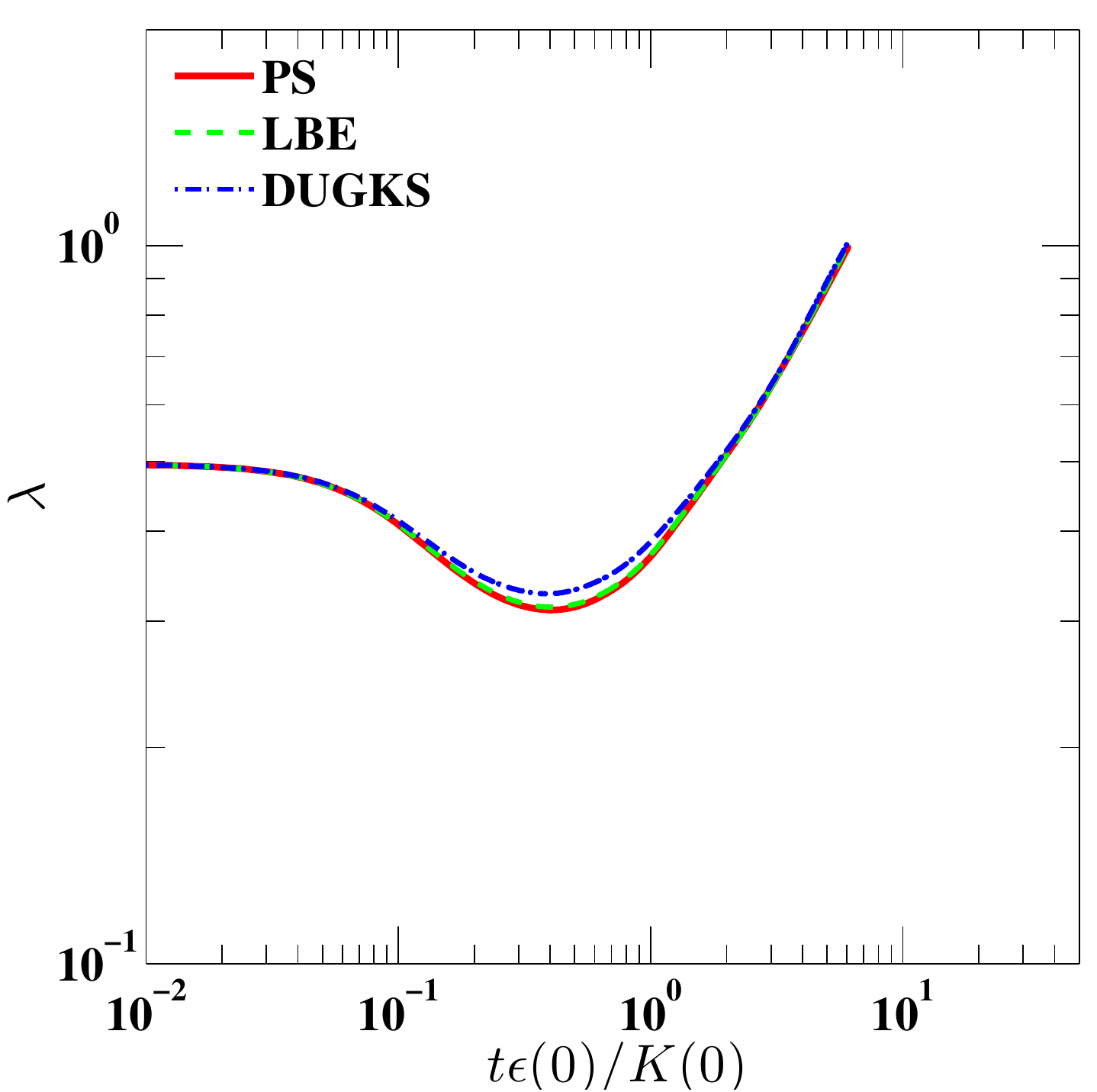}
\caption{$\text{Re}_{\lambda}=52.12£¬N^3=128^3$}
\end{subfigure}~
\begin{subfigure}[b]{0.48\textwidth}
\includegraphics[width=\textwidth]{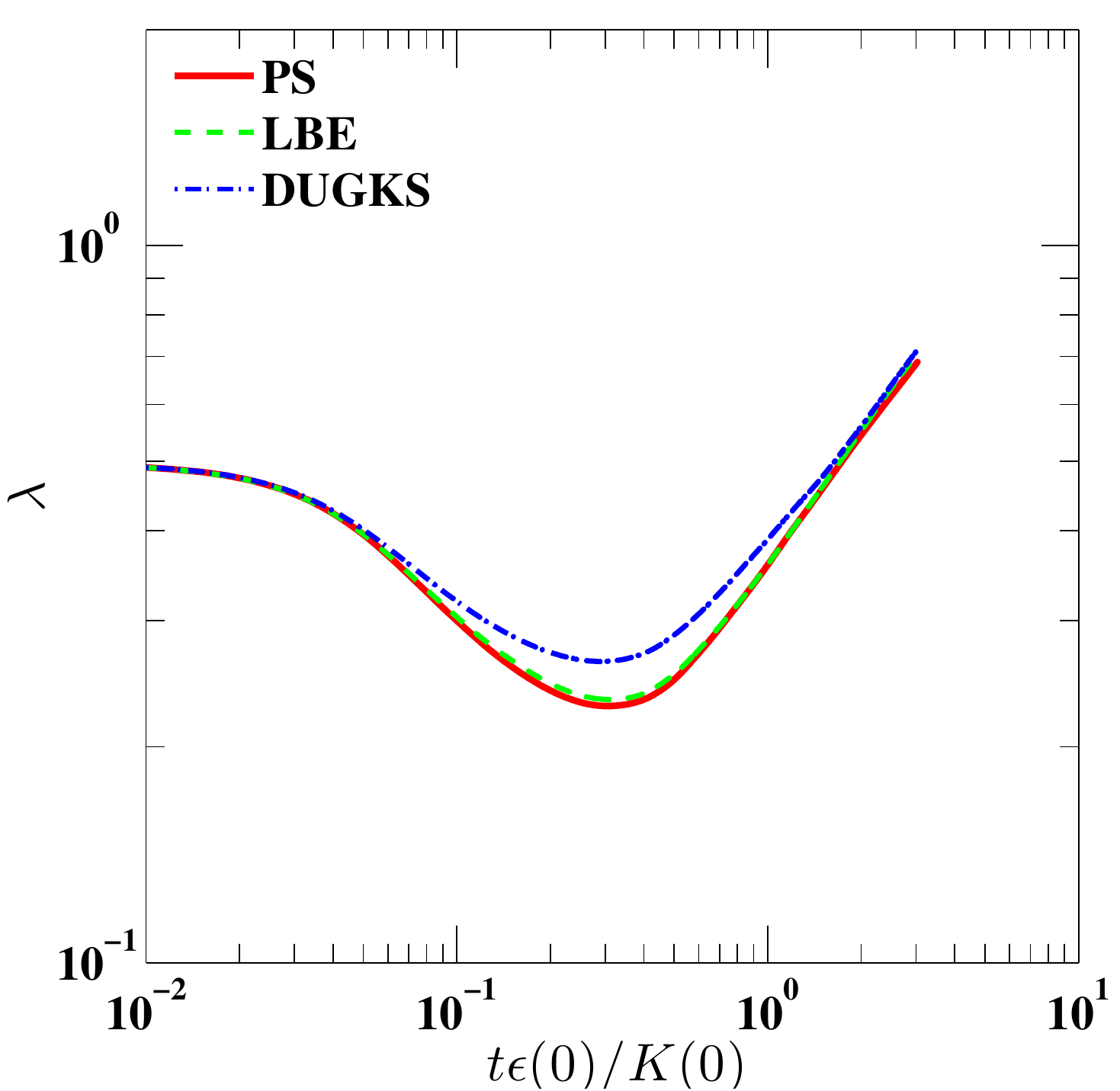}
\caption{$\text{Re}_{\lambda}=104.24, N^3=128^3$}
\end{subfigure}
\caption{Evolutions of the Taylor microscale length $\lambda$ at different $\text{Re}_{\lambda}$.} \label{fig:taylors}
\end{figure}

The similar results are also obtained from the evolutions of the Kolmogorov length $\eta$ and the Taylor microscale length $\lambda$, which are shown in Figs~\ref{fig:komos} and ~\ref{fig:taylors}, respectively. We observe that the maximum deviation appears around the minimums of $\eta$ or $\lambda$, where the adequate spatial resolution in DUGKS and LBE is most likely not met. It can be clearly found that the LBE is more accurate than the DUGKS in capturing both scales due to the lower numerical dissipation in LBE.

Upon the above observations, we conclude that the LBE gives more accurate results than the DUGKS at both $\text{Re}_{\lambda}$; with the fixed resolution of $128^3$, the flow fields can be reasonably resolved by the LBE method, but are not adequately resolved by the DUGKS, particularly in the high wavenumber region which represents the small-scales turbulent eddies. This means that the DUGKS has relatively larger numerical dissipation than the LBE.

It is interesting to figure out the reasoning behind the more dissipative nature of the DUGKS than the LBE. One of the major reason is that as a finite volume scheme, additional numerical dissipation is introduced in the DUGKS in the the initial data reconstruction. It should be noted that although the DUGKS is more dissipative than the LBE method, we argue that the coupled collision and transport mechanism in the flux reconstruction can ensure that the DUGKS still has relatively low numerical dissipation when compared with the direct upwinding reconstruction of the original distribution function without considering the collision effects \cite{ohwada2002construction,chen2015comparative}.

\subsection{Computational efficiency and numerical stability}

  Finally, we compare the computational efficiency of the LBE and DUGKS methods on a fixed mesh of ${128}^3$. For each iteration, the CPU time costs of the LBE and DUGKS are 0.666s and 0.911s, respectively, where both codes run on 16 cores based on the message passing interface (MPI) using two dimensional domain decomposition. Therefore, the LBE method is about $36.8\%$ faster than the DUGKS per time step. But, owing to the different time-steps used in the two methods, in our simulations two DUGKS time steps are equivalent to one LBE time step .

In terms of the numerical stability, we compute the maximum stable Taylor microscale Reynolds number of both LBE and DUGKS codes on a mesh of ${128}^3$. In the simulations, we set the CFL number to be $0.9$ in the DUGKS in order to make a fair comparison with the LBE in which the CFL number equals $1.0$. Without considering the accuracy, the LBE code blows up when the Taylor microscale Reynolds number reaches $\text{Re}_{\lambda}=26060$, while the DUGKS is still stable at such $\text{Re}_{\lambda}$. Therefore, the DUGKS is more stable than the LBE methods, which is consistent with the previous study \cite{wang2015comparative}.

\section{Discussions and Conclusions}

In this work, we present a comparative study of two kinetic approaches, the LBE and DUGKS methods, for direct numerical simulation of the decaying homogeneous isotropic turbulence, by comparing the results with those from the pseudo-spectral (PS) method.  Although the DNS of DHIT is easily achievable, it is the first and essential step to validate the DUGKS method before it is used to simulate more complex turbulent flows.

In our study, we first perform the DNS of DHIT using LBE, DUGKS and PS methods at two mesh resolutiones ($128^3$ and $256^3$) at $\text{Re}_{\lambda}=26.06$. In terms of accuracy, we first compare the instaneous flow fields. It is found that the instaneous velocity and vorticity fields predicted by the both the LBE and DUGKS methods are very similar to each other and agree reasonably well with the PS results. In addition, we compare some key statistic quantities, and find that both methods perform an accurate prediction on all the quantities of interest due to their low numerical dissipation. We also note that the DUGKS with a coarse mesh of $128^3$ underestimates the energy and dissipation rate spectra in the high wavenumber region, and yet, these discrepancies vanish with the fine mesh of $256^3$. This indicates that the DUGKS has a relatively large numerical dissipation compared with the LBE method, which can be attributed to the central difference employed in DUGKS to approximate the gradient at the cell interface. However, as the numerical results shown, this feature has little impact on the average kinetic energy and dissipate rate. Moreover, we observe that the results of skewness and flatness obtained by the DUGKS have high frequency oscillations due to the acoustic waves in the system.

The performance of the two methods at higher Reynolds numbers are also compared. Some key statistic quantities obtained by LBE and DUGKS methods are compared with those from the PS method.  The results show that good agreements are achieved between the LBE and the PS methods at both $\text{Re}_{\lambda}$,  but there are noticeable discrepancies between the results of DUGKS and PS methods due to the insufficient mesh resolution, which also indicates that the DUGKS is more dissipative than the LBE method.

In terms of the computational efficiency, the LBE method is about $36.8\%$ faster than the DUGKS per time step. It should be noted that although the DUGKS is less efficient than the LBE method on the same uniform mesh, as a finite volume method, the DUGKS can use non-uniform meshes without additional efforts for wall-bounded turbulence flows, such as a channel flow and pipe flow. For such flows, the mesh can be clustered near the walls where large flow gradients exist, and the computational efficiency can be largely improved, which will be presented in our subsequent work.  We also assess the numerical stability of the LBE and DUGKS methods by computing the maximum stable Taylor microscale Reynolds number on a fixed mesh without considering the accuracy. The results show that the DUGKS has a better numerical stability than the LBE method, which is consistent with the pervious results of laminar flows \cite{wang2015comparative}.

In conclusion, the LBE and DUGKS methods have similar accuracy for DNS of DHIT when the mesh resolution is sufficient to resolve the flow field, and the DUGKS has relatively larger numerical dissipation than the LBE; in addition, the DUGKS is less efficient than the LBE method with the same regular uniform mesh, but superior to the LBE method in terms of the numerical stability. This comparative study clearly demonstrated that the DUGKS method can serve as a viable kinetic method for DNS of turbulent flows. It must be emphasized that the main advantage of the DUGKS compared with the LBE method is that it can be implemented on non-uniform meshes easily, which we shall demonstrate in the subsequent study of  wall-bounded turbulent flows.

\section*{ACKNOWLEDGMENT}

The authors acknowledge the support by the National Natural Science Foundation of China (Grant No. 51125024).

\bibliography{DHIT}

\end{document}